\documentclass[11pt,letterpaper]{article}

\usepackage{geometry}
\usepackage{bbm}
\usepackage[toc,page]{appendix}
\usepackage{amsmath, amsthm, amsfonts, amssymb}
\usepackage{mathrsfs} % for calligraphic C
\usepackage{hyperref}
\usepackage{cleveref}
\usepackage{slashed}
\usepackage{dsfont}
\usepackage{wrapfig}
\usepackage{minibox}
\usepackage{graphicx}
\usepackage{graphics}
\usepackage{epsfig}
\usepackage{xcolor}
\usepackage{array}
\usepackage{tikz}
\usetikzlibrary{arrows,calc}
\usepackage{relsize}
\usetikzlibrary{decorations.pathmorphing,calc}
\usepackage{authblk}
\usepackage[labelfont=bf]{caption}
\usepackage{caption}
\usepackage{subcaption}

\def\a{\alpha}
\def\b{\beta}
\def\g{\gamma}

\def\d{\delta}
\def\ve{\varepsilon}
\def\m{\mu}
\def\n{\nu}

\def\L{\Lambda}
\def\r{\rho}
\def\s{\sigma}
\def\thintablerule{\hrule height0.4pt}

\def\bb{\boldsymbol \beta}

\def\gg{\boldsymbol \gamma}

\newcommand{\be}{\begin{equation}}
\newcommand{\ee}{\end{equation}}
\newcommand{\bea}{\begin{eqnarray}}
\newcommand{\eea}{\end{eqnarray}}
\newcommand{\bal}{\begin{aligned}}
\newcommand{\eal}{\end{aligned}}
\newcommand{\eq}[1]{Eq.~(\ref{#1})}
\newcommand{\fig}[1]{Fig.~\ref{#1}}

\newcommand{\sect}[1]{Section~\ref{#1}}
\newcommand{\tr}{\mathrm{tr}\,}

\numberwithin{equation}{section}

\allowdisplaybreaks

\begin{document}

%---------------------------------------------------------------------------------------------------------------------------------------------------
\tikzset{
    photon/.style={decorate, decoration={snake}, draw=black},
    electron/.style={draw=black, postaction={decorate},
        decoration={markings,mark=at position .55 with {\arrow[draw=black]{>}}}},
    gluon/.style={decorate, draw=black,
        decoration={coil,amplitude=4pt, segment length=5pt}} 
}
%---------------------------------------------------------------------------------------------------------------------------------------------------

\centerline{\LARGE RG flows in Non-Perturbative Gauge-Higgs Unification I}
\vskip .5cm
%\centerline{\LARGE }
%\vskip .5cm

\vskip 2 cm
\centerline{\large Nikos Irges and Fotis Koutroulis}
\vskip 1cm
\centerline{\it Department of Physics}
\centerline{\it National Technical University of Athens}
\centerline{\it Zografou Campus, GR-15780 Athens, Greece}
\centerline{\it e-mail: irges@mail.ntua.gr, fkoutroulis@central.ntua.gr}

\vskip 2.2 true cm
\thintablerule
\vskip 2.0ex

\centerline{\bf Abstract}
We initiate the continuum description of a non-perturbative 5d lattice Yang-Mills model with 4d boundaries using the $\ve$-expansion.
In its simplest version classically the bulk has an $SU(2)$ gauge symmetry and on the boundary there is an Abelian-Higgs system
with zero scalar potential.
In part I we compute the Renormalization Group flows and related quantities in a limit where the boundary decouples from the bulk.

\vskip 1.0ex\noindent
\vskip 2.0ex
\thintablerule

%\keywords{Renormalization Group; Field Theories in Higher Dimensions; Lattice Quantum Field Theory}

\newpage

\tableofcontents

\pagebreak

%---------------------------------------------------------------------------------------------------------------------------------------------------

%---------------------------------------------------------------------------------------------------------------------------------------------------
\section{Introduction}
%---------------------------------------------------------------------------------------------------------------------------------------------------

Non-Perturbative Gauge-Higgs Unification (NPGHU) is most simply realized in a five-dimensional (5d) 
Yang-Mills model defined on a hypercubic orbifold lattice, anisotropic in the fifth dimension
that we believe to represent a quite general version of quantum Brout-Englert-Higgs mechanisms.
It also involves some novel physics related to the behaviour of relativistic Quantum Field Theories in the vicinity of quantum phase transitions.
In its non-perturbative realization the lattice is periodic, practically infinite in the four-dimensional (4d) sense with lattice spacing $a_4$ and it has orbifold boundary conditions 
along the anisotropic direction, with lattice spacing $a_5$. The classical anisotropy parameter then is $\g=a_4/a_5$.
The resulting geometry consists of a 5d bulk with two 4d boundaries, located at the end-points of the fifth dimension.
The boundary conditions at the same time are such that the bulk $SU(2)$ (${\cal G}$ in general) gauge symmetry is broken at the boundaries down 
to a $U(1)$ (${\cal H}\subset {\cal G}$ in general) subgroup and a complex scalar $\phi$ (scalars in ${\cal G}/{\cal H}$ in some representation of ${\cal G}$ in general).
In addition, the lattice has a reflection symmetry around the midpoint of the fifth dimension about which it can be folded and then the limit
of an infinite fifth dimension can be taken. By taking the infinite volume limit we exclude phase transitions of the finite temperature type.
The dynamics is determined by the Wilson lattice plaquette action and by gauge invariant lattice operators.
In \cite{Maurizio} the phase diagram of this model was determined. It has three phases: a Confined phase, a Higgs phase and a Hybrid phase,
all separated by first order, quantum phase transitions. The Confined phase is the strong coupling phase in its 5d version.
The Hybrid phase is one where the boundary is deconfined while the bulk is confined.
The interesting thing about the Higgs phase is that the scalar potential responsible for spontaneous gauge symmetry breaking 
develops entirely as a quantum effect and the associated Higgs mass seems to be protected against uncontrollable quantum corrections.
Non-perturbatively in all phases the boundary remains coupled to the bulk even though this coupling may be weak in certain limits.
Our ultimate goal is to analyze the Renormalization Group (RG) flows involved in this setup, in a continuum language.
This is a hard task to do all at once and in part I of this work we are forced to take a certain limit in which the boundary decouples from the bulk.
As we will show this is a natural step when the lattice action is expanded classically in small lattice spacings and the expansion is truncated to lowest non-trivial order.
As a result, in the bulk we have a 5d $SU(2)$ Yang-Mills system and on the boundary a 4d, 
massless and free scalar QED (SQED) model that can be analyzed separately.
The phase diagrams for both cases is known. The boundary is expected to possess only 
a Gaussian fixed point while the bulk phase diagram for general $\g$ was determined 
via Monte Carlo (MC) simulations in \cite{KnechtliRago}:
it has only two phases, a Confined and a Coulomb phase separated again by a line of first order quantum phase transitions. 
A noteworthy property that the anisotropy brings
is that in the $\g < 1$ regime inside the Confined phase the lattice decomposes into weakly interacting 4d planes \cite{FuHolger}. This regime is sometimes called layered
and we will refer to it as "the Confined-layered phase", even though it is not strictly a different phase, it is part of the Confined phase. 

From the continuum point of view the boundary theory is more straightforward to describe, as it is a renormalizable theory, even though in a rather unusual limit where 
the scalar potential vanishes. Essentially it is the zero potential limit of the 4d Abelian-Higgs model
and usual regularization methods are trustworthy, such as Dimensional Regularization (DR). 
Due to the vanishing of the scalar potential though the loops of the boundary theory computed in DR are without scale
and further regularization is necessary. Here useful is the knowledge of the gauge invariant computation of the 1-loop Higgs potential in the Abelian-Higgs model \cite{IrgesFotis},
whose massless and free limit, after renormalization, defines the regulator for the massless, free SQED of the boundary.
The bulk is trickier because it involves a perturbatively non-renormalizable
5d gauge theory. Nonetheless, since the model can be put on a lattice where it seems to make perfect sense \cite{Orblat} especially in the vicinity of the phase transitions, it
is natural to ask if some continuum regularization can reproduce its observed properties. If this is possible then not only the elusive concept of non-perturbative renormalizability could be understood better but perhaps also new insights, hard o obtain from the lattice, could be perhaps gained.
One of the biggest obstacles in this program originates from the fact that the phase transitions involved are of first order.
To appreciate the issue we remind that continuum approaches allowing for systematic computations are 
often designed to be sensitive to critical surfaces on the phase diagram, thus to
second order phase transitions. This seems to suggest that the lattice and continuum results may be totally disconnected.
For concreteness we consider a main continuum tool in this respect, that of the $\ve$-expansion. In this approach, one computes in the context of
DR in $d = 4 -\ve$ dimensions and sets at the end $\ve=-1$ for $d=5$, $\ve=-2$ for $d=6$ and so forth.
In this sense this method lies somewhere between perturbation theory and non-perturbative methods and the hope is that it gives results that are not fake.
The reason why this method is interesting is that in its context a general 1-loop $\b$-function for 
a coupling $g$ has the form $\b = - c_t \ve g - c_q g^3$ with $c_t$ a tree level constant
and with $c_q$ its quantum counter-part. Then a balance between the tree level and quantum terms possibly arises and for the value
$g_*^2=-\ve c_t/c_q$ the 1-loop $\b$-function vanishes, indicating a point in the interior of the phase diagram where the system may become scale invariant.
Such points are known as Wilson-Fisher (WF) fixed points.
The usual attitude then is that in order that the system becomes exactly scale invariant on a WF point, the $\b$-function 
should vanish to all orders. Thus the 1-loop condition for $g_*$
has been extended to higher orders \cite{Morris}. This can be done without much effort because the quantum 
coefficients $c_q$ are independent of $d$ and the 4d coefficients are known 
at least up to four loops. Including these corrections modifies the 1-loop results numerically by approximately 30$\%$ but does not change the physics.
This is part of a program known as Asymptotic Safety \cite{Sannino}.

Our attitude here is slightly different. We take the vanishing of the 1-loop $\b$-function as an indication of the fact that certain quantum systems tend towards
scale invariance as a quantum phase transition is approached but we do not impose exact scale invariance, due to the non-perturbative fact that
the phase transitions involved are not necessarily of second order. If this is the case, as a quantum phase transition of first order is approached, from either side of
the phase transition the theory still tends to become scale invariant \cite{IrgesCorfu16} a fact reflected by the vanishing of its $\b$-functions to a certain order, 
but this does not mean that the $\b$-functions have to vanish to all orders. Thus in general no continuum limit can be taken 
and near the phase transition the effective theory is one with a finite cut-off.\footnote{We expect it to have a radiatively broken 
conformal symmetry as the mother 5d theory does not have any classical mass scale apart from the one associated 
with its gauge coupling $g_5$.} In addition, a finite cut-off could be independently imposed on the effective action by 
the presence of a Landau pole, as it is the case in a Higgs phase for example.
In the spirit of these two possible obstructions for exact scale invariance, we call our approach one of 'Weak Asymptotic Safety'.
In part I we are blind to the Higgs and Hybrid phases of the fully non-perturbative model of NPGHU and to the strict necessity for a finite cut-off on the phase transitions.
We see instead a limit where the Hybrid and Confined phases merge into a single Confined phase and the Higgs phase degenerates to a Coulomb phase,
with no Landau pole generated. The Weak Asymptotic Safety scenario, if realised, is due to the vanishing of the $\b$-function to a limited order.
Under its assumptions we will be able to compare the predictions from the $\ve$-expansion with the lattice MC results.
In the bulk and since in the Confined-layered phase the quantum evolution of the system is four-dimensional to a good approximation, 
we will be able to match two RG flow lines, one on each side of the phase transition, both ending on the same point on the line of phase transitions. 

At the technical level we define our classical effective Lagrangian as the naive continuum limit of the orbifold lattice constructed in \cite{Orblat}.
Starting from the lattice action we expand in small $a_4$ and $a_5$. This expansion generates an infinite number of terms, multiplied by increasing powers
of the lattice spacings. In part I of this work we truncate the expansion to the lowest non-trivial order that generates only classically marginal operators both
in the bulk and on the boundary. Then we quantize the Lagrangian using DR and the $\ve$-expansion. By setting $\ve=0$ on the boundary and $\ve=-1$ in the bulk
we will be able to compute $\b$-functions, anomalous dimensions and critical exponents and then the desired RG flows. 
In fact, the bulk phase diagram obtained from the $\ve$-expansion has the same qualitative features as the one obtained from lattice Monte Carlo simulations.
The attempt to match the phase diagrams more precisely is unlocked by setting the DR scale $\m$ equal to $F/a_4$. This is obviously correct dimensionally with the
non-trivial information about the relation between lattice and continuum scales hidden in the dimensionless quantity $F$ that non-perturbatively can be a complicated function of all the dimensionless couplings. 
Here we will assume that at leading order, to a good approximation $F$ is a (potentially $d$-dependent) constant. We argue that our results justify this assumption.
We interpret the shape of the line of phase transitions obtained by MC simulations to imply that the anisotropy $\g$ does not get renormalized by much and this will be our other working assumption.

As already implied above, some of the details of the calculations performed here could have been 
omitted at the cost of obscuring the important role of the anisotropy parameter and of the orbifold boundary conditions.
For instance, the derivation of the phase diagram in the context of the $\ve$-expansion in the presence of anisotropy has not been attempted before.
Another reason we present our calculations in some detail is that they set us up for part II of this work where 
the next to leading order, classically irrelevant operators generated by the lattice spacing expansion will be added. 
No already available results for $\b$-functions etc. exist in this case.
Then the boundary will not completely decouple from the bulk and the real orbifold phases, including the Higgs phase, will hopefully emerge \cite{Fotis2}. 

In section 2 we derive the classical continuum action from the lattice orbifold action. In section 3 we compute 1-loop diagrams
in Dimensional Regularization and in the $\xi=1$ gauge and then
extract $\b$-functions, anomalous dimensions and critical exponents. In section 4 we discuss the RG flows and attempt to match our results to the lattice.
In section 5 we compute and renormalize the Stress-Energy tensors. In section 6 we review our results.

%---------------------------------------------------------------------------------------------------------------------------------------------------
\section{The classical continuum action from the lattice action}\label{FOT}
%---------------------------------------------------------------------------------------------------------------------------------------------------

We define our classical, continuum action as the truncated expansion in a small lattice spacing of the orbifold lattice action defined in \cite{Orblat}, which we briefly review here.
We start by considering a Euclidean five-dimensional (5d) periodic hypercubic lattice, with an $SU(2)$ gauge symmetry. 
Then, the circle of the fifth dimension is projected by the discrete group $\mathbb{Z}_2$. This projection
identifies the upper semicircle with the lower semicircle and turns the circle into an (discretized) interval. The projection is also embedded non-trivially into the Lie group
so that at the endpoints of the interval only a $U(1)$ subgroup and a complex scalar from the 5d components of the $SU(2)$ gauge field survive. 
Thus, we obtain an orbifold lattice with an $SU(2)$ symmetry in the bulk and with the degrees of freedom of an Abelian-Higgs system living
on the 4d boundaries.

In the following we use capital Latin letters $M, N,\cdots = 0, 1, 2, 3, 5$ to denote the five-dimensional 
Euclidean or Minkowski index and small Greek letters $\m,\n,\cdots = 0,1,2,3$ to denote the four-dimensional part. 
We define the lattice coordinates as $n_M = (n_\m, n_5 )$ with $n_\m = 1,\cdots, L $ and $n_5 = 0,1,\cdots,N_5$. 
The values $n_5 =0, N_5$ correspond to the orbifold's fixed points.
The lattice gauge variables consist of the links $U({n_M,N}) \in SU(2)$
\bea\label{link1}
U(n_M,N) = e^{i a_N g_5 {\bf A}_N(n_M)}
\eea
located at $n_M$ and pointing in the direction $N$. 
The $a_N$ corresponds to the lattice spacing in the direction $N$. We will take $a_\m = a_4$ for all $\m$ but generically $a_5\ne a_4$.
$g_5$ is the 5d dimensionful continuum gauge coupling with mass dimension $[g_5]=-1/2$ and ${\bf A}_M \equiv A_M^A T^A$ 
is the Lie algebra valued gauge potential carrying the adjoint index $A$. The normalization of the generators is the usual ${\rm tr} \{T^A T^B\} = \frac{1}{2} \d^{AB}$.
We will take $L$ practically infinite and allow the numbers of lattice nodes 
$N_5 = \frac{\pi R}{a_5}$ in the fifth direction ($R$ is the radius of the projected parent circle) to be either finite or infinite.
To the order we are working here $R$ does not appear anywhere so we can think of it being infinite.

The orbifold condition on the links
\bea\label{orb.cond.}
(1 - \Gamma) U({n_M,N}) = 0, \hskip .25 cm \Gamma \equiv {\cal R} {\cal T}_g\, ,
\eea
is implemented on the periodic lattice by the reflection operator $\cal R$ and the group conjugation operator ${\cal T}_g$.
The combined projector satisfies $\Gamma^2=1$ and is a $\mathbb{Z}_2$ element.
The reflection operation acts on the nodes as
\bea
{\cal R} n_M = \bar n_M \equiv (n_\m, - n_5)
\eea
and on the gauge links as
\bea\label{R}
{\cal R} U({n_M,\n}) &=& U({\bar n_M,\n})  \nonumber\\
{\cal R} U({n_M,5}) &=& U({\bar n_M- a_5 \hat 5,5})
\eea
while the conjugation operator acts only on the gauge links, as
\bea\label{Tg}
{\cal T}_g U({n_M,N}) = g U({n_M,N}) g^{-1}
\eea
where $g^2$ is an element of the centre of $SU(2)$. We shall take $g = -i \sigma^3$. 
In order to ensure that $\Gamma^2 =1$ these operators must satisfy $[{\cal R},{\cal T}_g] =0 $ and ${\cal R}^2 = {\cal T}_g^2 =1$.

The orbifold lattice has a mirror symmetry around the central point in the fifth dimension, therefore it is sufficient to restrict ourselves on half of the lattice. 
Thus, we will deal with only one of the boundaries, and specifically the one at the $n_5 = 0$ fixed point.
On the boundary, the action of the reflection operator on the coordinates is trivial,
\bea
{\cal R} (n_\m, 0) = \bar n_M \equiv (n_\m, 0)\, ,
\eea
so that the orbifold condition \eq{orb.cond.} reads
\bea\label{f.p.orb.cond.}
U({(n_\m,0),\n}) = {\cal T}_g U({(n_\m,0),\n}) = g U({(n_\m,0),\n}) g^{-1} .
\eea
It is clear that generally the above constraints break a bulk gauge group $SU(N_C)$ with generators $T^A$ to a subgroup $\cal H$ on the boundary. 
Denoting the unbroken generators's index by $\a$ and those of broken by $\hat \a$, \eq{f.p.orb.cond.} implies that $[T^\a, g] = 0$. 
For $SU(2)$ which has the three generators $T^{A}=\frac{1}{2}\s^A, A=1,2,3$ with $\s^A$ the Pauli matrices 
we see that $g$ commutes only with $T^{\a = 3}$, thus at the boundary we are left with a $U(1)$ gauge symmetry. 
The two remaining generators $T^{\hat \a = 1,2}$ are broken.

We take for our lattice action the simple Wilson plaquette action, appropriately generalized to implement the structure of the orbifold lattice \cite{Orblat}. 
We denote by $U_{\m\n}^{U(1)}$ the 4d boundary plaquettes, constructed from links lying on the boundary 
and by $U_{\m5}^H$ the so called Hybrid plaquettes, with two links along the fifth dimension with one end at a fixed point and the other in the bulk. 
We denote by $U_{\m\n}^{SU(2)} \equiv U_{\m\n} $ and $U_{\m5}^{SU(2)} \equiv U_{\m5} $ the bulk plaquettes with their links lying only in the bulk.
Therefore, the anisotropic orbifold Wilson action can be separated into two parts, one that contains the boundary and hybrid plaquettes and one that contains the bulk plaquettes, according to 
\bea\label{orbifold_action}
S_{S^1/\mathbb{Z}_2} = S^{\rm b-H}_{S^1/\mathbb{Z}_2} + S^{\rm bulk}_{S^1/\mathbb{Z}_2} 
\eea
with 
\bea\label{b-H_action}
S^{\rm b-H}_{S^1/\mathbb{Z}_2} = \frac{1}{2N_C} \sum_{n_\m} \Biggl [ \b_4 \sum_{\m < \n} \frac{1}{2} \tr {\{ 1 - U_{\m\n}^{U(1) }(n_\m,0)\} } +  \b_5 \sum_{\m} \tr {\{ 1 - U_{\m5}^H(n_\m,0) \} }    \Biggr ]
\eea
and 
\bea\label{bulk_action}
S^{\rm bulk}_{S^1/\mathbb{Z}_2} = \frac{1}{2N_C} \sum_{n_\m,n_5} \Biggl [ \b_4 \sum_{\m < \n} \tr {\{ 1 - U_{\m\n}(n_\m,n_5)\} } +  \b_5 \sum_{\m} \tr {\{ 1 - U_{\m5}(n_\m,n_5) \} }    \Biggr ]
\eea
where $\b_4$ and $\b_5$ are dimensionless lattice couplings. 
Note here that a factor of $\frac{1}{2}$ in \eq{b-H_action}, will be cancelled by an extra factor of 2 due to the folding of the lattice about the midpoint of the extra dimension.
In the following we will often combine $\b_4$ and $\b_5$ and use an equivalent pair of couplings $\b$ and $\g$ determined as
\bea\label{b4b5}
\b_4 &=& \frac{2 N_C a_5}{g_5^2} = \frac{\b}{\g} \nonumber\\
\b_5 &=& \frac{2 N_C a_4^2}{a_5 g_5^2} =  \b \g \, 
\eea
for $SU(N_C)$. Even though we will mostly be concerned with $N_C=2$, when possible we will present expressions for general $N_C$. 
The alternative parametrization introduces the anisotropy parameter $\g=\sqrt{\b_5/\b_4}$ (whose classical value is $\g = \frac{a_4}{a_5}$ as already mentioned) and the
lattice coupling $\b = \sqrt{\b_5 \b_4} = \frac{2 N_C a_4}{g_5^2}$.\footnote{A warning concerning the lattice versus continuum notation: the anisotropy parameter $\g$ should not be confused with the
anomalous dimension of fields and operators, also denoted by $\g$. Also, the various lattice couplings denoted traditionally by $\b$ should not be confused with the notation for the continuum $\b$-functions.
To reduce the possibility of confusion we will denote $\b$-functions with the bold character $\bb$ and anomalous dimensions with $\gg$.
For example, in the next sections we will meet $\bb_{\b_{4,5}}$, the beta function of the lattice coupling $\b_{4,5}$ etc.}
%Also, in the above action there is a conventional relative weight $\frac{1}{2}$ for boundary plaquettes
%which will disappear once we fold the lattice about the central point of the fifth dimension.
Finally we stress that in \eq{orbifold_action} and \eq{bulk_action} there are three kinds of gauge links:
$U(1)$ boundary links, $SU(2)$ bulk links and hybrid links with one end on the boundary and the other in the bulk, 
with the unconventional gauge transformation $U \to \Omega^{U(1)}U\Omega^{\dag SU(2)}$ from where
their name originates. 

The next step is to expand \eq{b-H_action} and \eq{bulk_action} for small lattice spacings which after a truncation to the leading non-trivial order will lead us to a continuum classical action. 
Some of the following steps are standard in 4d lattice gauge theories but we review them anyway because eventually particularities of our 5d lattice will appear.

%---------------------------------------------------------------------------------------------------------------------------------------------------
\subsection{Boundary-Hybrid action}\label{b-H-action}
%---------------------------------------------------------------------------------------------------------------------------------------------------

$S^{\rm b-H}_{S^1/\mathbb{Z}_2}$ in \eq{b-H_action} can be split further into a boundary part and a hybrid part as
\bea
S^{\rm b-H}_{S^1/\mathbb{Z}_2} = S^b + S^H
\eea
with
\bea\label{b_action}
S^b =  \frac{\b_4}{2N_C} \sum_{n_\m} \sum_{\m < \n} \frac{1}{2} \tr {\{ 1 - U_{\m\n}^{U(1) }(n_\m,0)\} } 
\eea
and
\bea\label{H_action}
S^H = \frac{\b_5}{2N_C} \sum_{n_\m} \sum_{\m} \tr {\{ 1 - U_{\m5}^H(n_\m,0) \} }\, .
\eea
Let us first consider \eq{b_action} that contains links $U((n_\m,0),\n) \equiv U(n_\m,\n) = e^{ia_4 g_5 \bold{A_\n}(n_\m)}$
with $\bold{A_\m} \equiv A_\m^\a T^\a = A_\m^3 T^3$ and plaquettes of the form
\bea
U^{U(1)}_{\m\n}(n_\m) &=& U(n_\m,\m) U(n_\m + a_4 \hat \m,\n) U^\dag(n_\m + a_4 \hat \n,\m)U^\dag(n_\m,\n) \nonumber\\
&=& e^{ia_4 g_5 \bold{A_\m}(n_\m)} e^{ia_4 g_5 \bold{A_\n}(n_\m + a_4 \hat \m)} e^{-ia_4 g_5 \bold{A_\m}(n_\m + a_4 \hat \n)} e^{-ia_4 g_5 \bold{A_\n}(n_\m)}\, .
\eea
Using the Campbell-Baker-Housdorff formula we first obtain
\bea
U^{U(1)}_{\m\n}(n_\m) &=& \exp \Biggl[ i g_5 a_4 \Bigl \{ \bold{A_\m}(n_\m) + \bold{A_\n}(n_\m + a_4 \hat \m) - \bold{A_\m}(n_\m + a_4 \hat \n) - \bold{A_\n}(n_\m)     \Bigr\}      \nonumber\\
&-& \frac{a_4^2 g_5^2}{2}  \Bigl \{ [\bold{A_\m}(n_\m), \bold{A_\n}(n_\m + a_4 \hat \m)] + [\bold{A_\m}(n_\m), - \bold{A_\m}(n_\m + a_4 \hat \n)] \nonumber\\
&+& [\bold{A_\m}(n_\m), -\bold{A_\n}(n_\m )] + [ \bold{A_\n}(n_\m + a_4 \hat \m), - \bold{A_\m}(n_\m + a_4 \hat \n)] \nonumber\\
&+& [ \bold{A_\n}(n_\m + a_4 \hat \m), -\bold{A_\n}(n_\m )]  + [\bold{A_\m}(n_\m + a_4 \hat \n), \bold{A_\n}(n_\m )]   \Bigr\}   \Biggr]\, ,
\eea
which for small $a_4$ expands as 
\bea\label{UmnU1a}
U^{U(1)}_{\m\n}(n_\m) &=& \exp \Biggl[ i g_5 a_4 \Bigl \{ \bold{A_\m}(n_\m) + \bold{A_\n}(n_\m) + a_4 \hat \Delta_\m\bold{A_\n}(n_\m) - \bold{A_\m}(n_\m) \nonumber\\
&-& a_4 \hat \Delta_\n \bold{A_\m}(n_\m) - \bold{A_\n}(n_\m)     \Bigr\}   \nonumber\\
&-& \frac{a_4^2 g_5^2}{2}  \Bigl \{ [\bold{A_\m}(n_\m), \bold{A_\n}(n_\m)] + [\bold{A_\m}(n_\m), - \bold{A_\m}(n_\m)] \nonumber\\
&+& [\bold{A_\m}(n_\m), -\bold{A_\n}(n_\m )] + [ \bold{A_\n}(n_\m), - \bold{A_\m}(n_\m)] \nonumber\\
&+& [ \bold{A_\n}(n_\m), -\bold{A_\n}(n_\m )]  + [\bold{A_\m}(n_\m), \bold{A_\n}(n_\m )]   \Bigr\}  + O(a_4^2)  \Biggr] \nonumber\\
&=& \exp \Biggl[ i g_5 a_4^2 \Bigl \{ \hat \Delta_\m\bold{A_\n}(n_\m) - \hat \Delta_\n \bold{A_\m}(n_\m)   \Bigr\}  - a_4^2 g_5^2 [\bold{A_\m}(n_\m), \bold{A_\n}(n_\m)] + O(a_4^3)  \Biggr] \nonumber\\
&=& \exp \Biggl[ i g_5 a_4^2 \Bigl \{ \hat \Delta_\m\bold{A_\n}(n_\m) - \hat \Delta_\n \bold{A_\m}(n_\m)   \Bigr\} + O(a_4^3)  \Biggr]\, .
\eea
In the above we used that $\bold {A_\n}(n_\m + a \hat \m) \equiv  \bold{A_\n}(n_\m) + a \hat \Delta_\m\bold{A_\n}(n_\m)$ 
with $\hat \Delta_\m\bold{A_\n}(n_\m) = \frac{1}{a} [\bold {A_\n}(n_\m + a \hat \m) -  \bold{A_\n}(n_\m) ]$ a discrete derivative 
and that $[A_\m^3 T^3,A_\n^3 T^3] = 0$. Keeping only terms that are second order in $a_4$, \eq{UmnU1a} becomes 
\bea\label{UmnU1f}
U^{U(1)}_{\m\n}(n_\m) = e^{i g_5 a_4^2 \bold{F_{\m\n}}(n_\m) }
\eea
with $\bold{F_{\m\n}} = \hat \Delta_\m\bold{A_\n}(n_\m) - \hat \Delta_\n \bold{A_\m}(n_\m)$,
where $\bold{F_{\m\n}} = F^3_{\m\n} T^3$.
Substituting \eq{UmnU1f} into \eq{b_action} and expanding once more in small $a_4$ we obtain (after the folding that cancels a factor of 2)
\bea
S^b = \sum_{n_\m} \sum_{\m < \n}  \frac{\b_4}{2 N_C} \tr {\{  -i g_5 a_4^2 \bold{F_{\m\n}} + \frac{g_5^2 a_4^4}{2} \bold{F^2_{\m\n}} + O(a_4^8)+ \cdot \cdot \cdot    \} }\, .
\eea
After the trace and using \eq{b4b5} the action \eq{b_action} becomes
\bea\label{b_action_f}
S^b =\sum_{n_\m} a_4^4 \sum_{\m,\n}  \frac{a_5}{4} F^3_{\m\n}F^3_{\m\n} \, .
\eea
In the limit $a_4 \to 0$ the sum $\sum_{n_\m} a_4^4 \to \int d^4x$ and up to the multiplicative $a_5$ factor
we have a usual 4d continuum $U(1)$ gauge field action in infinite volume.

The procedure for $S^H$ is similar, with a difference originating from the definition of the links that it contains. 
In particular, a hybrid plaquette contains one $U(1)$ link (we denote the corresponding potential as ${\bf A_\m^a}$ below), one $SU(2)$ link and two hybrid (also $SU(2)$ valued)
links:
\bea\label{H_plaq.1}
U^H_{\m5} &=& e^{ia_4 g_5 \bold{A^a_\m}(n_\m,0)} e^{ia_5 g_5 \bold{A}_5(n_\m + a_4 \hat \m,0 )} e^{-ia_4 g_5 \bold{A}_\m(n_\m,a_5 \hat 5)} e^{-ia_5 g_5 \bold{A}_5(n_\m,0)} \nonumber\\
&=& \exp \Biggl [  i a_5 g_5 \Bigl \{  \g \bold{A^a_\m}(n_\m,0) +  \bold{A}_5(n_\m + a_4 \hat \m,0 ) - \g \bold{A_\m}(n_\m,a_5 \hat 5) - \bold{A}_5(n_\m,0)     \Bigr \} \nonumber\\
&-& \frac{g_5}{2} \Bigr \{ [a_4 \bold{A^a_\m}(n_\m,0), a_5  \bold{A}_5(n_\m + a_4 \hat \m,0 ) ] + [a_4 \bold{A^a_\m}(n_\m,0), -a_4 \bold{A_\m}(n_\m,a_5 \hat 5)] \nonumber\\
&+& [a_4 \bold{A^a_\m}(n_\m,0), - a_5 \bold{A}_5(n_\m,0)]  + [a_5 \bold{A}_5(n_\m + a_4 \hat \m,0 ), -a_4 \bold{A_\m}(n_\m,a_5 \hat 5)] \nonumber\\
&+& [a_5 \bold{A}_5(n_\m + a_4 \hat \m,0 ) , - a_5 \bold{A}_5(n_\m,0) ]  + [a_4 \bold{A_\m}(n_\m,a_5 \hat 5), a_5 \bold{A}_5(n_\m,0) ]    \Bigl\}            \Biggr ]\, .
\eea
As before, displaced fields are expanded as
\bea\label{expan.1}
\bold{A}_5(n_\m + a_4 \hat \m,0 ) \to \bold{A}_5(n_\m,0 ) + a_4 \hat \Delta_\m \bold{A}_5(n_\m,0 ) \, , \nonumber\\
\bold{A_\m}(n_\m,a_5 \hat 5) \to \bold{A_\m}(n_\m,0) + a_5 \hat \Delta_5 \bold{A_\m}(n_\m,0)\, .
\eea
To complete this calculation we have to express \eq{f.p.orb.cond.} at the gauge field level. By doing so, we obtain (we follow the notation of \cite{Mariano}) 
\bea\label{f_bound_cond.}
{\cal R}  \bold{A_\m} &\equiv& \a_\m A^A_\m T^A = \a_\m \eta^A A^A_\m T^A \Leftrightarrow \nonumber\\
A^A_\m T^A &=& \a_\m \eta^A A^A_\m T^A\, , \nonumber\\
{\cal R} \bold{A_5} &\equiv& \a_5 A^A_5 T^A = \a_5 \eta^A A^A_5 T^A \Leftrightarrow \nonumber\\
A^A_5 T^A &=& \a_5 \eta^A A^A_5 T^A
\eea
where $\a_\m = +1$, $\a_5 = -1$ are the parities of the $A_\m$ and $A_5$ fields respectively. Moreover, we have used the relation 
$ g T^A g^{-1} = \eta^A T^A$ (no sum on $A$) where $\eta^A = \pm 1$ is the parity of the generators with $\eta^a =1 $ and $\eta^{\hat a} = -1$. 
Thus, the Dirichlet boundary conditions \eq{f_bound_cond.} imply that on the boundary
\be
A^{1,2}_\m = 0 \hskip 0.5cm {\rm and}  \hskip 0.5cm A^3_5 = 0.
\ee
We also obtain the Neumann boundary conditions
\bea
\hat \Delta_5 A^3_\m =0, \hskip .1 cm \hat \Delta_5 A^{1,2}_5 =0\, 
\eea
and now \eq{expan.1} becomes
\bea\label{expan.2}
{A^A_5}(n_\m + a_4 \hat \m,0 ) &\to& {A^{1,2}_5}(n_\m,0 ) + a_4 \hat \Delta_\m {A^{1,2}_5}(n_\m,0 ) \, , \nonumber\\
{A^A_\m}(n_\m,a_5 \hat 5) &\to& {A^3_\m}(n_\m,0)\, .
\eea
Substituting these into \eq{H_plaq.1} and keeping only terms of order $O(a_4 a_5)$, we get
\bea
U^H_{\m5} = e^{ia_4 a_5 g_5 \sum_{\hat \a}{F^{\hat \a}_{\m5}}T^{\hat \a}}
\eea
with ${F^{1,2}_{\m5}} T^{1,2} = \hat \Delta_\m {A^{1,2}_5 T^{1,2}} + i g_5 [{A^3_\m T^3}, {A^{1,2}_5 T^{1,2}} ]$.
Defining a complex scalar field as $\phi = \frac{1}{\sqrt{2}}
(A_5^1 + i A_5^2)$ we get that 
\bea
{\rm tr} \sum_{\hat \a}{F^{\hat \a}_{\m5}}T^{\hat \a}{F^{\hat \a}_{\m5}} T^{\hat \a} \equiv 
 | \hat D_\m \phi|^2 \, , \hskip 0.5cm \hat D_\m = \hat \Delta_\m + i g_5 A_\m^3 \, .
\eea
Setting the above relations into the action \eq{H_action}, using \eq{b4b5} and keeping terms of order $O(a_4 a_5)$, the Hybrid action becomes
\bea\label{H_action_f}
S^H &=&  \sum_{n_\m} \frac{\b_5}{2N_C} \sum_{\m} a_4^2 a_5^2 g_5^2
|\hat D_\m \phi |^2 \Leftrightarrow \nonumber\\
S^H &=& \sum_{n_\m} a_4^4  \sum_{\m} a_5  |\hat D_\m \phi |^2\, .
\eea
The Hybrid action, similarly to \eq{b_action_f}, contains an extra multiplicative, dimensionful $a_5$ factor.
Adding \eq{b_action_f} and \eq{H_action_f}, the boundary-hybrid action $S^{\rm b-H}_{S^1/\mathbb{Z}_2}$ takes the final form 
\bea
S^{\rm b-H}_{S^1/\mathbb{Z}_2} = \sum_{n_\m} a_4^4  a_5 \Biggl [ \frac{1}{4} \sum_{\m,\n}   F^3_{\m\n}F^3_{\m\n}  +  \sum_{\m} |\hat D_\m \phi |^2  \Biggr ]\, , \nonumber
\eea 
to leading order in the lattice spacing expansion.
Note that the 4d scalar field $\phi$, which is a combination of the
components of the 5d gauge field $A_M$, plays the role of a Higgs-like field for a boundary observer.
In fact, the above action %\eq{b_H_action}
 is just the massless, free scalar QED (SQED) in 4d. 
Clearly, in the absence of a potential for $\phi$ there is no Higgs mechanism at the classical level.

Finally, notice that $S^{\rm b-H}_{S^1/\mathbb{Z}_2}$ contains both $a_4$ and $a_5$. As we have already mentioned, in the limit $a_4 \to 0$ the 
sum $\sum_{n_\m} a_4^4 \to \int d^4x$. On the other hand, there is no summation over $n_5$. 
Since at the expansion level that we work here bulk and boundary are decoupled, we can replace $a_5$ 
in $S^{\rm b-H}_{S^1/\mathbb{Z}_2}$%\eq{b_H_action}
 by $a_5 = \frac{\pi R}{N_5}$, independently of the bulk action.
 Recall also that the radius of the
fifth dimension is $R \to \infty$. Therefore, we can choose $N_5 \to (0, \rm finite \hskip .15cm \rm or \hskip 
.15cm \infty)$. Now, choosing $N_5 \to \infty$ indicates that $a_5 \to \hskip .1cm a_5^f $, where $a_5^f$ is an arbitrary
 finite and non-zero constant of mass dimension $[a_5^f] = -1$. The action $S^{\rm b-H}_{S^1/\mathbb{Z}_2}$ 
 therefore becomes
 
\bea\label{b_H_action}
S^{\rm b-H}_{S^1/\mathbb{Z}_2} = a_5^f \sum_{n_\m} a_4^4 \Biggl [ \frac{1}{4} \sum_{\m,\n}   F^3_{\m\n}F^3_{\m\n}  +  \sum_{\m} |\hat D_\m \phi |^2  \Biggr ]\, .
\eea

%---------------------------------------------------------------------------------------------------------------------------------------------------
\subsection{Bulk action}\label{Bulk_action}
%---------------------------------------------------------------------------------------------------------------------------------------------------

If it were not for the anisotropy there would be no extra calculation necessary for \eq{bulk_action} to leading order in the lattice spacing expansion.
Because of the presence of the anisotropy, we have to go through a few short steps.
It is easy to see that the bulk action can be separated further into a four-dimensional part and a five-dimensional part  
\bea
S^{\rm bulk}_{S^1/\mathbb{Z}_2} = S_{4d}  +  S_{5d}
\eea
with
\bea
S_{4d} = \frac{\b_4}{2N_C} \sum_{n_\m,n_5} \sum_{\m < \n} \tr {\{ 1 - U_{\m\n}(n_\m,n_5)\} } 
\eea
and 
\bea
S_{5d} = \frac{\b_5}{2N_C} \sum_{n_\m,n_5}  \sum_{\m} \tr {\{ 1 - U_{\m5}(n_\m,n_5) \} }  \, .
\eea
The 4d part is trivial as it is just the usual Yang-Mills action in 4d, multiplied by $\b_4$.
The plaquette is then 
\bea\label{Umn_f}
U_{\m\n}(n_\m,n_5) = e^{i g_5 a_4^2 \bold{F_{\m\n}}(n_\m,n_5) }\, ,
\eea  
with $\bold{F_{\m\n}} = F^A_{\m\n} T^A = \hat \Delta_\m\bold{A_\n}(n_\m,n_5) - \hat \Delta_\n \bold{A_\m}(n_\m,n_5) + i g_5 [\bold{A_\m},\bold{A_\n} ] $
and the action
\bea\label{bulk_action_f1}
S_{4d} = \frac{1}{2} \sum_{n_\m} a_4^4 \sum_{n_5} a_5  \sum_{\m,\n}  \frac{1}{2} F^A_{\m\n}F^A_{\m\n} \, .
\eea
Now in the continuum limit where the lattice spacings go to zero we obtain a five dimensional integral
since $\sum_{n_\m} a_4^4 \sum_{n_5} a_5 \to \int d^5 x$.

The plaquette along the fifth dimension is 
\bea
U_{\m5}(n_\m,n_5) =  e^{ia_4 g_5 \bold{A_\m}(n_\m,n_5)} e^{ia_5 g_5 \bold{A_5}(n_\m + a_4 \hat \m,n_5 )} e^{-ia_4 g_5 \bold{A_\m}(n_\m,n_5 + a_5 \hat 5)} e^{-ia_5 g_5 \bold{A_5}(n_\m,n_5)}\, , \nonumber
\eea
that finally leads to
\bea\label{bulk_action_f2}
S_{5d} = \frac{1}{2} \sum_{n_\m} a_4^4 \sum_{n_5} a_5  \sum_{\m}  \frac{1}{2} F^A_{\m5}F^A_{\m5}\, .
\eea
As expected, the combination of \eq{bulk_action_f1} and \eq{bulk_action_f2} is 5d covariant and the sum reconstructs to leading order into a 5d bulk action
\bea\label{f_bulk_action}
S^{\rm bulk}_{S^1/\mathbb{Z}_2} = \sum_{n_\m} a_4^4 \sum_{n_5} a_5  \sum_{M,N}  \frac{1}{4} F^A_{MN}F^A_{MN}\, .
\eea 
We note however that if there was some kind of dimensional reduction to 4d at work, we could have expressed \eq{bulk_action_f2}
as $\frac{1}{2}(D_\m \Phi)^A (D_\m \Phi)^A$ with $D_\m$ the $SU(2)$ gauge covariant derivative. Together with \eq{bulk_action_f1} this term
could be interpreted as a 4d gauge-adjoint Higgs system, with no classical potential though.

%---------------------------------------------------------------------------------------------------------------------------------------------------
\subsection{The leading order continuum action}\label{tloa}
%---------------------------------------------------------------------------------------------------------------------------------------------------

The classical scaling dimensions of the coupling and fields can be found in Appendix \ref{dim.analy.}. 
Before we take the continuum limit of \eq{b_H_action} and \eq{f_bulk_action} we perform the rescaling
\bea\label{rescaling}
\{ \bold{A}_\mu, \bold{A}_5 \} \to \frac{1}{\sqrt{a_5^f}} \{ \bold{A}_\mu, \bold{A}_5 \}%\bold{A}_M \to \frac{1}{\sqrt{a_5}} \bold{A}_M
\eea  
only on the boundary-Hybrid action, and now $\{ \bold{A}_\mu, \bold{A}_5 \}$ have dimension $[\bold{A}_\mu] = [\bold{A}_5] = 1$ as a usual gauge field in 4d. In particular, we get that%and now $\bold{A}_M$ has dimension $[\bold{A}_M] = 1$ as a usual gauge field in 4d. For the boundary-Hybrid action we get that 
\bea
(F^3_{\m\n})^2 \to \frac{1}{a_5^f} (F^3_{\m\n})^2, \hskip 1cm |\hat D_\m \phi|^2 \to  \frac{1}{a_5^f} |\hat D_\m \phi|^2 \nonumber
\eea
where now the covariant derivative takes the form $ \hat D_\m = \hat \Delta_\m + i \frac{g_5}{\sqrt{a_5^f}} A_\m^3 \equiv \hat \Delta_\m + i g \sqrt{\g} A_\m^3$
with the dimensionless coupling and the anisotropy factor %with the dimensionless coupling 
\be\label{g-g5}
g = \frac{g_5}{\sqrt{a_4}} \hskip .3 cm {\rm and } \hskip .3 cm \g = \frac{a_4}{a_5^f} \, ,
\ee
We can then bring the rescaled $S^{\rm b-H}_{S^1/\mathbb{Z}_2}$ action into a 5d action by
\bea\label{rescaled_b-H}
S^{\rm b-H}_{S^1/\mathbb{Z}_2} &=& \sum_{n_\m} a_4^4  {\cal L}_{\rm bound} \Leftrightarrow \nonumber\\
&=& \sum_{n_\m} a_4^4 \sum_{n_5} \frac{a_5}{a_5} \d(n_5) {\cal L}_{\rm bound} \Leftrightarrow \nonumber\\
S^{\rm b-H}_{S^1/\mathbb{Z}_2} &=& \sum_{n_\m} a_4^4 \sum_{n_5} a_5 \d(a_5 n_5) {\cal L}_{\rm bound}\, .
\eea
As we have already mentioned the bulk action remains untouched under the above rescaling with the field strength reading 
\be
F^A_{MN} = \hat \Delta_M A_N - \hat \Delta_N A_M- g_5 f^{ABC}A^B_M A^C_N\, ,
\ee 

and the bulk action still reading 
\bea\label{rescaled_bulk}
S^{\rm bulk}_{S^1/\mathbb{Z}_2} = \sum_{n_\m} a_4^4 \sum_{n_5} a_5 \sum_{M,N}  \frac{1}{4} F^A_{MN}F^A_{MN}\, .
\eea 

Now we are ready to take the naive continuum limit in \eq{rescaled_b-H} and \eq{rescaled_bulk} using $\sum_{n_\m} a_4^4 \sum_{n_5} a_5 \to \int d^5 x$, 
$\d (a_5 n_5) \to \d(x_5)$, $ \hat \Delta_M \to \partial_M $, $ \hat D_M \to D_M $ and $\hat p_M = \frac{2}{a} \sin \frac{a p_M}{2} \to p_M$. 
Switching to Minkowski space we finally obtain the five-dimensional continuum orbifold action
\bea\label{orb.cond.ac}
S_{S^1/\mathbb{Z}_2} = \int d^5 x \Biggl [  P(x_5) {\cal L}_{\rm bulk} + \d(x_5) {\cal L}_{\rm bound}   \Biggr ] \, ,
\eea 
where ${\cal L}_{\rm bulk} = -\frac{1}{4} F^A_{MN}F^A_{MN}  $, ${\cal L}_{\rm bound} =  - \frac{1}{4} F^3_{\m\n}F^3_{\m\n}  + |D_\m \phi |^2 $
and $ P (x_5) =  1- \delta(x_5) $.
The projection operators act in such a way that for $x_5=0$ they allow only ${\cal L}_{\rm bound}$ and in the bulk only $ {\cal L}_{\rm bulk}$ to survive.
The classical action \eq{orb.cond.ac} is invariant under the global symmetries Parity (P), Charge conjugation (C) and "Stick" symmetry ($\cal S$).
For the action of these symmetries see \cite{Symmetries}. In Appendix \ref{G.T.C.O.} we work out their action on the continuum fields.
This is the action that we will quantize in this paper.
Due to the continuum limit, at this order, the boundary action even though otherwise completely decoupled from, is suppressed with respect to the bulk action.
%Finally, we can move the $\L_5$ factor between ${\cal L}_{\rm bulk}$ and ${\cal L}_{\rm bound}$ at will by the rescaling \eq{rescaling} and its inverse,
%provided that we keep track of the resulting modifications in the dimensional analysis.

%---------------------------------------------------------------------------------------------------------------------------------------------------
\section{Quantum effects}
%---------------------------------------------------------------------------------------------------------------------------------------------------

In the following we quantize the classical action and compute the 1-loop corrections with the goal of extracting its $\b$-functions and anomalous dimensions.
We stress that we are in a 'zero temperature' context with no compact extra dimension and associated Kaluza-Klein states.
The gauge fixed action, in the $R_\xi$ gauge, is
\bea\label{g.f.orb.ac.}
S_{S^1/\mathbb{Z}_2} &=& \int d^5 x  \Biggl [   P (x_5) \Biggl \{  -\frac{1}{4} F^A_{MN}F^A_{MN} - \frac{1}{2\xi} (\partial_M A_M^A)^2 + \partial_M \bar c^C D^{CB}_M c^B \Biggr \} \nonumber\\
&+& \d(x_5)   \Biggl \{  - \frac{1}{4} F^3_{\m\n}F^3_{\m\n}  + |D_\m \phi |^2   - \frac{1}{2\xi} (\partial_\m A^3_\m)^2 + \partial_\m \bar c^3 \partial_\m c^3   \Biggr \} \Biggr ] \, .
\eea
In the bulk, $D_M^{CB} = \delta^{CB}\partial_M + g_5 f^{CBA}A_M^A$ and $c^B$, $ \bar c^C$ are the non-Abelian 
ghost and anti-ghost respectively. On the boundary only the third component of $c^A $ and $\bar c^A $ survive since
ghosts should respect the orbifold boundary conditions which read 
\bea
c^A(x_\m,-x_5) = \eta^A c^A(x_\m,x_5)  \nonumber
\eea
and at the boundary, at $x_5 = 0$, they give
\bea
c^a(x_\m,0) &=& + c^a(x_\m,0) = c^3(x_\m,0) \, , \nonumber\\
c^{\hat a}(x_\m,0) &=& - c^{\hat a}(x_\m,0) = - c^{1,2}(x_\m,0) = 0\, .
\eea
The gauge fixing term should also respect the corresponding boundary conditions giving 
\bea
\partial_M A_M^A (x_\m,-x_5) &=& \a_M \a_M \eta^A \partial_M A_M^A (x_\m,x_5) \nonumber
\eea
which means that 
\bea
\partial_\m A_\m^3 (x_\m,0) \ne 0, \hskip .2cm
\partial_5 A_\m^3 (x_\m,0) = 0\, . \nonumber
\eea
Therefore, it is not necessary to use separate gauge fixing and ghosts for the boundary action, since $\partial_\m A_\m^3$ and $c^3$, $\bar c^3$ are 
components of the bulk gauge fixing and ghosts terms.%\\Finally a useful relation for the following section is the bulk gauge fixed action, before the rescaling of the gauge field, which reads
%\bea\label{beforeresc.}
%S^{bulk} = \int d^5 x  \Biggl [  P (x_5) \Biggl \{  -\frac{1}{4} F^A_{MN}F^A_{MN} - \frac{1}{2\xi} (\partial_M A_M^A)^2 + \partial_M \bar c^C D^{CB}_M c^B \Biggr \} \Biggr ]
%\eea
%with $ F^A_{MN} = \partial_M A_N^A - \partial_N A_M^A - g_5 f^{ABC} A^B_M A^C_N  $ and $D^{CB}_M = \delta^{CB}\partial_M + g_5 f^{CBA}A_M^A $. A dimensional analysis for both \eq{g.f.orb.ac.} and \eq{beforeresc.} is performed in Appendix \ref{dim.analy.}.

The Feynman rules deriving from \eq{g.f.orb.ac.} are computed in Appendix \ref{F.rules}. 

%---------------------------------------------------------------------------------------------------------------------------------------------------
\subsection{1-loop diagrams}\label{CQE}
%---------------------------------------------------------------------------------------------------------------------------------------------------

At 1-loop level there are four sets of possible diagrams corresponding to one-, two-, three- and four-point functions named ${\cal T}^i_j$, ${\cal M}^i_j$, ${\cal K}^i_j$ and ${\cal B}^i_j$ respectively. 
Here, the superscript $i = A_M, G, A_\m, \phi$ indicates the field(s) running in the loop and the subscript $j = A_M, G, A_\m, \phi$ the external fields. 
Note that the label for the boundary gauge field is $A_\m^3\equiv A_\m$ and the label for a generic ghost is $G$.
The notation for the loop and external momenta is 
$k_M = \{ q_\m,k_5 \}$ and $p_M = \{ l_\m,p_5 \}$ respectively.
With these in mind, we present the 1-loop integrals obtained from the 5d orbifold action \eq{g.f.orb.ac.}. 
It is convenient to separate them to a boundary and a bulk part and evaluate them separately. In particular, the bulk 1-loop integrals will be functions of the coupling 
$g_5$, while the boundary ones will contain the coupling $g \sqrt{\g}$. %In particular, regarding 
%the bulk 1-loop integrals it is convenient to calculate them in the basis $ A'_M = \sqrt{\L_5} A_M $ and to use the relation of \eq{g-g5} to write final expressions as functions of the coupling 
%$g_5$ instead of $g$. 
The momentum space projector that appears in the following expressions is
\be
P(p_5) = 1 - \d_{p_5,0}\, .
\ee
We perform all calculations in the $\xi=1$ gauge.

The 1-point function (Tadpole) vanishes both in the bulk and the boundary by Lorentz and gauge invariance.
The next set of one-loop diagrams are the two-point functions ${\cal M}^i_j$.
The first diagram here is a two-leg Tadpole due to the gauge field self interaction:   
%-------------------------------------
\vskip .5cm
\begin{center}
\begin{tikzpicture}[scale=0.7]
\draw [photon] (0,0)--(1.8,0);
\draw [photon] (0,0.9) circle [radius=0.9];
\draw [photon] (-1.8,0)--(-0,0);
\node at (-2.5,0) {$R$};
\node at (2.5,0) {$S$};
\node at (-1.3,2) {$M$};
\node at (1.35,2) {$N$};
\node at (5,0) {$= i {\cal M}^{A_M}_{A_M,RS}$\, .};
\end{tikzpicture}
\end{center}
%-------------------------------------
It evaluates to
\bea\label{MAA1}
i{\cal M}^{A_M}_{A_M,RS} &=& i g^2 \g \int{ \frac{d^5 k}{(2 \pi)^5}} S^1_{{\cal M}} \Bigg \{ - P(k_5) K^{BCDE}_{MNRS} + 2 g_{\rho\sigma} \d_{k_5,0} \Delta_{MNRS}   \Biggr \} \times \nonumber\\
&&  i \Biggl \{ P(k_5) \frac{ \delta^{BC}}{ k^2} \Bigl (- g^{MN}  +  (1- \xi)\frac{k^M k^N}{k^2} \Bigr ) \nonumber\\
&+& \frac{\delta_{k_5,0}\delta_{M,\m}\delta_{N,\n}}{q^2} \Bigl (- g^{\m\n}  +  (1- \xi)\frac{q^\m q^\n}{q^2} \Bigr )+ \frac{\delta_{k_5,0} \delta_{M,5}\delta_{N,5}}{ q^2}         \Biggr \} \, ,
\eea
where the bulk/boundary symmetry factor is given by $S^1_{{\cal M}} = \frac{P(k_5)}{2}  + \delta_{k_5,0} $.

Taking $k_5 = 0$ (then $P(0) = 0$) projects the computation on the boundary.
On the boundary there are two possible external fields, $A_\m^3$ and $\phi$, so we have two different cases.
The first case corresponds to the choice $S=\sigma$, $R = \rho$, $M=5$ and $N=5$, giving  
%-------------------------------------
\vskip .5cm
\begin{center}
\begin{tikzpicture}[scale=0.7]
\draw [photon] (0,0)--(1.8,0);
\draw [dashed] (0,0.9) circle [radius=0.9];
\draw [photon] (-1.8,0)--(0,0);
\node at (0,1.8) {$ > $};
\node at (4,0) {$= i {\cal M}^{\phi}_{A_\m,\rho\sigma}$};
\end{tikzpicture}
\end{center}
%-------------------------------------
so according to that, \eq{MAA1} becomes  
\bea\label{MTDR}
{\cal M}^{\phi}_{A_\m,\rho\sigma} &=&- 2 g^2 \g g_{\rho\sigma} \int{ \frac{d^4 q}{i(2 \pi)^4}}  \Delta_{55\rho\sigma}  \frac{1}{ q^2} 
= - 2 g^2 \g g_{\rho\sigma}\int{ \frac{d^4 q}{i(2 \pi)^4}}  \frac{1 }{ q^2} \label{MphiA}
\eea
This is a massless Tadpole, zero in DR. Nevertheless, it is useful for the following section to convert it to a usual vacuum polarization integral. 
We have that
\bea
{\cal M}^{\phi}_{A_\m,\rho\sigma} &=& - 2 g^2 \g g_{\rho\sigma}\int{ \frac{d^4 q}{i(2 \pi)^4}}  \frac{q^2 }{ q^2(q+l)^2} \label{MphiA} 
= - 2 g^2 \g g_{\rho\sigma} g_{\m\n} B^{\m\n}(q,q+l) \label{MphiA}
\eea
Its contracted version 
\bea\label{MTL}
{\cal M}^{\phi}_{A_\m} &=& \frac{1}{3} \Bigl ( - g^{\rho \sigma} + \frac{l^\rho l^{\sigma}}{l^2}  \Bigr) {\cal M}^{\phi}_{A_\m,\rho\sigma} = 2 g^2 \g g_{\m\n} B^{\m\n}(q,q+l)
\eea
will be also useful below. The second case corresponds to $ S=5$, $R = 5$, $M=\m$ and $N=\n$:
%-------------------------------------
\vskip .5cm
\begin{center}
\begin{tikzpicture}[scale=0.7]
\draw [dashed] (0,0)--(1.8,0);
\draw [photon] (0,0.9) circle [radius=0.9];
\draw [dashed] (-1.8,0)--(0,0);
\node at (0.9,0) {$ > $};
\node at (-0.9,0) {$ > $};
\node at (4,0) {$= i {\cal M}^{A_\m}_{\phi}$};
\end{tikzpicture}
\end{center}
%-------------------------------------
and is equal to
\bea\label{MphiT}
{\cal M}^{A_\m}_{\phi} =- 2 g^2 \g g_{\m\n} \int{ \frac{d^4 q}{i(2 \pi)^4}}  \Delta_{\m\n55}  \frac{- g^{\m\n}}{ q^2} =  2 d g^2 \g \m^{4-d} \int{ \frac{d^d q}{i(2 \pi)^d}}  \frac{1}{ q^2} \, .
\eea
It is also a massless tadpole and thus zero in DR. Again, we convert it to a usual two-point function diagram
\bea\label{MphiL}
{\cal M}^{A_\m}_{\phi} = 8 g^2 \g  \int{ \frac{d^4 q}{i(2 \pi)^4}}  \frac{q^2}{ q^2(q+l)^2}=  8 g^2 \g g_{\m\n} B^{\m\n}(q,q+l)\, .
\eea
Next, we turn to the bulk where $k_5 \ne 0$ and $P(k_5 ) = 1$. The contributing diagram is
%-------------------------------------
\vskip .5cm
\begin{center}
\begin{tikzpicture}[scale=0.7]
\draw [photon] (0,0)--(1.8,0);
\draw [photon] (0,0.9) circle [radius=0.9];
\draw [photon] (-1.8,0)--(-0,0);
\draw [->]  (-1.7,0.25)--(-1,0.25);
\node at (-1.2,-0.4) {$p$};
\node at (-2.5,0) {$S,E$};
\node at (2.5,0) {$R,D$};
\node at (-1.3,2) {$M,B$};
\node at (1.35,2) {$N,C$};
\draw [->]  (0.45,0.6) arc [start angle=-45, end angle=-135, radius=0.6cm];
\node at (0,0.8) {$k$};
\draw [->]  (1.7,0.25)--(1,0.25);
\node at (1.2,-0.4) {$p$};
\node at (5,0) {$= i {\cal M}^{A_M}_{A_M,RS}$};
\end{tikzpicture}
\end{center}
%-------------------------------------
that evaluates to
\bea\label{MAA2}
{\cal M}^{A_M}_{A_M,RS} = - \frac{g_5^2}{2} \int{ \frac{d^5 k}{i(2 \pi)^5}} K^{BCDE}_{MNRS} \frac{ \delta^{BC}}{ k^2}  g^{MN}\, ,
\eea 
where here and in the following we use \eq{g-g5} combined with $ \g = \frac{a_4}{a_5^f} $. %and $\L_5 = \frac{1}{a_5}$.
\\From \eq{KBCDE} the contraction $K^{BCDE}_{MNRS}$ with the propagator gives 
\bea
- \d^{BC} K^{BCDE}_{MNRS}   g^{MN} =  2 {\cal C}_A \d^{DE} (1-d) g_{RS}  \nonumber
\eea 
where we have used that $f^{ABD}f^{ABE} =  {\cal C}_A \d^{DE}$.
So, \eq{MAA2} becomes 
\bea\label{MAA3}
{\cal M}^{A_M}_{A_M,RS} &=& g_5^2 {\cal C}_A (1-d) g_{RS} \m^{5-d}\int{ \frac{d^d k}{i(2 \pi)^d}}  \frac{ \delta^{DE}}{k^2} \, .
\eea
This is a vanishing in DR massless tadpole. In other regularization schemes though it may be non-zero. To obtain a useful for such a case expression, the common trick is 
to multiply the integrand by $ \frac{(k+p)^2}{(k+p)^2} $: 
\bea
{\cal M}^{A_M}_{A_M,RS} &=& g_5^2 {\cal C}_A (1-d) g_{RS} \m^{5-d}\int{ \frac{d^d k}{i(2 \pi)^d}}  \frac{ \delta^{DE}(k^2 + 2k \cdot p + p^2)}{k^2(k+p)^2}\nonumber\\
&=& g_5^2 {\cal C}_A (1-d) g_{RS} \delta^{DE} \Bigl [ g_{MN} B^{MN} (k, k + p) + 2 p_M B^M (k, k + p) + p^2 B_0(k, k + p)    \Bigr ] \, . \nonumber
\eea
Then using Passarino-Veltman (PV) reduction formulae and setting $d=5$ it becomes
\bea\label{MA3fin.}
{\cal M}^{A_M}_{A_M,RS} &=& - 4 g_5^2 {\cal C}_A g_{RS} \delta^{B B'} g_{MN} B^{MN} (k, k + p) \, .
\eea
The next contribution to the two-point function is  
%-------------------------------------
\vskip .5cm
\begin{center}
\begin{tikzpicture}[scale=0.7]
\draw [photon] (1,0)--(3,0);
\draw [photon] (0,0) circle [radius=1];
\draw [photon] (-3,0)--(-1,0);
\node at (-1.3,1.3) {$M$};
\node at (1.3,1.3) {$M'$};
\node at (-1.3,-1.3) {$R$};
\node at (1.3,-1.3) {$R'$};
\node at (-3,0.4) {$N$};
\node at (3,0.4) {$N'$};
\node at (5.5,0) {$=\, \, i{\cal M}^{A_M A_M}_{A_M,N N'}$};
\end{tikzpicture}
\end{center}
%-------------------------------------
given by  
\bea\label{MAAA1}
i {\cal M}^{A_M A_M}_{A_M,N N'} &=& g^2 \g \int{ \frac{d^5 k}{(2 \pi)^5}} S^2_{{\cal M}} \Bigg [ P(k_5) L^{ABC}_{MNR} + i \d_{k_5,0} Q_{MNR}   \Biggr ]  \Bigg [ P(k_5) L^{A'B'C'}_{M'N'R'} + i \d_{k_5,0} Q_{M'N'R'}   \Biggr ] \times \nonumber\\
&&i \Pi_{A A'}^{M M'}(k_M,q_\m) i \Pi_{C C'}^{R R'}(k_M+ p_M,q_\m + l_\m)\, , \nonumber\\
\eea
with $S^2_{{\cal M}} = \frac{P(k_5)}{2} + \d_{k_5,0}$.
It is convenient to separate at this point the boundary from the bulk. 

Starting from the boundary where $P(k_5 = 0) = 0$, \eq{MAAA1} gives
\bea\label{MAAA2}
{\cal M}^{boun.}_{N N'} = g^2 \g \int{ \frac{d^4 q}{(2 \pi)^4 i}} Q_{MNR}  Q_{M'N'R'}  \Pi_{\m \m'}(q_\n) \Pi_{\rho \rho'}(q_\n + l_\n)\, .
\eea
\eq{MAAA2} corresponds to further two cases. 

The first is when $(M ,N, R) = (5 ,\n, 5) $, $(M' ,N', R') = (5 ,\n', 5) $, corresponding to the diagram
%-------------------------------------
\vskip .5cm
\begin{center}
\begin{tikzpicture}[scale=0.7]
\draw [photon] (1,0)--(3,0);
\draw [dashed] (0,0) circle [radius=1];
\draw [photon] (-3,0)--(-1,0);
\node at (0,1) {$ > $};
\node at (0,-1) {$ < $};
\node at (5.5,0) {$=\, i{\cal M}^{\phi \phi}_{A_\m, \n \n'}$\, .};
\end{tikzpicture}
\end{center}
%-------------------------------------
This diagram is the vacuum polarization of $A_\m$ in massless SQED and it is non-zero zero.
However, this is not a quantum correction to the mass of the gauge boson but a correction to the gauge coupling. 
Using \eq{QMNR} that gives $Q_{5 \n 5} = 2 q_\n + l_\n $, $Q_{5 \n' 5} = 2 q_{\n'} + l_{\n'} $, \eq{MAAA2} becomes
\bea\label{MphiphiA}
{\cal M}^{\phi \phi}_{A_\m, \n \n'} &=& g^2 \g \int{ \frac{d^4 q}{(2 \pi)^4 i}} \frac{( 2 q_\n + l_\n )}{q^2} \frac{( 2 q_{\n'} + l_{\n'} )}{(q + l)^2}\, .
\eea
This integral can be simplified by using the massless limit of standard PV formulae, see for example the Appendix of \cite{IrgesFotis}.
After the reduction, we obtain the expression
\bea\label{MphiphiA2}
{\cal M}^{\phi \phi}_{A_\m, \n \n'} &=& g^2 \g \Bigl [ 4 B_{\n \n'}(q, q+l) + 2 l_\n B_{\n'}(q, q+l) + 2 l_{\n'} B_{\n}(q, q+l) + l_{\n} l_{\n'} B_{0}(q, q+l)  \Bigr ]\, . \nonumber\\ 
\eea
 Finally, the contracted version of \eq{MphiphiA2} is given by   
\bea\label{MphiphiAf}
{\cal M}^{\phi \phi}_{A_\m} &=& \frac{1}{3} \Bigl ( - g^{\n \n'} + \frac{l^\n l^{\n'}}{l^2}  \Bigr) {\cal M}^{\phi \phi}_{A_\m, \n \n'} = \frac{l^2 g^2 \g}{3} B_0(q, q+ l) -  \frac{4}{3} g_{\m\n} B^{\m\n}(q,q+l) \, , \nonumber\\
\eea
where $ g_{\m\n} B^{\m\n}(q,q+l) $ corresponds to a massless tadpole integral, vanishing in DR.

The second case is when $(M ,N, R) = (5 ,5, \rho) $, $(M' ,N', R') = (5 ,5, \rho) $, where now the relevant diagram is
%-------------------------------------
\vskip .5cm
\begin{center}
\begin{tikzpicture}[scale=0.7]
\draw [dashed] (-2.3,0)--(-1,0);
\draw [dashed] (1,0)--(2.3,0);
\draw [dashed] (-1,0)--(1,0);
\node at (-1.6,0) {$ > $};
\node at (1.6,0) {$ > $};
\node at (0,0) {$ > $};
\node at (4.5,0) {$=\, i {\cal M}^{\phi A_\m}_\phi $\, . };
\draw  [photon] (-1,0) .. controls (-1,0.555) and (-0.555,1) .. (0,1)
.. controls (0.555,1) and (1,0.555) .. (1,0);
\end{tikzpicture}
\end{center}
%-------------------------------------
Here, \eq{QMNR} gives $Q_{5 5 \rho} = 2 l_\rho + q_\rho $, $Q_{5 5 \rho} = 2 l_{\rho'} + q_{\rho'} $ and \eq{MAAA2} becomes
\bea\label{MphiAphi}
{\cal M}^{\phi A_\m}_\phi &=& g^2 \g \int{ \frac{d^4 q}{(2 \pi)^4 i}} \frac{ -g^{\rho \rho'} }{q^2} \frac{( 2 l_\rho + q_\rho )( 2 l_{\rho'} + q_{\rho'})}{(q + l)^2}\, .
\eea
It can be reduced as 
\bea\label{MphiAphif.}
{\cal M}^{\phi A_\m}_\phi &=& 4 g^2 \g \Bigl [ - l^2 B_0 (q, q+l)  - l^\m B_\m (q, q+l) - \frac{1}{4} g^{\m\n}B_{\m\n}(q, q+l)   \Bigr]\nonumber\\
&=& l^2 g^2 \g \Bigl [ - 2 B_0 (q, q+l) - g_{\m\n} B^{\m\n}(q,q+l) \Bigr]\nonumber\\
&=& l^2 g^2 \g M_\phi(l^2)  \, , 
\eea
where $M_\phi(l^2)  =  - 2 B_0 (q, q+l) - g_{\m\n} B^{\m\n}(q,q+l) $ and with $g_{\m\n} B^{\m\n}(q,q+l)$ vanishing in DR.
A comment before we move on is that on boundary there is no contribution from the ghost fields since they are decoupled. 
It is straightforward to check that the scalar field stays massless and the above correction contributes only to the anomalous dimension of $\phi$.

In the bulk $P(k_5) =1$ and there is only one diagram:
%-------------------------------------
\vskip .5cm
\begin{center}
\begin{tikzpicture}[scale=0.7]
\draw [photon] (1,0)--(3,0);
\draw [photon] (0,0) circle [radius=1];
\draw [photon] (-3,0)--(-1,0);
\draw [->]  (-1.9,0.2)--(-1.2,0.2);
\node at (-1.7,-0.4) {$p_M$};
\node at (1.7,-0.4) {$p_M$};
\draw [<-]  (0.45,0.75) arc [start angle=45, end angle=135, radius=0.6cm];
\node at (0,0.55) {$k+p$};
\draw [->]  (0.45,-0.55) arc [start angle=-45, end angle=-135, radius=0.6cm];
\node at (0,-0.4) {$k$};
\node at (-1.3,1.3) {$M,A$};
\node at (1.3,1.3) {$M',A'$};
\node at (-1.3,-1.3) {$R,C$};
\node at (1.3,-1.3) {$R',C'$};
\node at (-3,0.4) {$N,B$};
\node at (3,0.4) {$N',B'$};
\draw [<-]  (1.9,0.2)--(1.2,0.2);
%\node at (2.6,0.2) {$-p$};
\node at (5.5,0) {$=\, \, i{\cal M}^{A_M A_M}_{A_M,N N'}$};
\end{tikzpicture}
\end{center}
%-------------------------------------
\eq{MAAA1} in this case reads
\bea\label{MAAA3}
{\cal M}^{A_M A_M}_{A_M,N N'} &=& - \frac{g_5^2}{2} \int{ \frac{d^5 k}{(2 \pi)^5 i}} L^{ABC}_{NRM} L^{A'B'C'}_{N'R'M'} \Pi_{A A'}^{M M'}(k_M,q_\m) \Pi_{C C'}^{R R'}(k_M+ p_M,q_\m + l_\m) \nonumber\\
&=& - \frac{g_5^2}{2} \int{ \frac{d^5 k}{(2 \pi)^5 i}} f^{ABC} f^{A'B'C'} \d_{A A'} \d_{C C'}  G_{NRM}  G_{N'R'M'} \frac{- g^{M M'} }{k^2} \frac{- g^{R R'} }{( k + p )^2} \, .\nonumber\\
\eea 
Identifying $G_{MNR}$ from \eq{GMNR} for $ p_{12} = p - k $, $ p_{23} = 2 k + p $ and $ p_{31} = - k - 2 p $ we have the identity
\bea
G_{NRM}  G_{N'R'M'} g^{M M'}  g^{R R'}  &=& - (-6 +4 d) k_{N}k_{N'} - (-3 +2 d) k_{N}p_{N'} - (-3 +2 d) p_{N}k_{N'}  \nonumber\\
&-& (-6 + d) p_{N}p_{N'} - 2 g_{N N'} k^2 - 2 g_{N N'} k \cdot p - 5 g_{N N'} p^2  \nonumber
\eea 
and by following the reduction program \eq{MAAA3} takes the final form
\bea\label{MAAAf}
{\cal M}^{A_M A_M}_{A_M,N N'} &=& \frac{g_5^2}{2} {\cal C}_A \d^{B B'} \Bigl [ (-6 + 4 d) B_{N N'}(k,k+p) + (-3 +2 d) p_{N'} B_{N}(k, k+ p)  \nonumber\\
&+& (-3 +2 d) p_{N} B_{N'}(k, k+ p) + (-6 + d) p_N p_{N'} B_0(k, k + p) + 2 g_{N N'} g_{AB} B^{AB}(k, k + p) \nonumber\\
&+& 2 g_{N N'} p_{M} B^{M}(k, k + p) + 5 g_{N N'} p^2 B_0(k, k + p)    \Bigr ] \, ,
\eea
where again we have used that $f^{ABC} f^{A'B'C'} \d_{A A'} \d_{C C'} = {\cal C}_A \d^{B B'}$.
Of course, the vacuum polarization diagrams of the bulk are not yet complete since
they admit contributions from the ghost fields. The ghost contribution to the vacuum polarization is
%-------------------------------------
\vskip .5cm
\begin{center}
\begin{tikzpicture}[scale=0.7]
\draw [photon] (1,0)--(3,0);
\draw [] (0,0) circle [radius=1];
\draw [photon] (-3,0)--(-1,0);
\node at (0,1.02) {$ > $};
\node at (0,-0.95) {$ < $};
\draw [->]  (-1.9,0.2)--(-1.2,0.2);
\node at (-1.7,-0.4) {$p_M$};
\node at (1.7,-0.4) {$p_M$};
\node at (0,0.5) {$k+p$};
\node at (0,-0.4) {$k$};
\node at (-1.1,1.2) {$A$};
\node at (1.1,1.2) {$A'$};
\node at (-1.1,-1.2) {$C$};
\node at (1.1,-1.2) {$C'$};
\node at (-3,0.4) {$N,B$};
\node at (3,0.4) {$N',B'$};
\draw [<-]  (1.9,0.2)--(1.2,0.2);
%\node at (2.6,0.2) {$-p$};
\node at (5.5,0) {$=\, \, i{\cal M}^{G G}_{A_M,N N'}$};
\end{tikzpicture}
\end{center}
%-------------------------------------
and it evaluates to
\bea
i{\cal M}^{G G}_{A_M,N N'} &=& (- 1 ) g_5^2  f^{BAC} f^{B' C' A'} \int{ \frac{d^5 k}{(2 \pi)^5 }} S^3_{\cal M} \frac{i k_N \d_{A A'}}{ k^2} \frac{i ( k + p)_{N'} \d_{C C'}}{ (k + p)^2}\, ,
\eea
where the $(-1)$ factor comes from the fact that ghosts are anti-commuting fields. Notice here that the symmetry factor is $S^3_{\cal M}  =1 $ for 
$k_5 \ne 0 $ and that $f^{BAC} f^{B' C' A'} \d_{A A'} \d_{CC'}= f^{BAC} f^{B' C A} = - f^{BAC} f^{B' A C} = - {\cal C}_A \d^{BB'}$.
The final form of the above integral is
\bea\label{MGGAf}
{\cal M}^{G G}_{A_M,N N'} &=& - g_5^2 {\cal C}_A \d^{B B'} \Bigl [ B_{N N'}(k,k+p) + p_{N'} B_{N}(k,k+p)  \Bigr ]\, .
\eea
The complete vacuum polarization in the bulk is given by the sum of \eq{MA3fin.}, \eq{MAAAf} and \eq{MGGAf}:
\bea\label{MAfin.1}
{\cal M}_{A, MN} &=&  g_5^2  {\cal C}_A \d^{A B} \Bigl [ (-4 + 2d ) B_{M N}(k,k+p) + \frac{(- 5 + 2d )}{2} p_{N} B_{M}(k, k+ p)  \nonumber\\
&+& \frac{(- 3 + 2d )}{2} p_{M} B_{N}(k, k+ p) + \frac{(-6 + d )}{2} p_M p_{N} B_0(k, k + p) + 5 g_{M N} g_{AB} B^{AB}(k, k + p) \nonumber\\
&+& g_{M N} p_{A} B^{A}(k, k + p) + \frac{5}{2} g_{M N} p^2 B_0(k, k + p)    \Bigr ] \, 
\eea
which for $d = 5$ becomes
\bea\label{MAfin.}
{\cal M}_{A, MN} &=&  g_5^2  {\cal C}_A \d^{A B} \Bigl [ 6 B_{M N}(k,k+p) + \frac{5}{2} p_{N} B_{M}(k, k+ p)  \nonumber\\
&+& \frac{7}{2} p_{M} B_{N}(k, k+ p) - \frac{1}{2} p_M p_{N} B_0(k, k + p) + 5 g_{M N} g_{AB} B^{AB}(k, k + p) \nonumber\\
&+& g_{M N} p_{A} B^{A}(k, k + p) + \frac{5}{2} g_{M N} p^2 B_0(k, k + p)    \Bigr ] \, .
\eea
The last one-loop two-point function diagram in the bulk corresponds to the vacuum polarization of the ghost propagator given by
%-------------------------------------
\vskip .5cm
\begin{center}
\begin{tikzpicture}[scale=0.7]
\draw [] (-2.3,0)--(-1,0);
\draw [] (1,0)--(2.3,0);
\draw [] (-1,0)--(1,0);
\node at (-1.6,0) {$ > $};
\node at (1.6,0) {$ > $};
\node at (0,0) {$ > $};
\draw [->]  (-1.9,0.3)--(-1.2,0.3);
\node at (-1.7,-0.4) {$p_M$};
\node at (1.7,-0.4) {$p_M$};
\node at (0,0.7) {$k$};
\node at (0,-0.4) {$k+p$};
\node at (-1.1,1.5) {$M,A$};
\node at (1.1,1.5) {$N,A'$};
\node at (-1,-1.2) {$C$};
\node at (1,-1.2) {$C'$};
\node at (-3,0.4) {$B$};
\node at (3,0.4) {$B'$};
\draw [<-]  (1.9,0.3)--(1.2,0.3);
\node at (5,0) {$=\, i {\cal M}^{G A_M }_G $\, . };
\draw  [photon] (-1,0) .. controls (-1,0.555) and (-0.555,1) .. (0,1)
.. controls (0.555,1) and (1,0.555) .. (1,0);
\end{tikzpicture}
\end{center}
%-------------------------------------
Of course ghosts are not physical degrees of freedom nevertheless this contribution will play a role in the renormalization program 
of the bulk Lagrangian, performed in the next section. This diagram evaluates to
\bea
i{\cal M}^{G A_M }_G &=& g_5^2 f^{ABC} f^{A' B' C'} \int{ \frac{d^5 k}{(2 \pi)^5 }} S^4_{\cal M} p_{M}(k+p)_N\frac{- i g^{M N} \d_{A A'}}{ k^2} \frac{i \d_{C C'}}{ (k + p)^2} \, ,
\eea
with symmetry factor $S^4_{\cal M} = 1$ for $k_5 \ne 0 $.
Finally, the above integral admits the reduced form
\bea\label{MGAGf}
{\cal M}^{G A_M }_G &=& g_5^2 {\cal C}_A \d^{B B'} \frac{p^2}{2} B_0(k, k + p)\, .
\eea
With this contribution the calculation of the ${\cal M}^i_j$ diagram set ends.

Now, let us move on to the ${\cal K}^i_j$ one-loop diagrams which correct the cubic vertices of the boundary and the bulk. 
In general there are two types, corresponding to reducible and irreducible diagrams called Triangles. The two types have the form
\bea
%-------------------------------------
\begin{tikzpicture}[scale=0.7]
\draw [] (1,0)--(3,-1.5);
\draw [] (1,0)--(3,1.5);
\draw [] (0,0) circle [radius=1];
\draw [] (-3,0)--(-1,0);
\end{tikzpicture}
%-------------------------------------
\, , \hskip .3 cm
%-------------------------------------
\begin{tikzpicture}[scale=0.7]
\draw [] (0.9,-0.5)--(3,-1.5);
\draw [] (0.9,0.5)--(3,1.5);
\draw [] (0,0) circle [radius=1];
\draw [] (-3,0)--(-1,0);
\end{tikzpicture}
%-------------------------------------
\nonumber
\eea
Let us first discuss the boundary. 
Diagrams with three $A_\m^3$ external fields should be all zero by gauge invariance. We have checked that this is indeed the case.
Another important case on the boundary is when we have Triangle diagrams with three external $\phi$ fields.
Such a diagram, if non-zero, would be a quantum contribution to the Higgs potential that is absent at the classical level. 
Appendix \ref{F.rules} shows that there is no possible way to connect the Feynman rules so as to obtain a Triangle diagram contributing to a scalar cubic vertex. 
Actually, this was expected since the existence of a vertex with three complex scalars as external legs, would violate charge conservation. 
%Thus, it is forbidden at the quantum level for a cubic term to appear in the effective scalar potential of massless SQED.

There are two non-zero Triangle diagrams on the boundary correcting the gauge-$\phi$ vertex. The first one is given by  
%-------------------------------------
\vskip .5cm
\begin{center}
\begin{tikzpicture}[scale=0.7]
\draw [dashed] (-2.3,0)--(-1,0);
\draw [dashed] (1,0)--(2.3,1);
\draw [photon] (1,0)--(2.3,-1);
\draw [dashed] (-1,0)--(1,0);
\node at (2,1.5) {$ l_1 $};
\node at (2,-1.5) {$ l_2 $};
\node at (-1.9,0.5) {$ l $};
\node at (0,1.5) {$q+l$};
\node at (0,-0.4) {$q$};
\draw [->, very thick ] (-1.6,0) -- (-1.5,0);
\draw [->, very thick ] (0,0) -- (0.2,0);
\draw [->, very thick ] (1.65,0.45)--(1.85,0.65);
\node at (4.5,0) {$=\, i {\cal K}^{\phi A}_{\phi A \phi,\m} $\, , };
\draw  [photon] (-1,0) .. controls (-1,0.555) and (-0.555,1) .. (0,1)
.. controls (0.555,1) and (1,0.555) .. (1,0);
\end{tikzpicture}
\end{center}
%-------------------------------------
where $ l + l_1 + l_2 = 0$.
It is equal to
\bea
i {\cal K}^{\phi A}_{\phi A \phi,\m} &=& 2 i g^3 \g^{3/2} i g^{\m\n} \int{ \frac{d^4 q}{(2 \pi)^4}} i \frac{ 2 l_{\n'} + q_{\n'} }{q^2} \frac{- i g_{\n \n'}}{(q+l)^2}\, .
\eea
Following the appropriate reduction formulae it can be written as
\bea\label{KphiAf}
{\cal K}^{\phi A}_{\phi A \phi,\m} &=& - 3 g^3 \g^{3/2} l_\m B_0(q, q + l)\, .
\eea
The second Triangle diagram correcting the gauge-scalar vertex is given by
%-------------------------------------
\vskip .5cm
\begin{center}
\begin{tikzpicture}[scale=0.7]
\draw [dashed] (-1,0) -- (1,1);
\draw [dashed] (-1,0) -- (1,-1);
\draw [photon] (1,1) -- (1,-1);
%%%
\draw [photon] (-2,0) -- (-1,0);%p1
\draw [dashed] (1,1) -- (2.5,1);%p3
\draw [dashed] (1,-1) -- (2.5,-1);%p4
%%%
\node at (1.7,1.5) {$l_2$};
%\draw [<-] (1.3,0.8) -- (1.9,0.8);
\node at (1.7,-1.5) {$l$};
%\draw [<-] (1.3,-0.8) -- (1.9,-0.8);
\node at (-1.9,0.5) {$l_1$};
%\draw [->] (-1.6,0.2) -- (-1.1,0.2);
%%%
%\draw [->] (-0.4,0.1) -- (0.2,0.4);
\node at (-0.4,1) {$q + L_1$};
%\draw [<-] (-0.4,-0.1) -- (0.2,-0.4);
\node at (-0.1,-1) {$q$};
%\draw [<-] (0.8,-0.3) -- (0.8,0.3);
\node at (2.2,0) {$q + L_2$};
%%%
\draw [->, very thick] (0.2,0.6) -- (0,0.5);
\draw [<-, very thick] (0.2,-0.6) -- (0,-0.5);
\draw [->, very thick] (1.7,1) -- (1.5,1);
\draw [->, very thick] (1.5,-1) -- (1.7,-1);
\node at (6,0) {$=\,\, i {\cal K}^{\phi A \phi }_{\phi A \phi,\m}  $\, ,};
\end{tikzpicture}
\end{center}
%-------------------------------------
where $ l_1 + l_2 + l = 0$ and $L_1 = l_1 $, $ L_2 = l_1 +l_2 $.
Its explicit expression is
\bea
i {\cal K}^{A \phi \phi }_{A \phi \phi,\m} &=& - i g^3 \g^{3/2} \int{ \frac{d^4 q}{(2 \pi)^4 }} i  \frac{ (2q +  l_1 )_\m}{q^2} \frac{ (q + l_1 +l_2  )_\n }{(q + L_1)^2} \frac{ (q +  l )_\r }{( q + L_2)^2} g^{\n\r}\, .\nonumber\\
\eea
Using momentum conservation to simplify the calculation its reduced form is
\bea\label{KAphiphif}
{\cal K}^{A \phi \phi }_{A \phi \phi,\m} &=& g^3 \g^{3/2} \Bigl [ l_\m B_0(q, q+l_2) - 2 l^2 C_\m(q,q+ L1,q+ L_2 ) - l_{1,\m} l^2 C_0(q,q+ L1,q+ L_2 ) \Bigr ]\, . \nonumber\\
\eea
The total correction to the gauge-scalar vertex is given by the sum of \eq{KphiAf} and \eq{KAphiphif}:
\bea\label{Kfin.a}
{\cal K}_{A \phi \phi,\m} &=& g^3 \g^{3/2} \Bigl [ - 3 l_\m B_0(q, q + l) + l_\m B_0(q, q+l_2) \nonumber\\
&-&  2 l^2 C_\m(q,q+ L1,q+ L_2 ) -  l_{1,\m} l^2 C_0(q,q+ L1,q+ L_2 ) \Bigr ]\, .
\eea

Next we discuss the bulk, where things are quite different. In particular, recall that in the bulk lies a non-Abelian gauge theory, 
so there is a non-trivial classical cubic gauge-field vertex. Likewise there are Triangle diagrams with external gauge fields which are non-zero and correct this vertex. 
Notice also that all the vertices of the bulk action \eq{g.f.orb.ac.} contain a single coupling constant. Therefore, the renormalization program 
of the bulk can be done by evaluating only the Triangle diagrams that contribute as quantum corrections to the gauge-ghost vertex. 
This is technically simpler and 1-loop corrections come only from two irreducible diagrams. For loop momenta, we adopt the notation of \cite{IrgesFotis} 
where $P_1 = p_1$, $P_2 = p_1 + p_2$ and $ p_1 + p_2 + p = 0 $.
The first diagram
%-------------------------------------
\vskip .5cm
\begin{center}
\begin{tikzpicture}[scale=0.7]
\draw [] (-1,0) -- (1,1);
\draw [] (-1,0) -- (1,-1);
\draw [photon] (1,1) -- (1,-1);
%%%
\draw [photon] (-3,0) -- (-1,0);%p1
\draw [] (1,1) -- (3,1);%p3
\draw [] (1,-1) -- (3,-1);%p4
%%%
\node at (2.1,1.5) {$p_2$};
%\draw [<-] (1.3,0.8) -- (1.9,0.8);
\node at (2.1,-1.5) {$p$};
%\draw [<-] (1.3,-0.8) -- (1.9,-0.8);
\node at (-1.9,0.5) {$p_1$};
\node at (-3.3,0.4) {$A,M$};
\node at (3.3,1.4) {$B$};
\node at (3.3,-1.4) {$C$};
\node at (1,1.4) {$N$};
\node at (1,-1.4) {$R$};
%\draw [->] (-1.6,0.2) -- (-1.1,0.2);
%%%
%\draw [->] (-0.4,0.1) -- (0.2,0.4);
\node at (-0.4,1) {$k + P_1$};
%\draw [<-] (-0.4,-0.1) -- (0.2,-0.4);
\node at (-0.1,-1) {$k$};
%\draw [<-] (0.8,-0.3) -- (0.8,0.3);
\node at (2.2,0) {$k + P_2$};
\draw [->, very thick] (0.2,0.6) -- (0,0.5);
\draw [<-, very thick] (0.2,-0.6) -- (0,-0.5);
\draw [->, very thick] (1.7,1) -- (1.5,1);
\draw [->, very thick] (1.5,-1) -- (1.7,-1);
%%%
\draw [->] (1.7,1) -- (1.5,1);
\draw [->] (1,-1) -- (1.7,-1);
\node at (6,0) {$=\,\, i {\cal K}^{GAG }_{AGG,M}  $};
\end{tikzpicture}
\end{center}
%-------------------------------------
is equal to
\bea
i {\cal K}^{GAG }_{AGG,M} &=& - g_5^3 f^{AED} f^{BB'F} f^{CC'F'} \int{ \frac{d^5 k}{(2 \pi)^5  }} k_M (k + P_1)_N p_R \frac{i \d^{EB'}}{k^2} \frac{i \d^{DC'}}{(k + P_1)^2} \frac{i \d^{F F'}}{(k+P_2)^2} g^{NR}\Rightarrow \nonumber\\
{\cal K}^{GAG }_{AGG,M} &=& g_5^3 \frac{{\cal C}_A}{2} f^{ABC} \int{ \frac{d^5 k}{(2 \pi)^5 i }} \frac{k_M ( k \cdot p + p \cdot p_1 ) }{k^2 (k+P_1)^2 (k+P_2)^2}\, ,
\eea
where we have used the identity $i f^{AB'D} i f^{BB'F} i f^{CDF} = i \frac{{\cal C}_A}{2} f^{ABC}$.
Using the usual reduction formula the above integral takes the final form
\bea\label{KGAGf}
{\cal K}^{GAG}_{AGG,M} &=&  g_5^3 \frac{{\cal C}_A}{2} f^{ABC} \Bigl[ p^N C_{MN}(k, P_1,P_2 ) + p \cdot p_1 C_M (k,P_1,P_2)   \Bigr]\, .
\eea 
The other diagram that corrects the gauge-ghost vertex is 
%-------------------------------------
\vskip .5cm
\begin{center}
\begin{tikzpicture}[scale=0.7]
\draw [photon] (-1,0) -- (1,1);
\draw [photon] (-1,0) -- (1,-1);
\draw [] (1,1) -- (1,-1);
%%%
\draw [photon] (-3,0) -- (-1,0);%p1
\draw [] (1,1) -- (3,1);%p3
\draw [] (1,-1) -- (3,-1);%p4
%%%
\node at (2.1,1.5) {$p_2$};
\node at (2.1,-1.5) {$p$};
\node at (-1.9,0.5) {$p_1$};
\node at (-0.4,1) {$k + P_1$};
\node at (-0.1,-1) {$k$};
\node at (2.2,0) {$k + P_2$};
\node at (-3.3,0.4) {$A,M$};
\node at (3.3,1.4) {$B$};
\node at (3.3,-1.4) {$C$};
\node at (1,1.4) {$N$};
\node at (1,-1.4) {$R$};
%%%
\draw [->, very thick] (1,0.2) -- (1,0);
\draw [->, very thick] (1.7,1) -- (1.5,1);
\draw [->, very thick] (1.5,-1) -- (1.7,-1);
\node at (6,0) {$=\,\, i {\cal K}^{AGA }_{AGG,M}  $\, ,};
\end{tikzpicture}
\end{center}
%-------------------------------------
and its explicit form is
\bea
i {\cal K}^{AGA }_{AGG,M} &=& g_5^3 f^{AED} f^{BD'F} f^{CE'F'} \int{ \frac{d^5 k}{(2 \pi)^5  }} G_{MNR} (k - p)_{N'} p_{R'}   \frac{-i g^{R R'} \d^{EE'}}{k^2} \nonumber\\
&\times& \frac{-i g^{N N'} \d^{DD'}}{(k + P_1)^2} \frac{i \d^{F F'}}{(k+P_2)^2} \Rightarrow \nonumber\\
{\cal K}^{AGA }_{AGG,M} &=& - g_5^3 \frac{{\cal C}_A}{2} f^{ABC} \int{ \frac{d^5 k}{(2 \pi)^5 i }} \frac{ G_{MNR} (k - p)^N p^R }{k^2 (k+P_1)^2 (k+P_2)^2}\, ,
\eea
where here $G_{MNR}$ is given in \eq{GMNR} for $ p_{12} = p_1 - k $, $ p_{23} = 2 k + p_1 $ and $ p_{31} = - 2 p_1 - k $.
Its reduced form is
\bea\label{KAGAf}
{\cal K}^{AGA}_{AGG,M} &=& - g_5^3 \frac{{\cal C}_A}{2} f^{ABC} \Bigl[ p^N C_{MN}(k, P_1,P_2 )  - p_M B_0(k, k + p_2 ) \nonumber\\
&+& ( p \cdot p_1 -2 p^2 )C_M (k,P_1,P_2) + 2 p_M ( p_N -  p_{1,N} )C^N (k,P_1,P_2) \nonumber\\
&+& p_{1,M} p_N C^N(k,P_1,P_2) + ( p \cdot p_1 p_M  - p^2 p_{1,M} )C_0 (k,P_1,P_2)   \Bigr]\, .
\eea  
The total contribution to the gauge-ghost vertex is the sum of \eq{KGAGf} and \eq{KAGAf}:
\bea\label{Kfin.}
{\cal K}_{AGG,M} &=& g_5^3 \frac{{\cal C}_A}{2} f^{ABC} \Bigl[  p_M B_0(k, k + p_2 ) + 2 p^2 C_M (k,P_1,P_2) - 2 p_M ( p_N -  p_{1,N} )C^N (k,P_1,P_2) \nonumber\\
&-&  p_{1,M} p_N C^N(k,P_1,P_2) - ( p \cdot p_1 p_M  - p^2 p_{1,M} )C_0 (k,P_1,P_2)   \Bigr]\, .
\eea
\eq{Kfin.} along with \eq{KphiAf} and \eq{KAphiphif} are the contributions needed 
to carry out the renormalization program of both the boundary and bulk Lagrangians. 

There are non-vanishing Box diagrams both on the boundary and in the bulk.
The role of these boxes can be quite subtle in some cases and for this reason we compute them explicitly on the boundary, 
in a separate Appendix \ref{BDS}. We will make some related comments in the following.

%---------------------------------------------------------------------------------------------------------------------------------------------------
\subsection{Renormalization, $\b$-functions and anomalous dimensions}\label{ROOA}
%---------------------------------------------------------------------------------------------------------------------------------------------------

Boundary and bulk are decoupled to the order that we are working and renormalization can be carried out separately for the two sectors.
The process needs some care in both cases because on the boundary we have a massless, free SQED and in the bulk a 5d $SU(2)$ gauge theory which
is perturbatively non-renormalizable. In particular for the latter an issue is that Gamma functions have no poles in odd dimensions which makes 
dimensional regularization delicate. A way around this is to perform the calculations in $d=4-\ve$ dimensions and set $\ve =-1$ in the end. This version of DR
is called $\ve$-expansion, whose validity is not always guaranteed. What makes this approach interesting is that we know from lattice Monte Carlo 
simulations the phase diagram of the bulk theory \cite{KnechtliRago} so we have a robust check. On the other hand, massive, free SQED has issues of renormalizability due to the
divergent Box diagrams that we compute in Appendix \ref{BDS}. In the massive case we do not know of a smooth resolution to the problem of absorbing these
divergences in counter-terms, except from the ad hoc counter-term proposed by Salam in \cite{salam}. For the massless limit however that is realized 
on the orbifold's boundary the dangerous integrals are scaleless of the $0/0$ type, whose limit is regularizable. We define this limit so that the resulting
renormalized theory coincides with the massless, free limit of the gauge invariant 1-loop Abelian-Higgs model presented in \cite{IrgesFotis}.

Having assumed that the anisotropy parameter $\g$ does not get renormalized at this order, 
for both the boundary and the bulk actions we have to renormalize only one coupling and the corresponding fields. 
In the following to simplify some expressions, following \cite{Morris}, we will use the auxiliary dimensionless 
couplings\footnote{From this point on several old and new (not necessarily independent) couplings will appear. To facilitate the reader we list them in Appendix \ref{AppCoupl}.}
\be\label{a-g2}
\a_{4,0} \equiv  \frac{1}{(4 \pi)^2} \m^{d-4} g_0^2
\ee 
on the boundary and
\be\label{a-g52}
\a_{5,0} \equiv \frac{2 N_C}{(4 \pi)^2}  \m^{d-4} g_{5,0}^2
\ee 
in the bulk. We introduce the counterterm  
\bea\label{g0}
g_0 &=& g + \d g \Rightarrow \nonumber\\
g_0 &=& g ( 1 + \frac{\d g}{g} ) = ( 1+ \d_{g} ) g = Z_{g} g 
\eea
%with $ g' = g \sqrt{\g} $, 
for the boundary gauge coupling and
\bea\label{g50}
g_{5,0} &=& g_5 + \d g_5 \Rightarrow \nonumber\\
g_{5,0} &=& g_5 ( 1 + \frac{\d g_5}{g_5} ) = ( 1+ \d_{g_5} ) g_5 = Z_{g_5} g_5 \, ,
\eea
for the bulk gauge coupling. In $d$-dimensions the scale independence relations of the bare couplings
\bea
\m \frac{d g_0 }{d \m } &=& \m \frac{d ( 4 \pi \m^{ \frac{4-d}{2} } \sqrt{\a_{4,0}} ) }{d \m } = 0 \label{bG4} \nonumber\\
\m \frac{d g_{5,0} }{d \m } &=& \m \frac{d ( 4 \pi \m^{ \frac{4-d}{2} } \sqrt{\frac{ \a_{5,0}}{2 N_C} } )}{d \m } = 0 \label{bG5}
\eea
generate the $\b$-function equations.
Finally, for the anomalous dimensions of the fields we define
\bea
\phi_0  &=& \sqrt{Z_\phi} \phi = \sqrt{1 + \d_\phi} \phi \label{phi0}\\
A_{\m,0}  &=& \sqrt{Z_{A_\m}} A_\m = \sqrt{1 + \d_{A_\m}} A_\m  \label{Am0}\\
A_{M,0}  &=& \sqrt{Z_{A_M}} A_M = \sqrt{1 + \d_{A_M}} A_M  \label{AM0} \\
c^A_0  &=& \sqrt{Z_{c^A}} c^A = \sqrt{1 + \d_{c^A}} c^A \label{cA0}\\
\bar c^A_0  &=& \sqrt{Z_{c^A}} \bar c^A = \sqrt{1 + \d_{c^A}} \bar c^A \label{barcA0} 
\eea
where the subscript $0$ indicates bare quantities.

Let us start from the bare boundary Lagrangian given by
\bea\label{Lb.b.}
{\cal L}_{\rm bound,0} &=& -\frac{1}{4}F^3_{\m\n,0 } F^3_{\m\n,0} + \partial_\m \bar \phi_0 \partial_\m  \phi_0   - \frac{1}{2} (\partial_\m A^3_{\m,0})^2 + \partial_\m \bar c_0^3 \partial_\m c_0^3  \nonumber\\
&+& i g_0 \sqrt{\g} A_{\m,0}^3  \Bigl (  \phi_0 \partial_\m \bar \phi_0  -   \bar \phi_0  \partial_\m \phi_0  \Bigr) + g_0^2 \g (A_{\m,0}^3)^2  \bar \phi_0 \phi_0 
\eea
where substituting \eq{g0}, \eq{phi0} and \eq{Am0} we obtain
\bea
{\cal L}_{\rm bound,0} = {\cal L}_{R} + {\cal L}_{\rm count.} \nonumber
\eea 
with $ {\cal L}_{R}$ the renormalized Lagrangian
\bea
{\cal L}_R &=& -\frac{1}{4}F^3_{\m\n} F^3_{\m\n} + \partial_\m \bar \phi \partial_\m  \phi   - \frac{1}{2} (\partial_\m A^3_\m)^2 + \partial_\m \bar c^3 \partial_\m c^3  \nonumber\\
&+& i g \sqrt{\g} A_\m^3  \Bigl (  \phi \partial_\m \bar \phi  -   \bar \phi  \partial_\m \phi  \Bigr) + g^2 \g (A_\m^3)^2 \bar \phi \phi \, ,
\eea
and $ {\cal L}_{\rm count.} $ the counter-term Lagrangian 
\bea\label{Lcount.}
 {\cal L}_{\rm count.} &=& \frac{1}{2} \Bigl \{ - \d_{A_\m} g_{\m\n} l^2 + {\cal M}^{\phi\phi}_{A_\m,\m\n}  \Bigr \} A^{3,\m} A^{3,\n} + \Bigl \{ \d_\phi l^2 + {\cal M}^{\phi A_\m}_{\phi}  \Bigr \} \phi \bar \phi \nonumber\\
&+& \Bigl \{g \sqrt{\g} \d_3 l_\m  + {\cal K}_{A \phi \phi,\m}  \Bigr \} A^3_\m \phi \bar \phi + \Bigl \{ \d_4 g^2 \g + {\cal B}_{A \phi A \phi}  \Bigr \} (A_\m^3)^2 \phi \bar \phi \, .
\eea
For the gauge-scalar three- and four-point vertices the following relations hold:
\bea\label{d3}
Z_3 &=& Z_{g}  Z_\phi \sqrt{Z_{A_\m}}  \Rightarrow \nonumber\\
1 + \d_3 &=& ( 1 + \d_{g} ) ( 1 + \d_\phi ) ( 1 + \frac{1}{2} \d_{A_\m} ) \Rightarrow \nonumber\\
\d_3 &=& \d_{g} + \d_\phi + \frac{1}{2} \d_{A_\m}
\eea
and 
\bea\label{d4}
Z_4 &=& Z_{g}^2  Z_\phi Z_{A_\m}  \Rightarrow \nonumber\\
1 + \d_4 &=& ( 1 + 2 \d_{g} ) ( 1 + \d_\phi ) ( 1 + \d_{A_\m} ) \Rightarrow \nonumber\\
\d_4 &=& 2 \d_{g} + \d_\phi + \d_{A_\m} 
\eea
respectively. Ghosts are completely decoupled in SQED, thus there is no need to renormalize them.
The Feynman rules for the counter-terms deriving from \eq{Lcount.} are
\begin{itemize}
\item Gauge boson 2-point function
%-------------------------------------
\begin{center}
\begin{tikzpicture}
\draw[photon] (-3,0)--(-0.5,0) ;
\draw [thick] [fill=black] (-1.7,0) circle [radius=0.1];
\node at (2,0) {$=\displaystyle
 - i \d_{A_\m} g_{\m\n} l^2$};
\end{tikzpicture}
\end{center}
%------------------------------------- 
\item Scalar 2-point function
%-------------------------------------
\begin{center}
\begin{tikzpicture}
\draw[dashed] (-1,0)--(1.5,0) ;
\draw [thick] [fill=black] (0.25,0) circle [radius=0.1];
\node at (4,0) {$=\displaystyle i \d_\phi l^2 $};
\end{tikzpicture}
\end{center}
%------------------------------------- 
\item The $A_\m$-$\phi$-$\bar \phi$ counterterm vertex
%-------------------------------------
\begin{center}
\begin{tikzpicture}[scale=0.7]
\draw [dashed] (-2.5,1.5)--(-1,0);
\draw [dashed] (-2.5,-1.5)--(-1,0);
\draw[photon] (-1,0)--(1,0);
\draw [thick] [fill=black] (-1,0) circle [radius=0.1];
\draw [<-, very thick] (-1.5,0.55) -- (-1.7,0.75);
\draw [->, very thick] (-1.5,-0.55) -- (-1.7,-0.75);
\node at (4,0) {$=  \displaystyle i g \sqrt{\g} \d_3 l_\m $};
\end{tikzpicture}
\end{center}
%------------------------------------- 
\item The $A_\m$-$A_\n$-$\phi$-$\bar \phi$ vertex counterterm
%-------------------------------------
\begin{center}
\begin{tikzpicture}[scale=0.7]
\draw [dashed,thick] (0,0)--(1.5,1.4);
\draw [dashed,thick] (0,0)--(1.5,-1.4);
\draw [photon] (-1.5,1.4)--(0,0);
\draw [photon] (-1.5,-1.4)--(0,0);
\draw [thick] [fill=black] (0,0) circle [radius=0.1];
\draw [<-, very thick] (0.65,0.55) -- (0.85,0.75);
\draw [->, very thick] (0.65,-0.55) -- (0.85,-0.75);
\node at (4,0) {$=\displaystyle 2 i g_{\m\n} \d_4 {g}^2 \g $\, .};
\end{tikzpicture}
\end{center}
\end{itemize}
The renormalization conditions needed to make the theory finite at 1-loop are in order. 
For the gauge boson propagator, diagrammatically, we have that  
\be
%-------------------------------------
\begin{tikzpicture} [scale=0.9]
\draw [photon,thick] (-2.3,0)--(-1.2,0);
\draw [thick] [fill=gray] (-0.5,0) circle [radius=0.8];
\draw [photon,thick] (0.3,0)--(1.3,0);
\node at (2,0) {$+$};
\draw [photon,thick] (2.5,0)--(3.6,0);
\draw [thick] [fill=black] (3.7,0) circle [radius=0.1];
\draw [photon,thick] (3.8,0)--(4.9,0);
\node at (6,0) {$=\, \, 0$};
\end{tikzpicture}
%------------------------------------- 
\nonumber
\ee
This implies that the contracted gauge propagator satisfies
\bea\label{R.c.da}
- \frac{1}{3} \left(g_{\m\n} -\frac{l_\m l_\n}{l^2}   \right) ( -  \d_{A_\m} g_{\m\n} l^2) + {\cal M}^{\phi\phi}_{A_\m} = 0 \, . 
\eea
The second condition demands that 
\be
%-------------------------------------
\begin{tikzpicture} [scale=0.9]
\draw [dashed] (-2.3,0)--(-1.2,0);
\draw [thick] [fill=gray] (-0.5,0) circle [radius=0.8];
\draw [dashed] (0.3,0)--(1.3,0);
\node at (2,0) {$+$};
\draw [dashed] (2.5,0)--(3.6,0);
\draw [thick] [fill=black] (3.7,0) circle [radius=0.1];
\draw [dashed] (3.8,0)--(4.9,0);
\node at (5.8,0) {$=\, \, 0 $};
\end{tikzpicture}
%------------------------------------- 
\nonumber
\ee
which, as equation, reads
\bea\label{R.c.dphi}
\d_\phi l^2 + {\cal M}^{\phi A_\m}_{\phi} = 0 \, .
\eea 
Finally, the condition for the three-point vertex demands
\be
%-------------------------------------
\begin{tikzpicture} [scale=0.9]
\draw [dashed] (0.3,0.4)--(1.1,0.9);
\draw [dashed] (0.3,-0.4)--(1.1,-0.9);
\draw [photon] (-1.3,0)--(-2.3,0);
\draw [<-, very thick] (0.65,0.55) -- (0.85,0.75);
\draw [->, very thick] (0.65,-0.55) -- (0.85,-0.75);
\draw [thick] [fill=gray] (-0.5,0) circle [radius=0.8];
\node at (1.9,0) {$+$};
\draw [dashed] (3.8,0.0)--(5.0,0.8);
\draw [dashed] (3.8,0.0)--(5.0,-0.8);
\draw [photon] (3.7,0)--(2.7,0);
\draw [<-, very thick] (4.55,0.45) -- (4.75,0.65);
\draw [->, very thick] (4.55,-0.45) -- (4.75,-0.65);
\draw [thick] [fill=black] (3.8,0) circle [radius=0.1];
\node at (6,0) {$=\,\,  {0}$};
\end{tikzpicture}
%-------------------------------------
\nonumber
\ee
or
\bea\label{R.c.d3}
g \sqrt{\g} \d_3 l_\m  + {\cal K}_{A \phi \phi,\m} = 0\, . 
\eea
Regarding the bulk, following similar arguments as for the boundary, we have the bare bulk Lagrangian
\bea
{\cal L}_{\rm bulk,0} &=& -\frac{1}{4}\Bigl ( \partial_M A_{N,0}^A - \partial_N A_{M,0}^A \Bigr )^2 - \frac{1}{2} (\partial_M A_{M,0}^A)^2 
+  \partial_M \bar c_0^B \partial_M c_0^B \nonumber\\
&-&  g_{5,0} f^{ABC} \partial_M A_{N,0}^A  A^B_{M,0} A^C_{N,0} - \frac{1}{4} g_{5,0}^2 \g (f^{ABC} A^B_{M,0} A^C_{N,0})(f^{ADE} A^D_{M,0} A^{E}_{N,0})  \nonumber\\
&+&  g_{5,0} f^{CBA} \partial_M \bar c_0^C  c_0^B A_{M,0}^A \, .
\eea 
Substituting \eq{g50}, \eq{AM0}, \eq{cA0} and \eq{barcA0} we get that 
\bea
{\cal L}_{\rm bulk,0} &=& {\cal L}_R + {\cal L}_{\rm count.} \nonumber
\eea
with ${\cal L}_R $ given by 
\bea
{\cal L}_R &=& -\frac{1}{4}\Bigl ( \partial_M A_{N}^A - \partial_N A_{M}^A \Bigr )^2 - \frac{1}{2} (\partial_M A_{M}^A)^2 
+ \L_5 \partial_M \bar c^B \partial_M c^B \nonumber\\
&-&  g_5 f^{ABC} \partial_M A_{N}^A  A^B_{M} A^C_{N} - \frac{1}{4} g_5^2 (f^{ABC} A^B_{M} A^C_{N})(f^{ADE} A^D_{M} A^{E}_N)  \nonumber\\
&+& g_{5,0} f^{CBA} \partial_M \bar c^C  c^B A_{M}^A 
\eea 
and ${\cal L}_{\rm count.} $ by 
\bea
{\cal L}_{\rm count.}  &=& \frac{1}{2}  \Bigl \{ - \d^{AB} g_{MN} p^2  \d_{A_M}  + {\cal M}_{A,MN}  \Bigr \} A^A_M A_N^B + \Bigl \{ -  \d^{AB}  p^2 \d_{c^A}  + {\cal M}^{G A_M}_{G}  \Bigr \} c^A \bar c^B \nonumber\\
&+&  \Bigl \{ -  g_5 p_M f^{ABC} \d_{A_3}  + {\cal K}_{ A_M,M}  \Bigr \} A_N^A  A^B_M A^C_N  \nonumber\\ 
&+& \Bigl \{ - g_5^2 g_{MN}g_{RS} f^{ABC} f^{ADE} \d_{A_4} + {\cal B}_{ A_M, MNRS}  \Bigr \} (A^B_M A^C_N)( A^D_R A^{E}_S) \nonumber\\
&+&  \Bigl \{ -i g_5 f^{ABC} p_M \d_1 + {\cal K}_{AG G,M}  \Bigr \} c^A \bar c^B A_M\, ,
\eea
where we have defined
\bea\label{d1}
Z_1 &=& Z_{g_5}  Z_{c^A} \sqrt{Z_{A_M}}  \Rightarrow \nonumber\\
1 + \d_1 &=& ( 1 + \d_{g_5} ) ( 1 + \d_{c^A} ) ( 1 + \frac{1}{2} \d_{A_M} ) \Rightarrow \nonumber\\
\d_1 &=& \d_{g_5} + \d_{c^A} + \frac{1}{2} \d_{A_M}\, 
\eea
and $ \d_{A_3} = \frac{3}{2} \d_{A_M} $, $ \d_{A_4} = 2 \d_{A_M} $.
In order to renormalize the above Lagrangian we need three renormalization conditions. The Feynman rules for the counter-terms are
\begin{itemize}
\item Gauge boson 2-point function
%-------------------------------------
\begin{center}
\begin{tikzpicture}
\draw[photon] (-3,0)--(-0.5,0) ;
\draw [thick] [fill=black] (-1.7,0) circle [radius=0.1];
\node at (2,0) {$=\displaystyle
 - i \d^{AB} g_{MN} p^2  \d_{A_M} $};
\end{tikzpicture}
\end{center}
%------------------------------------- 
\item Ghost 2-point function
%-------------------------------------
\begin{center}
\begin{tikzpicture}
\draw[] (-1,0)--(1.5,0) ;
\draw [thick] [fill=black] (0.25,0) circle [radius=0.1];
\node at (4,0) {$=\displaystyle - i  \d^{AB}  p^2 \d_{c^A} $};
\end{tikzpicture}
\end{center}
%------------------------------------- 
\item The $A_M$-$c$-$\bar c$ counterterm vertex
%-------------------------------------
\begin{center}
\begin{tikzpicture}[scale=0.7]
\draw [] (-2.5,1.5)--(-1,0);
\draw [] (-2.5,-1.5)--(-1,0);
\draw[photon] (-1,0)--(1,0);
\draw [thick] [fill=black] (-1,0) circle [radius=0.1];
\draw [<-, very thick] (-1.5,0.55) -- (-1.7,0.75);
\draw [->, very thick] (-1.5,-0.55) -- (-1.7,-0.75);
\node at (4,0) {$=  \displaystyle g_5 \d_1 f^{ABC} p_M \, . $};
\end{tikzpicture}
\end{center}
%------------------------------------- 
\end{itemize}
The renormalization conditions for the bulk Lagrangian are next in order. For the gauge boson propagator, diagrammatically, we have
\be
%-------------------------------------
\begin{tikzpicture} [scale=0.9]
\draw [photon,thick] (-2.3,0)--(-1.2,0);
\draw [thick] [fill=gray] (-0.5,0) circle [radius=0.8];
\draw [photon,thick] (0.3,0)--(1.3,0);
\node at (2,0) {$+$};
\draw [photon,thick] (2.5,0)--(3.6,0);
\draw [thick] [fill=black] (3.7,0) circle [radius=0.1];
\draw [photon,thick] (3.8,0)--(4.9,0);
\node at (6,0) {$=\, \, 0$};
\end{tikzpicture}
%------------------------------------- 
\nonumber
\ee
which yields
\bea\label{R.c.dA}
- \frac{1}{d-1} \left(g_{MN} -\frac{p_M p_N}{p^2}   \right) \Bigl [ - \d^{AB} g_{MN} p^2 \d_{A_M} + {\cal M}_{A,MN} \Bigr ] = 0 \, .
\eea
The second condition demands that 
\be
%-------------------------------------
\begin{tikzpicture} [scale=0.9]
\draw [] (-2.3,0)--(-1.2,0);
\draw [thick] [fill=gray] (-0.5,0) circle [radius=0.8];
\draw [] (0.3,0)--(1.3,0);
\node at (2,0) {$+$};
\draw [] (2.5,0)--(3.6,0);
\draw [thick] [fill=black] (3.7,0) circle [radius=0.1];
\draw [] (3.8,0)--(4.9,0);
\node at (5.8,0) {$=\, \, 0 $};
\end{tikzpicture}
%------------------------------------- 
\nonumber
\ee
which, in equation form is
\bea\label{R.c.dc}
-  \d^{AB}  p^2 \d_{c^A}  + {\cal M}^{G A_M}_{G} = 0 \, .
\eea 
The third condition involves the three-point vertex and requires
\be
%-------------------------------------
\begin{tikzpicture} [scale=0.9]
\draw [] (0.3,0.4)--(1.1,0.9);
\draw [] (0.3,-0.4)--(1.1,-0.9);
\draw [photon] (-1.3,0)--(-2.3,0);
\draw [<-, very thick] (0.65,0.6) -- (0.75,0.7);
\draw [->, very thick] (0.65,-0.6) -- (0.75,-0.7);
\draw [thick] [fill=gray] (-0.5,0) circle [radius=0.8];
\node at (1.9,0) {$+$};
\draw [] (3.8,0.0)--(5.0,0.8);
\draw [] (3.8,0.0)--(5.0,-0.8);
\draw [photon] (3.7,0)--(2.7,0);
\draw [<-, very thick] (4.55,0.45) -- (4.65,0.6);
\draw [->, very thick] (4.55,-0.45) -- (4.65,-0.6);
\draw [thick] [fill=black] (3.8,0) circle [radius=0.1];
\node at (6,0) {$=\,\,  {0}$};
\end{tikzpicture}
%-------------------------------------
\nonumber
\ee
translating into
\bea\label{R.c.d1}
g_5 f^{ABC} p_M \d_1 + {\cal K}_{AG G,M} = 0 \, .
\eea 
Next we evaluate the 1-loop integrals in Dimensional Regularization.

Let us start with the boundary where things are more straightforward. There, the corresponding diagrams are four-dimensional 
PV scalar and tensor integrals and their values in DR is standard. 
The complete vacuum polarization diagram is given by \eq{MphiphiAf} and in DR by
\bea\label{dr.int.ma}
{\cal M}^{\phi \phi}_{A_\m} &=& \frac{l^2 g^2 \g}{3} B_0(q, q+ l) \nonumber\\
&=& \frac{1}{16 \pi^2} \Bigl [ \frac{2 l^2 g^2 \g}{3} \frac{1}{\ve}  \Bigr ] + ( {\cal M}^{\phi \phi}_{A_\m})_f\, ,
\eea  
where $( {\cal M}^{\phi \phi}_{A_\m})_f$ corresponds to the finite part.
Recall that on the boundary lies a massless SQED so the pole of the propagators is at $l^2 = 0$. In the on-shell renormalization 
scheme we demand that the external momenta are equal to zero and also that the subtraction point is at $\m = 0$. Therefore the finite parts
of the one-loop diagrams, proportional to $\ln \frac{\m^2}{l^2}$ can be made to vanish in the on-shell limit. 
Substituting \eq{dr.int.ma} in the condition of the gauge field renormalization, \eq{R.c.da}, we obtain
\bea\label{deltaA}
\d_{A_\m} l^2 &=& - {\cal M}^{\phi\phi}_{A_\m} \Rightarrow \nonumber\\
\d_{A_\m} &=& \frac{1}{16 \pi^2} \Bigl [  - \frac{2g^2 \g }{3} \frac{1}{\ve}  \Bigr ]\, . %+ ( {\cal M}^{\phi\phi}_{A_\m} )_f
\eea
The scalar propagator is given by \eq{MphiAphif.}, which in DR reads 
\bea
{\cal M}^{\phi A_\m}_\phi &=&- 2  l^2 g^2 B_0 (q, q+l) \nonumber\\
&=& \frac{1}{16 \pi^2} \Bigl [ - 4 l^2 g^2 \g \frac{1}{\ve}  \Bigr ] +  ({\cal M}^{\phi A_\m}_\phi )_f \, . 
\eea
Substituting this in the condition \eq{R.c.dphi} we obtain 
\bea\label{deltaphi}
\d_\phi l^2 &=& - {\cal M}^{\phi A_\m}_{\phi}\Rightarrow \nonumber\\
\d_\phi &=& \frac{1}{16 \pi^2} \Bigl [  4 {g}^2 \g \frac{1}{\ve}  \Bigr ]\, .
\eea 
Finally, the one-loop contribution to the three-point vertex is given by \eq{Kfin.} and specifically in DR it is
\bea
{\cal K}_{A \phi \phi,\m} &=& {g}^3 \g^{3/2} \Bigl [ - 3 l_\m B_0(q, q + l) + l_\m B_0(q, q+l_2) \nonumber\\
&-&  2 l^2 C_\m(q,q+ L1,q+ L_2 ) -  l_{1,\m} l^2 C_0(q,q+ L1,q+ L_2 ) \Bigr ] \Rightarrow \nonumber\\
{\cal K}_{A \phi \phi,\m} &=& \frac{1}{16 \pi^2} \Bigl [ - 4 {g}^3 \g^{3/2} l_\m \frac{1}{\ve}  \Bigr ] + ( {\cal K}_{A \phi \phi,\m})_f \, .
\eea
Substituting in the vertex condition, \eq{R.c.d3}, we get 
\bea\label{delta3}
g \sqrt{\g} \d_3 l_\m  &= & - {\cal K}_{A \phi \phi,\m} \Rightarrow \nonumber\\
\d_3 &=& \frac{1}{16 \pi^2} \Bigl [  4 {g}^2 \g \frac{1}{\ve}  \Bigr ] - \frac{1}{g \sqrt{\g}} ( {\cal K}_{A \phi \phi,\m})_f\, .
\eea
Using the result $ \d_3 = \d_\phi $ along with \eq{d3} we have
\bea\label{deltag}
\d_{g} &=& -\frac{1}{2} \d_{A_\m} - \frac{1}{g \sqrt{\g} } ( {\cal K}_{A \phi \phi,\m})_f\Rightarrow \nonumber \\
\d_{g} &=& \frac{1}{16 \pi^2} \Bigl [  \frac{{g}^2 \g }{3} \frac{1}{\ve}  \Bigr ] - \frac{1}{g \sqrt{\g} } ( {\cal K}_{A \phi \phi,\m})_f \Rightarrow \nonumber \\
\d g &=& \frac{1}{16 \pi^2} \Bigl [  \frac{{g}^3 \g }{3} \frac{1}{\ve}  \Bigr ] - \frac{1}{ \sqrt{\g} }( {\cal K}_{A \phi \phi,\m})_f
\eea
and we see that the renormalization of the coupling comes only through the renormalization of the gauge field. 
Thus, we need only one counter-term for the $A_\m$ and $g$ renormalizations. 
Finally, knowing $\d_{g}$, $\d_\phi$ and $\d_{A_\m}$ we can fix from \eq{d4} the four-vertex counter-term:
\bea
\d_4 &=& \frac{1}{16 \pi^2} \Bigl [  4 {g}^2 \g \frac{1}{\ve}  \Bigr ] +( {\cal B}_{A \phi A \phi})_f \, ,
\eea
which is equal to $\d_\phi$ and $\d_3$ at the divergent part level.
Now we can write down the renormalized boundary Lagrangian in the on-shell scheme:
\bea\label{L.bound.R.}
{\cal L}_{\rm bound} &=&  -\frac{1}{4}F^3_{\m\n} F^3_{\m\n} + \partial_\m \bar \phi \partial_\m  \phi   - \frac{1}{2\xi} (\partial_\m A^3_\m)^2 + \partial_\m \bar c^3 \partial_\m c^3  \nonumber\\
&+& i g \sqrt{\g}  A_\m^3  \Bigl (  \phi \partial_\m \bar \phi  -   \bar \phi  \partial_\m \phi  \Bigr) + g^2 \g (A_\m^3)^2 \bar \phi \phi \, .
\eea
The 1-loop corrected action still has a vanishing potential after renormalization after regularizing the finite parts to zero.
In addition we are able to determine the $\beta$-function of the gauge coupling. We need \eq{g0} and \eq{a-g2} which in DR and around four-dimensions, imply that 
\be\label{BaG4}
 \bb_{\a_4} \equiv \m \frac{d \a_4(\m)}{d\m}  = - \ve \a_4 + \frac{2 \g }{3} \a_4^2\, ,
\ee
where the first and the second term correspond to the classical and the quantum parts of the $\b$-function. 
The classical contribution vanishing in 4d, the solution of the above RG equation is
\be\label{a.b.r.c.}
\a_4(\m) = \frac{3}{\g \ln{\frac{\m_{L4}^2}{\m^2}}}\, .
\ee
Here, 
\be
\m_{L4} = m e^{\frac{3}{2 \g \a_{4,m}}}
\ee 
is the Landau pole of the boundary gauge coupling with $m$ a reference mass scale where $\a_4(m)=\a_{4,m}$. Finally, notice that from \eq{a-g2} it holds that
\bea
 \frac{1}{g \m^{-\frac{\ve}{2}}} \bb_{g \m^{-\frac{\ve}{2}}}  &=&  \frac{1}{2 \a_4}  \bb_{\a_4} \Rightarrow \nonumber\\ 
\bb_{g \m^{-\frac{\ve}{2}}} &=& -\frac{\ve}{2} g \m^{\frac{-\ve}{2}} + \frac{1}{16 \pi^2}   \frac{g^3 \g \m^{\frac{-3 \ve}{2}} }{3} \label{bg-ba} \, .
\eea

We now move to the bulk where the computation in DR and the associated renormalization program are done in the $\ve$-expansion. In particular for $d=5$, 
we expand the theory around $d =4 - \ve $ and after renormalization we set $\ve = -1$.
In fact, in the bulk we have a 5d $SU(2)$ theory whose corresponding $\b$-functions and anomalous dimensions are known in the $\ve$-expansion since a long time 
(up to the $\ve$ factor they are identical to the $d=4$ data) and
in principle we could give directly those results. For future use however we go through some of the standard steps.
Starting from the vacuum polarization of the gauge field \eq{MAfin.1} and performing the Feynman parameterization on its contracted version we have in DR
\bea\label{MAe}
{\cal M}_A &=& \frac{1}{3} (-g^{MN} + \frac{p^Mp^N}{p^2}  ) {\cal M}_{A,MN} \nonumber\\
&=& \frac{p^2 g_5^2 {\cal C}_A \d^{AB} }{16 \pi^2} \Bigl [ - \frac{10}{3} \frac{1}{\ve}    \Bigr ] + ( {\cal M}_A )^{DR}_f \, ,
\eea 
where $( {\cal M}_A )^{DR}_f$ is the finite part and it is proportional to $ \ln \frac{\m^2}{p^2} $. 
Notice here that we have performed the contraction of $ {\cal M}_A $ in $4$-dimensions even though the bulk is five-dimensional.
Next follows the correction to the ghost propagator given by \eq{MGAGf}. This contribution reads 
\bea\label{MGe}
{\cal M}^{G A_M}_G = \frac{p^2 g_5^2 {\cal C}_A \d^{AB} }{16 \pi^2} \Bigl [ \frac{1}{\ve}   \Bigr ] + ( {\cal M}^{G A_M}_G )^{DR}_f \, ,
\eea
where $( {\cal M}^{G A_M}_G )^{DR}_f $ is proportional to $ \ln \frac{\m^2}{p^2} $.
Finally, the correction to the gauge-ghost vertex given by \eq{Kfin.} is equal to
\bea\label{KAe}
{\cal K}_{AGG} = \frac{p_M g_5^3 {\cal C}_A f^{ABC} }{16 \pi^2} \Bigl [ \frac{1}{\ve}   \Bigr ] +  ( {\cal K}_{AGG} )^{DR}_f \, ,
\eea  
where $ ( {\cal K}_{AGG} )^{DR}_f $ is the finite part.
We can now use these results in the bulk renormalization conditions in Sect.\ref{ROOA}. 
In the on-shell scheme where $p^2 = 0$ (and where the finite parts of \eq{MAe} and \eq{MGe} are zero) the first condition \eq{R.c.dA} gives 
\bea\label{deltaAMe}
\d^{AB} p^2 \d_{A_M} &=& - {\cal M}_{A} \Rightarrow \nonumber\\
\d_{A_M} &=& \frac{g_5^2 {\cal C}_A }{16 \pi^2} \Bigl [ \frac{10}{3} \frac{1}{\ve}   \Bigr ] \, . 
\eea
The second condition corresponding to the ghost propagator is given by \eq{R.c.dc} and reads
\bea\label{deltaghoste}
- \d^{AB}  p^2 \d_{c^A} &=& - {\cal M}^{G A_M}_{G} \Rightarrow \nonumber\\
\d_{c^A} &=& \frac{g_5^2 {\cal C}_A }{16 \pi^2 } \Bigl [ \frac{1}{\ve}  \Bigr ]\, .
\eea 
Finally, the third renormalization condition \eq{R.c.d1} combined with \eq{KAe} gives
\bea\label{delta1e}
g_5 f^{ABC} p_M \d_1 &=& - {\cal K}_{AG G,M} \Rightarrow \nonumber\\
\d_1 &=& \frac{g_5^2 {\cal C}_A }{16 \pi^2 } \Bigl [ - \frac{1}{\ve}   \Bigr ]   +  \frac{( {\cal K}_{AGG} )_f}{g_5}\, .
\eea 
Then the counter-term of the gauge coupling from \eq{d1} is
\bea\label{dge}
\d_{g_5} = \frac{g_5^2 {\cal C}_A }{16 \pi^2 } \Bigl [ - \frac{11}{3} \frac{1}{\ve}  \Bigr ] - \frac{( {\cal K}_{AGG} )_f}{g_5} \Rightarrow\nonumber\\
\d g_5 =  \frac{g_5^3 {\cal C}_A }{16 \pi^2 } \Bigl [ - \frac{11}{3} \frac{1}{\ve}  \Bigr ]  - ( {\cal K}_{AGG} )_f \, .
\eea
All the counterterms of the bulk theory are now fixed and we can write down the renormalized bulk Lagrangian:
\bea\label{R.e.Bulk}
{\cal L}_{\rm bulk} &=& -\frac{1}{4}\Bigl ( \partial_M A_{N}^A - \partial_N A_{M}^A \Bigr )^2  %- \frac{m}{2\xi} (\partial_M A_{M}^A)^2 
+  \partial_M \bar c^B \partial_M c^B \nonumber\\
&-&  g_5 f^{ABC} (\partial_M A_N^A)  A^B_M A^C_{N} - \frac{1}{4} g_5^2 (f^{ABC} A^B_M A^C_N)(f^{ADE} A^D_M A^{E}_N)  \nonumber\\
&+&  g_5 f^{CBA} (\partial_M \bar c^C)  c^B A_{M}^A \, .
\eea 
The $\b$-function can be determined from \eq{g50}, \eq{a-g52} and \eq{bG5} that in the $\ve$-expansion give
\be\label{BaG5}
\bb_{\a_5}\equiv \m \frac{d \a_5(\m)}{d \m} = - \ve \a_5 - \frac{11{\cal C}_A \a_5^2 }{3N_C} \, .
\ee
Notice that $\bb_{\a_5} $ consists of a classical and a quantum part, where the latter corresponds to the usual one-loop $\b$-function 
of an $SU(N)$ gauge theory in $d=4$, originally calculated in \cite{WGP}. We can also form a dimensionless in $d$-dimensions coupling, $g_5 \m^{-\frac{\ve}{2}}$, whose $\b$-function is
\bea\label{bg5-ba5}
 \frac{1}{g_5 \m^{-\frac{\ve}{2}}} \bb_{g_5 \m^{-\frac{\ve}{2}}}  &=&  \frac{1}{2 \a_5}  \bb_{\a_5} \Rightarrow \nonumber\\ 
\bb_{g_5 \m^{-\frac{\ve}{2}}} &=& -\frac{\ve}{2} g_5 \m^{\frac{-\ve}{2}} - \frac{1}{16 \pi^2} \frac{11  {\cal C}_A}{3} g_5^3 \m^{\frac{-3 \ve}{2}}  \, .
\eea
The solution of the RG equation \eq{BaG5} then is
\bea\label{a5RG}
\a_5(\m) = \frac{3N_C \ve M^\ve  }{11{\cal C}_A \a_{5, M} ( \m^\ve - M^\ve ) + 3N_C \ve \m^\ve  } \a_{5, M} 
\eea 
that satisfies $\a_5(M) = \a_{5,M}$ at some reference mass scale $M$. 
There is a potential Landau pole to this equation, at the scale $\m_{L5}$ where the denominator vanishes:
\be
\m_{L5} =  M \left(\frac{11{\cal C}_A \a_{5,M}}{11{\cal C}_A \a_{5,M} + 3N_C \ve}\right)^{1/\ve}\, .
\ee
The Landau pole above is real and positive when $\a_{5,M} >  -\frac{3 N_C}{11{\cal C}_A} \ve$. For $SU(2)$ (${\cal C}_A =2 $ and $N_C=2$) in $d=5$ this is $\a_{5,M} >  3/11$.
Then,
\be\label{LandauPole}
\m_{L5} = M \frac{11 \a_{5,M}-3}{11 \a_{5,M}}\, .
\ee
Suppose now that we pick a reference scale such that $\a_{5,M} <  -\frac{3 N_C}{11{\cal C}_A} \ve$.
The RG equation tells us that with this choice there can not be a Landau pole, as long as the 
value $\a_{5*} =  -\frac{3 N_C}{11{\cal C}_A} \ve$ is not crossed. In fact, as we will see in the next section
this is precisely the value of the coupling on the non-trivial fixed point, where $\m_*=\infty$ and there is a continuum limit.\footnote{When we say 
here and in the following "continuum limit", we really mean "continuum limit to 1-loop in the $\ve$-expansion", as explained in the Introduction.}
Thus, the Landau branch is disconnected from the branch with the continuum limit where $\a < \a_{5*}$. 
Finally if $\a_{5,M} = -\frac{3 N_C}{11{\cal C}_A} \ve$, then \eq{a5RG} indicates that the coupling freezes at $\m_*$ and does not run.

%---------------------------------------------------------------------------------------------------------------------------------------------------
\section{RG Flows and the Phase Diagram}\label{R.G.F.P.D}
%---------------------------------------------------------------------------------------------------------------------------------------------------

We now apply the above formalism to the boundary and bulk couplings and operators. The couplings and their associated operators of interest here 
are $g $ and $ {\cal O}_{A\phi \bar \phi} = A_\m \phi \partial_\m \bar \phi $ on the boundary and $g_5$ and $ {\cal O}_{AAA} = (\partial_M A_N) A_M A_N $ in the bulk.

%---------------------------------------------------------------------------------------------------------------------------------------------------
\subsection{Boundary}\label{Boundary Case}
%---------------------------------------------------------------------------------------------------------------------------------------------------

On the boundary there is a massless, free SQED with a $\b$-function
given by \eq{BaG4}. In terms of $g$, and using \eq{bg-ba} we can rewrite the $\b$-function according to \eq{gbf1} as
\bea\label{sdbg}
\bb_{g \m^{-\frac{\ve}{2}}} &=& ( d -\frac{\ve}{2} -d ) g \m^{\frac{-\ve}{2}} + \frac{1}{16 \pi^2}   \frac{g^3 \g \m^{\frac{-3 \ve}{2}} }{3} \nonumber\\
&=& ( d_{{\cal O}_{A\phi \bar \phi}} - d ) g \m^{\frac{-\ve}{2}} + \bb^1_{g}\, ,
\eea
with $d_{{\cal O}_{A\phi \bar \phi}} = d -\frac{\ve}{2} $ and $\bb^1_{g } = \frac{1}{16 \pi^2} \Bigl [  \frac{ g^3 \g \m^{\frac{-3\ve}{2}} }{3}  \Bigr ]$.
The anomalous dimensions of $A_\m$ and $\phi$ are given by %\eq{deltaA} and \eq{deltaphi} and they read
\bea
\gg_{A_\m} &=&  \frac{1}{16 \pi^2} \frac{2 g^2 \g \m^{-\ve} }{3} = \frac{2}{3} \a_4 \g \, , \\
\gg_{\phi} &=&- \frac{1}{16 \pi^2} 4 g^2 \g \m^{-\ve}  = -4 \a_4 \g
\eea
respectively. The DR here is for $d=4$ which corresponds to $\ve = 0$. 
The classical dimension of the marginal operator is $d_{{\cal O}_{A\phi \bar \phi}} = 4 = d $ and \eq{sdbg} becomes 
\bea
\bb_{g \m^{\frac{-\ve}{2}}} &=& \frac{1}{16 \pi^2}   \frac{{g}^3 \g }{3}  \, .
\eea
The $\b$-function vanishes for $g = 0$ where a Gaussian, IR fixed point $G$ is located.
At the Gaussian fixed point the anomalous dimensions
$\gg_{A_\m}$ and $\gg_\phi$ vanish and according to \eq{gOidOi}, $\Delta_{g {\cal O}_{A\phi \bar \phi}} = d_{{\cal O}_{A\phi \bar \phi}}$.
Now \eq{dsigma2} tells us that $g $ (or $\a_4$) is marginal. From \eq{a.b.r.c.} we see that $ \a_4(\m) \to 0 $ as $ \m \to 0$, i.e. it inherits the triviality of 
the 5d gauge coupling and becomes along with $g {\cal O}_{A\phi \bar \phi}$ marginally irrelevant in the IR.

\begin{figure}
\begin{center}
\begin{tikzpicture}
\draw[thick] (0,0) -- (6,0) node[anchor=north west] {$\a_4(\m)$};
%\draw[thick,->] (0,0) -- (0,4.5) node[anchor=south east] {$\b_b(a_b)$};
%%%
\foreach \x in {0}
\draw (\x cm,1pt) -- (\x cm,-1pt) node[anchor=north] {$\x$};
%\foreach \y in {}
%\draw (1pt,\y cm) -- (-1pt,\y cm) node[anchor=east] {$\y$};
%%%
%\draw (0,3) arc (75:0:3cm);
%\draw[dashed] (0,0) .. controls (1,3) and (2,3) .. (5,3.7);
\draw[ very thick,<-] (2,0) -- (2.05,0);
%\draw[dashed] (5.4,0) -- (5.4,4);
%\node at (5.8,3.7) { $\L_b$ };
\draw[->] (-1,0) -- (-0.3,0);
\node at (-1.5,0) {$G$};
\node at (-0.0,0.30) {IR};
\draw [thick] [fill=black] (0,0) circle [radius=0.1];
\draw [] [] (0,0) circle [radius=0.55];
%%%%
\draw[<-] (0.65,0.25) -- (1.4,0.55);
\node at (1.9,0.55) {$\d \sigma_{\a_4}$};
\node at (5.0,0.30) {UV};
%%%%
\end{tikzpicture}
\end{center}
%%%%%%%%%
\caption{The RG flow for the marginally irrelevant boundary coupling, $\a_4(\m)$. 
$G$ is the IR Gaussian fixed point where $\a_4 = 0$ and $\d \sigma_{\a_4}$ is a small region around $G$.}
\label{RGab}
\end{figure}
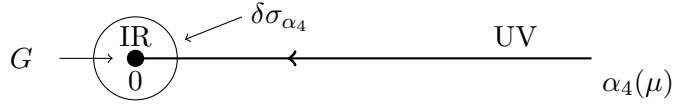

We conclude that the boundary theory flows from the UV to the IR and reaches $G$, where the theory becomes non-interacting and without a mass-gap:
the massless, free SQED breaks down to a free Maxwell theory and a free, massless scalar theory, both 4d CFT's.
A qualitative picture showing the one-dimensional direction of RG flow for the boundary coupling $\a_4(\m)$ is given in \fig{RGab}.
%%%%%%%%%%%%%%%%%%%%%%%%%%%%%%%%%%%%%%%%
 %%%%%%%%%%%%%%%%%%%%%%%%%%%%%%%%%%%%%%%%
 
 At this stage the boundary theory's quantum behaviour on (near) $G$ should better be analyzed
 by exact (softly deformed) CFT techniques.

%---------------------------------------------------------------------------------------------------------------------------------------------------
\subsection{Bulk}\label{Bulk Case}
%---------------------------------------------------------------------------------------------------------------------------------------------------

\begin{figure}
\begin{subfigure}{.4\textwidth}
\begin{center}
\begin{tikzpicture}[scale=0.8]
\draw[thick,->] (0,0) -- (7,0) node[anchor=north west] {};
\draw[thick,->] (0,0) -- (0,6) node[anchor=south east] {$\a_5(\m)$};
%%%
\foreach \x in {0}
\draw (\x cm,1pt) -- (\x cm,-1pt) node[anchor=north] {$\x$};
\foreach \y in {}
\draw (1pt,\y cm) -- (-1pt,\y cm) node[anchor=east] {$\y$};
%%%
%\draw (0,3) arc (75:0:3cm);
\draw[thick] (0.5,0.5) .. controls (1.5,3.5) and (2.5,3.5) .. (7,3.5);
\draw[red,thick] (0.5,6) .. controls (1.3,4.5) and (2,3.5) .. (4,3.5);
%%%%
\node at (-0.4,3.5) {$\a_{5*}$};
\node at (6.8,-0.4) {$\m_*$};
\draw[dashed] (0,3.5) -- (4.5,3.5);
\draw[dashed] (6.8,0) -- (6.8,3.5);
%%%%
\node at (2.2,5.5) {$\a_{5,M} > \a_{5*}$};
%%%%
\draw[dashed] (4,0) -- (4,3.5);
\node at (4,-0.4) {$\m_{L,5}$};
%%%%
\node at (-0.6,0.5) {$\a_{5,M}$};
\node at (0.5,-0.4) {$M$};
\draw[dashed] (0,0.5) -- (0.5,0.5);
\draw[dashed] (0.5,0) -- (0.5,0.5);
%%%%
\end{tikzpicture}
\end{center}
\caption{}
\end{subfigure}
%%%%%%%%%
\hskip 1.cm 
\begin{subfigure}{.4\textwidth}
\begin{center}
\begin{tikzpicture} [scale=0.7]
\draw[thick,->] (0,0) -- (8,0) node[anchor=north west] {$\a_5$};
\draw[thick,->] (0,0) -- (0,6) node[anchor=south east] {$\bb_{\a_5}$};
%%%
\foreach \x in {0}
\draw (\x cm,1pt) -- (\x cm,-1pt) node[anchor=north] {$\x$};
\foreach \y in {}
\draw (1pt,\y cm) -- (-1pt,\y cm) node[anchor=east] {$\y$};
%%%
\draw[thick] (0,0) .. controls (2,4) and (4,4) .. (7,-0.5);
%\draw[thick] (0,0) .. controls (3,3) and (4,3) .. (6,-0.5);
%\draw[thick] (0,0) .. controls (1,2) and (2,2) .. (5,-0.5);
%\node at (4.4,3) { $\g_3 < \g_2$ };
%\node at (3.8,1.5) { $\g_2 < \g_1$ };
%\node at (2,0.8) { $\g_1$ };
%%%%%%
\draw[->] (-1.3,0) -- (-0.3,0);
\node at (-2,0) {$$};
\node at (-1,0.5) {${\rm G}$};
\draw[<-] (7,0.5) -- (7.5,1);
\node at (8,1.5) {$\rm WF$};
%%%%%%%
\draw [thick] [fill=black] (0,0) circle [radius=0.1];
\draw [thick] [fill=black] (6.7,0) circle [radius=0.1];
%\draw [thick] [fill=black] (5.7,0) circle [radius=0.1];
%\draw [thick] [fill=black] (4.4,0) circle [radius=0.1];
%%%%
%\draw[ very thick,<-] (3.3,2.2) -- (3.4,2.2);
\draw[ very thick,<-] (0.5,1) -- (0.6,1.1);
\draw[ very thick,<-] (5.9,1) -- (6,0.9);
%\draw[ very thick,<-] (2.5,1.2) -- (2.6,1.2);
%%%
\node at (6.5,-0.4) {$\a_{5*}$};
%\node at (5.5,-0.4) {$g^2_{*}$};
%\node at (4.4,-0.4) {$g^1_{*}$};
%\node at (6.5,1) {$\cdot \cdot \cdot$};
%%%
\end{tikzpicture}
\end{center}
%%%%%%%%%
\caption{}
%%%%%%%%%
\end{subfigure}
\caption{(a): The RG flow of $\a_5(\m)$ as a function of the mass scale $\m$. There is a reference scale $M$ where $ \a_5(M) = \a_{5,M} $. 
The black (lower) curve corresponds to $\a_{5,M} < \a_{5*} $ where the system has a continuum limit at $\m_* = \infty$ 
and the coupling reaches a WF fixed point, $\a_{5*}$. The red (upper) curve corresponds to $\a_{5,M} > \a_{5*}$ 
where the theory has a Landau pole at $\m = \m_{L,5}$, with $ \m_{L,5} < \m_* $, and there is no continuum limit. This is the Landau branch.
The value of $\a_{5*}$ is independent of $\g$.
(b): The RG flow direction of $\bb_{\a_5}$ as a function of $\a_5$. There is an IR Gaussian fixed point (G) 
at $\a_5=0$ and a non-trivial UV fixed point (WF) at $\a_{5*} = \frac{3 N_C}{11 {\cal C}_A}$.%(For interpretation of the colours in the figure(s), the reader is referred to the web version of this article.)
}
\label{RGaB-nppd}
\end{figure}
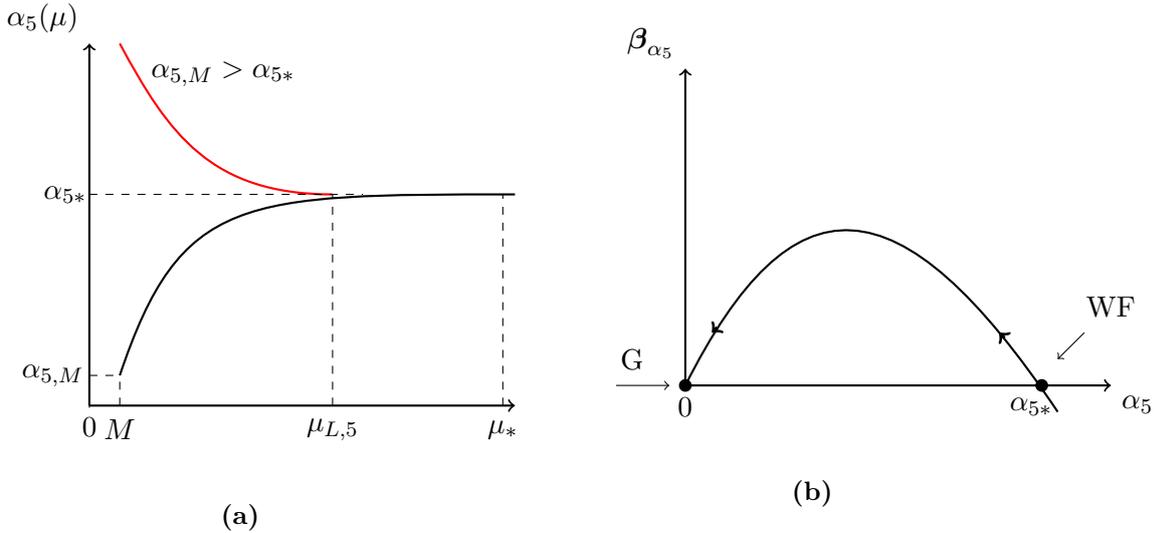

In $d$-dimensions $\bb_{\a_5}$ (or equivalently $\bb_{g_5 \m^{\frac{-\ve}{2}}}$), is given by \eq{BaG5} (or by \eq{bg5-ba5}).
According to \eq{gbf1} we have
\bea\label{bg53}
\bb_{g_5 \m^{\frac{-\ve}{2}}} &=& (d - \frac{\ve}{2} -d )g_5 \m^{\frac{-\ve}{2}} - \frac{11 {\cal C}_A}{3} \frac{g_5^3 \m^{\frac{-3 \ve}{2}}}{16 \pi^2}  \nonumber\\
&=& (d_{{\cal O}_{AAA}} -d )g_5 \m^{\frac{-\ve}{2}} + \bb^1_{g_5}(g_5) \, ,
\eea
with $ d_{{\cal O}_{AAA}} = d - \frac{\ve}{2} $ and $\bb^1_{g_5}(g_5) = - \frac{11  {\cal C}_A}{3} \frac{g_5^3 \m^{\frac{-3 \ve}{2}}}{16 \pi^2}$. 
The anomalous dimension of the gauge field $A_M$ is given by
\be\label{gAM}
\gg_{A_M} = - \frac{1}{16 \pi^2} \frac{10  {\cal C}_A g_5^2 \m^{-\ve}}{3} = - \frac{10  {\cal C}_A \a_5}{6N_C}\, .
\ee
Setting $\bb_{g_5 \m^{\frac{-\ve}{2}}} = 0$ we obtain a Gaussian fixed point G at $g_5 = 0$ but also a Wilson-Fisher fixed point WF at $g_{5*} = 4\pi \sqrt{\frac{- 3 \ve }{22  {\cal C}_A} } \m^{\frac{\ve}{2}}$
or equivalently at $ \a_5 = 0 $ and 
\be
\a_{5*} = \frac{3 N_C}{11 {\cal C}_A } (-\ve)
\ee
(that can be also obtained directly from \eq{BaG5}) respectively, in agreement with \cite{Peskin1}. 
On G the anomalous dimension of $A_M$ vanishes while on WF \eq{gAM} gives
\be\label{gAM2}
\gg_{A_M*} =  \frac{5}{11} \ve =  0.45 \ve\, .
\ee
For a five-dimensional bulk $d=5$ and $\ve = -1$ which 
gives $d_{{\cal O}_{AAA}} = 5.5 $, $\gg_{A_M*} = - 0.45$ and for the WF fixed point $g_{5*} = 4\pi \sqrt{\frac{3 }{22  {\cal C}_A}} \m^{-\frac{1}{2}}$ and 
$ \a_{5*} = \frac{3 N_C}{11  {\cal C}_A} $.
An important quantity is the mass scale where the bulk coupling reaches WF. 
This scale is denoted by $\m_*$ and its value is determined by demanding 
that $g_5(\m_*) = g_{5*}$ or $\a_5(\m_*) = \a_{5*}$. The latter, together with \eq{a5RG} for $\ve=-1$ determines
\be\label{mu*}
\m_* = \frac{\a_{5*}}{\a_{5,M}} \frac{ \Bigl (  11{\cal C}_A \a_{5,M} - 3 N_C   \Bigr ) }{\Bigl ( 11{\cal C}_A \a_{5*} - 3 N_C   \Bigr )} M \, .
\ee
Inserting the $\a_{5*}$ determined above into \eq{mu*} and if $\a_{5,M} \ne \a_{5*}$, the denominator vanishes and then $\m_* = \infty$. 
If $\a_{5,M} = \a_{5*}$ then there is a $0/0$ limit to be taken indicating that $\m_* = C_{\m_*} M $, with $C_{\m_*}$ some constant.
In the continuum branch where $ \a_{5,M} < \a_{5*} $, when the coupling reaches the WF point the theory becomes (1-loop) scale invariant, $\a_5(\m)$ stops running and obtains its maximum value, $\a_{5*}$.  
In the Landau branch on the other hand where $ \a_{5,M} >  \a_{5*} $ the running stops at $\m=\m_{5L}$ given by \eq{LandauPole}.
The RG flow of the coupling $\a_5(\m)$ is qualitatively shown in the left of  \fig{RGaB-nppd}.
Regarding the Landau branch one can see that it has a negative $\b$-function which means that it describes the RG flow beyond the WF point.
Now there are two possibilities: one is that the WF point coincides with a phase transition. We will see that in such a case beyond the WF
can only be a Confined phase in which case the 1-loop computations is not valid to begin. The other case is that the WF does not coincide with a 
phase transition, it just signals some qualitative change in the behaviour of the coupling inside the Coulomb phase and the appearance
of the Confined phase a bit further. In this case the Landau branch, viewed from beyond the WF point, is disconnected from the Gaussian fixed point   
which is the starting point of the RG flow as determined by the $\ve$-expansion. Also, according to the RG equation and the negative sign $\b$-function in this branch, the system tends to
become less scale invariant as it approaches the phase transition.
This is a too exotic scenario to accept it at face value because among other things a higher loop analysis could easily change it, so from now on
we concentrate only on the continuum branch.

Let us deal more thoroughly with the two fixed points of $g_5$. 
On the Gaussian fixed point we have $\gg_{A_M} = 0$ indicating that $\Delta_{{\cal O}_{AAA}} = d_{{\cal O}_{AAA}} = 5.5$. 
Then $\Delta_{{\cal O}_{AAA}} - d = 0.5 > 0$ and from  \eq{dsigma2} for a small deformation $\d \sigma_{g_5}$ of the coupling around the fixed point we have that
\bea\label{dsgfp}
\d \sigma_{g_5} (\m) = \Bigr (\frac{\m}{M} \Bigl )^{0.5} \d \sigma_{g_5} (M)\, 
\eea
with $M$ some fixed reference mass scale.
This shows that for the Gaussian fixed point $\d \sigma_{g_5} (\m) \to 0 $ as $\m \to 0$, implying that the small area around it 
decreases as $\m$ decreases. Since G is an IR fixed point, we conclude that $g_5$ is an IR irrelevant coupling flowing towards G, an IR attractive fixed point.
Correspondingly, $g_5 {\cal O}_{AAA}$ is marginally irrelevant in the IR.

When the coupling reaches the non-trivial UV fixed point on the other hand, a non-zero anomalous dimension develops. 
According to \eq{gOi} we have
\be
\gg_{{\cal O}_{AAA}*} = \partial_{g_5} \bb^1_{g_5}(g_{5*}) =  -1.5
\ee
and then
\be
\Delta_{{\cal O}_{AAA}*} = d_{{\cal O}_{AAA}} + \gg_{{\cal O}_{AAA}*} = 5.5-1.5 = 4\, < \, d=5\, ,
\ee
showing that $g_5$ is a relevant coupling in the vicinity of the WF fixed point.
Notice that $3 \gg_{A_M*} = 3\cdot (-0.45) = - 1.35$ which is a bit larger than $\gg_{{\cal O}_{AAA}*}=-1.5$. 
From \eq{dsigma2} we also have that
\bea\label{dsntfp}
\d \sigma_{g_5} (\m) = \Bigr (\frac{M}{\m} \Bigl ) \d \sigma_{g_5} (M)\, ,
\eea
which means that regarding WF, $\d \sigma_{g_5} (\m) \to 0\, (\infty)$ as $\m \to \infty\, (0)$, 
indicating that the small area around it decreases (increases) as $\m$ increases (decreases). 
In other words, $g_5 {\cal O}_{AAA}$ is indeed a relevant operator in the UV and the WF point is attractive at $g_5 = g_{5*}$.
The above for $\g=1$ reproduces earlier results that can be found for example in \cite{Morris, Sannino}.
A qualitative picture of the bulk $\b$-function and the direction of RG flow including the corresponding fixed points, 
in the limit where the bulk is decoupled from the boundary, is shown on the right of \fig{RGaB-nppd}.
We keep in mind for later that according to the $\ve$-expansion itself the RG flow of \fig{RGaB-nppd} lies effectively on a one-dimensional phase diagram 
parametrized by $\a_5$ because we have assumed that $\g$ is a classical quantity. 
From the point of view of the non-perturbative phase diagram parametrized by $\b_4$ and $\b_5$ we can think of our calculations projecting us on
constant $\g$ trajectories.  

To have a more quantitative picture of the behaviour of the theory near and on the fixed points
we evaluate the critical exponents $\n$ and $\eta$ connected with the correlation length $\xi$ as the WF fixed point is approached and the power 
law behaviour of a 2-point correlation function on the WF point. To evaluate the first 
critical exponent we use the linearized version of $\bb_{g_5 \m^{\frac{-\ve}{2}}}$ around the WF fixed point according to \eq{dsigma1}
\bea\label{a5l}
g_5 (\m) = \Bigr (\frac{M}{\m} \Bigl )^{\Delta_{{\cal O}_{AAA}} - d} g_5(M)\, .
\eea
Recalling from statistical physics that the characteristic range of correlations is given by $\xi \propto {1}/{\m}$ as $\m\to \m_*$
and combining with \eq{a5l} we obtain
\bea
\xi &\propto& \frac{1}{\m} = \left[ \frac{g_5(\m)}{g_5(M)} \right]^{ \frac{ 1}{\Delta_{{\cal O}_{AAA}} - d }  } M^{-1} \Rightarrow \nonumber\\
\xi &\propto& g_5(\m)^{ \frac{ 1}{ \Delta_{{\cal O}_{AAA}} - d } } = g_5(\m)^{-\n}
\eea
which fixes the critical exponent to $ \n = \frac{ 1}{d - \Delta_{{\cal O}_{AAA}} } = 1 $ for the five-dimensional bulk in agreement, to leading order, 
with the result obtained in Eq. (2.13) of \cite{Morris}.
Including corrections up to 4 loops, changes the exponent by approximately 30$\%$, without changing the qualitative picture.
Regarding $\eta$, the universal form of a two-point correlation function of a field $\Phi$ in $d$-dimensions follows the relation
\bea
\langle \Phi(x)\Phi(0)\rangle \sim \frac{1}{x^{d-2+\eta}}\, .
\eea
On the other hand, at the fixed point from CFT arguments it is also valid to express the 2-point correlation function as
\bea
\langle \Phi(x)\Phi(0)\rangle \sim \frac{1}{x^{2\Delta_\Phi}} = \frac{1}{x^{2 d_\Phi + 2 \gg_{\Phi*}}} \, ,
\eea
where $d_\Phi $ and $ \gg_\Phi $ are the classical and anomalous dimensions for the field $\Phi$ respectively. In our case we are interested 
in the 2-point function of the gauge field which, according to Appendix \ref{dim.analy.}, has classical dimension 
$d_{A_M} = \frac{d-2}{2}$ and an anomalous dimension given by \eq{gAM2}. Combining these relations we obtain
\be\label{eta}
d - 2 + \eta = 2 d_{A_M} + 2 \gg_{A_M*} \Rightarrow \eta = - 0.9 \, .
\ee

%%%%%%%%%%%%%%%%%%%%%%%%%%%%%%%%%%%
\subsubsection{Matching to the non-perturbative phase diagram}
%%%%%%%%%%%%%%%%%%%%%%%%%%%%%%%%%%%

\begin{figure}
\begin{subfigure}{.4\textwidth}
\begin{center}
\begin{tikzpicture} [scale=0.7]
\draw[thick,->] (0,0) -- (8,0) node[anchor=north west] {$\b_4$};
\draw[thick,->] (0,0) -- (0,6) node[anchor=south east] {$\b_5$};
%%%
\foreach \x in {0}
\draw (\x cm,1pt) -- (\x cm,-1pt) node[anchor=north] {$\x$};
\foreach \y in {0.5,1,1.5,2,2.5,3,3.5,4,4.5}
\draw (1pt,\y cm) -- (-1pt,\y cm) node[anchor=east] {$\y$};
%%%
\draw[thick] (0.4,4.3) .. controls (1.3,3.3) and (1.65,1.65) .. (6.65,0.5);
%\draw (1.5,3.5) .. controls (2.02,2) and (2.02,2) .. (2.02,0);
%%%%%%
\draw[dashed] (0.0,0.0) --(6.8,3.6);
\node at (5.2,2.3) {$\g = 1$}; 
%%%%%%%
%\draw [thick] [fill=black] (1.5,3.5) circle [radius=0.1];
%%%%
\draw[] (0.80,-0.05) --(0.80,0.05);
\draw[] (1.50,-0.05) --(1.50,0.05);
\draw[] (2.20,-0.05) --(2.20,0.05);
\draw[] (2.90,-0.05) --(2.90,0.05);
\draw[] (3.60,-0.05) --(3.60,0.05);
\draw[] (4.30,-0.05) --(4.30,0.05);
\draw[] (5.00,-0.05) --(5.00,0.05);
\draw[] (5.70,-0.05) --(5.70,0.05);
\draw[] (6.40,-0.05) --(6.40,0.05);
\draw[] (7.10,-0.05) --(7.10,0.05);
%%%%
%\node at (-0.50,-0.3) {$$};
%\node at (0.20,-0.3) {$0.8$};
\node at (0.80,-0.3) {$1.0$};
\node at (1.50,-0.3) {$1.2$};
\node at (2.20,-0.3) {$1.4$};
\node at (2.90,-0.3) {$1.6$};
\node at (3.60,-0.3) {$1.8$};
\node at (4.30,-0.3) {$2.0$};
\node at (5.00,-0.3) {$2.2$};
\node at (5.70,-0.3) {$2.4$};
\node at (6.40,-0.3) {$2.6$};
%%%%
%%%%
\node at (4,3.5) {{Coulomb}};
%\node at (6,0.8) {\textit{Hybrid}};
\node at (1.2,1.5) {{Confined}};
\end{tikzpicture}
\end{center}
%%%%%%%%%
\caption{}
%%%%%%%%%
\end{subfigure}
\hskip 1. cm
\begin{subfigure}{.4\textwidth}
\begin{center}
\begin{tikzpicture}[scale=0.7]
\draw[thick,->] (0,0) -- (8,0) node[anchor=north west] {};
\draw[thick,->] (0,0) -- (0,6) node[anchor=south east] {$\b_4$};
%%%
\foreach \x in {0}
\draw (\x cm,1pt) -- (\x cm,-1pt) node[anchor=north] {$\x$};
\foreach \y in {}
\draw (1pt,\y cm) -- (-1pt,\y cm) node[anchor=east] {$\y$};
%%%
\draw[red,thick] (0.6,4.3) .. controls (1.3,2.3) and (1.65,2.3) .. (6.65,2.3);
\draw[thick] (0.6,2.3) .. controls (1.3,1.3) and (1.65,1.3) .. (6.65,1.3);
\draw[green,thick] (0.6,1.3) .. controls (1.3,0.3) and (1.65,0.3) .. (6.65,0.3);
%%%%%%
\node at (5.5,2.65) {$\g < 1$}; 
\node at (5.5,1.65) {$\g = 1$}; 
\node at (5.5,0.65) {$\g > 1$}; 
%%%%%%%
%%%%
%%%%
\node at (6.65,-0.4) {$\m_*$};
\node at (8.35,-0.4) {$\m$};
\draw[dashed] (6.65,0) -- (6.65,3.3);
%%%%
\node at (-1.5,2.3) {$\b_{4*}(\g<1)$};
\node at (-1.5,1.3) {$\b_{4*}(\g=1)$};
\node at (-1.5,0.3) {$\b_{4*}(\g>1)$};
\node at (0.6,-0.4) {$M$};
\draw[dashed] (0,2.3) -- (3,2.3);
\draw[dashed] (0,1.3) -- (3,1.3);
\draw[dashed] (0,0.3) -- (3,0.3);
%\draw[dashed] (0.6,0) -- (0.6,4.3);
\draw[] (0.6,-0.05) -- (0.6,0.05);
%%%%
%%%%
%%%%
\end{tikzpicture}
\end{center}
\caption{}
\end{subfigure}
%%%%%%%%%
\caption{(a): The phase diagram of an anisotropic 5d $SU(2)$ Yang-Mills theory, according to lattice Monte Carlo simulations,
reproduced qualitatively from \cite{KnechtliRago}. The thick line indicates first order quantum phase transitions.
Its upward concavity suggests a mildly renormalized anisotropy factor.
The dashed line shows the isotropic lattice where $ \b_{4*} = \b_{5*} \simeq 1.65$.
(b): The RG flow of $\b_4(\m)$ as a function of the mass scale $\m$ for different values of anisotropy parameter $\g$ in the continuum branch. 
There is a reference scale $M$ where $ \b_4(M) = \b_{4,M} $. Here are depicted the WF fixed points, $\b_{4*}(\g)$ 
for $\g <1$, $\g=1$ and $\g>1$, when $\m = \m_*  = \infty$. In this basis $ \b_4(M) > \b_{4*}(\g) $. The behaviour of $\b_5(\m)$ is similar but with the $\g$-dependence reversed.}
\label{b4-mu}
\end{figure}
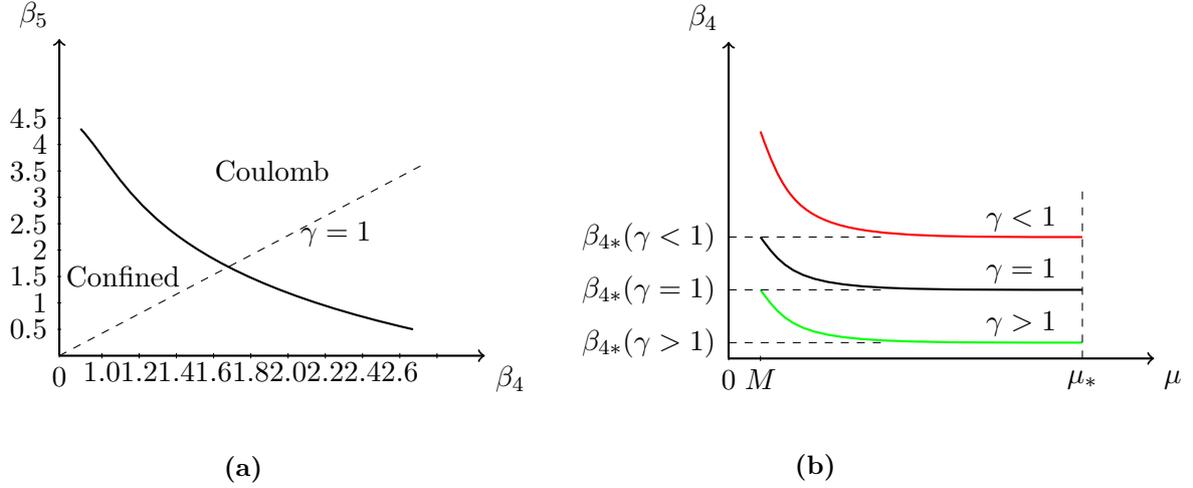

We would like to see now if there is a connection between the WF fixed points that the $\ve$-expansion produces
and the non-perturbative phase diagram that lattice simulations see.
The non-perturbative phase diagram for the bulk, reproduced on the left in Fig. \ref{b4-mu}, has been determined in \cite{KnechtliRago}.
It is constructed in the space of $(\b_4, \b_5)$ couplings, multiplying the 4d and extra dimensional plaquettes respectively. 
It exhibits two phases, a Coulomb and a Confined phase, separated by a line of first order phase transitions. 
The most interesting aspect of the role of $\g$ is that in the regime $\g < 1$ and in the Confined phase, the 
5d space becomes layered along the fifth dimension. The same approximately happens in the Coulomb phase near the phase transition. 
In order to construct the phase diagram of the bulk theory from the $\ve$-expansion and be able to compare it 
to the non-perturbative one, we should connect the bulk couplings $g_5$ (or $\b$) and $\g$ to the lattice couplings $\b_4$ and $\b_5$. 
We have already seen that the order to which we have truncated the naive lattice spacing expansion, among others,
projected us onto RG trajectories of constant $\g$. 
There is one conclusion that we can already draw if we bring $\g$ in the game, namely that we expect the curve of WF fixed points on the $\b_4-\b_5$ plane to be a parabola,
qualitatively the same as in Fig. \ref{b4-mu}.  
Conversely, if the non-perturbative phase diagram is a good fit to a parabola 
the anisotropy parameter is not expected to be renormalized by much.\footnote{This seems to imply that finite temperature phase transitions along the fifth dimension that are governed by $N_5$ but also by $\g$ 
are tied to a significant renormalization of $\g$ for given $N_5$.} 
This means that it is sufficient to concentrate on the RG flow of $\b_4$, the one of
$\b_5$ being determined by the relation $\b_5=\g^2 \b_4$. 
%%%%%%%%%%%%%%%%%%%%%%%%%%%%%%%%%%%%%%%%%%%%
%%%%%%%%%%%%%%%%%%%%%%%%%%%%%%%%%%%%%%%%%%%%

\begin{figure}
\begin{center}
\includegraphics[width=12cm]{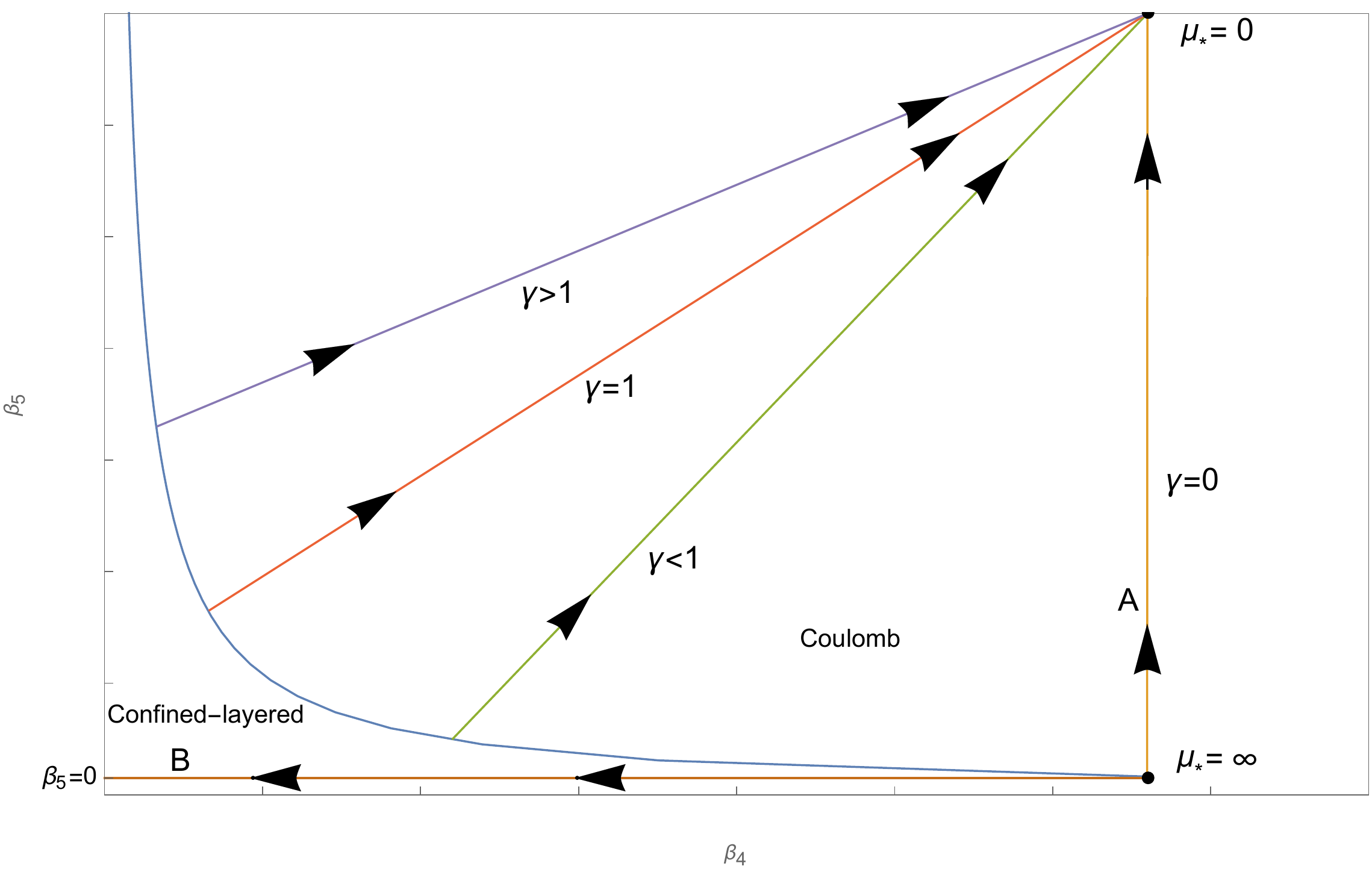}
\caption{\small The phase diagram and RG flows in the bulk according to the $\ve$-expansion. The (blue) curve represents WF fixed points.
The straight lines are $\g=$ const. RG flows. The flows labelled by $A$ and $B$ are correlated according to \eq{correlatedRG}.\label{PhDi}}
\end{center}
\end{figure}

For a more detailed comparison there is a price to be paid. Recall the definition of $\b_4 = \frac{2 N_C a_5}{g^2_5}$ and notice the
appearance of $a_5$. On the lattice $a_4$ and $a_5$ are independent quantities. Now recall also that in order to derive a classical, continuum action from the lattice spacing
expansion we took the continuum limit in $a_4$. This is why $a_4$ had disappeared from the Lagrangians that we quantized.
Quantization however re-introduces a scale $\m$ that can be thought of as a scale proportional to
a cut-off, a role that the lattice spacings had on the lattice, to begin. So in the context of the $\ve$-expansion we are in a basis where 
the dynamical scale is $\m$ and there is a classical parameter $\g$, while non-perturbatively we have the two dynamical scales $a_4$ and $a_5$
or $a_4$ and $\g$. If we assume that $\g$ is not renormalized, in both cases $\g=a_4/a_5={\rm const.}$
Therefore if we want to relate the two pictures we need a relation between $a_{4}$ and $\m$. It turns out that all we need to specify is
\be\label{mu}
\m = \frac{F(\b_4,\b_5)}{a_4} \, .
\ee
To justify \eq{mu} beyond the naive dimensional analysis which is obviously correct, we note that
$F$ could be a complicated, unknown dimensionless function of the couplings. 
In any case we can expand it around the WF point:
\be
\m = \frac{1}{a_4}\bigl[ F(\b_{4*},\b_{5*}) + O(a_4,a_5) )\bigr]
\ee
and notice that the series near the WF curve can be safely truncated to the first term in the expansion, $f\equiv F(\b_{4*},\b_{5*})$, with
$f$ a (possibly $d$-dependent) non-zero constant.\footnote{The assumption $f\ne 0$ can be justified if we keep $a_5={\rm const.}$ 
and take in the number of lattice nodes $L=l/a_4$, the physical size $l$ of the box very large. Then $F=f + O(1/L)$ and $\m = f/a_4 + O(1/l)$. 
This is equivalent to saying that to obtain the classical continuum action we have taken both $a_4\to 0$ and $l\to \infty$. Then quantum effects 
effectively re-introduce a scale in an infinitely sized box.}
Under this assumption we can set $a_5=\frac{1}{\g} \frac{f}{\m}$ and the rest is simple algebra.
For example we can express $\b_4$ as a function of the bulk coupling $\a_5$ and from that determine its $\b$-function and RG flows. We have
\be
\b_4 = \frac{N_C^2 }{4 \pi^2} \frac{\m^{-\ve} }{\a_5} a_5 = \frac{N_C^2}{4 \pi^2} \frac{f \m^{-(\ve + 1)} }{\g \a_5}
\ee
and inserting into \eq{BaG5} we obtain the RG equation and the $\b$-function of $\b_4$
\be\label{b4rge}
 \bb_{\b_4} \equiv \m \frac{d \b_4}{d \m} = \ve \b_4 + \frac{11 {\cal C}_A N_C }{12 \pi^2 } \frac{f}{\g} \m^{-(\ve + 1)} \, .
\ee
Regarding the fixed points of $\b_4$,
for the Gaussian fixed point situated at $g_5 = 0$ we get that $\b_4 = \infty$.
For the WF point we set $\bb_{\b_4} = 0$ and we get
\be\label{b4b5*e}
\b_{4*} = \frac{11 {\cal C}_A N_C}{12 \pi^2} \frac{f}{-\ve \g} \m^{-(\ve + 1)} \, .
\ee
Notice that substituting directly $g_{5*} = 4\pi \sqrt{\frac{-3\ve}{ 22 {\cal C}_A }} \m^{ \frac{\ve}{2}} $ into $\b_4$ gives the same result.
Solving the RGE gives for the running coupling $\b_4(\m)$
\bea\label{b4b5mu}
\b_4(\m) &=& \Bigl (  \frac{11 {\cal C}_A N_C }{12 \pi^2 }  \frac{f}{\g \ve} \m^{-(\ve + 1)} + \b_{4,M}  \Bigr ) \frac{\m^\ve}{M^\ve} - 
\frac{11 {\cal C}_A N_C }{12 \pi^2 }  \frac{f}{\g \ve} \m^{-(\ve + 1)}  \hskip .25 cm \, .
\eea
Here we have defined a reference scale $M$ where $\b_4(M) = \b_{4,M}$, which is the same scale where 
$\a_5(M) = \a_{5,M}$. Recall that the maximum value of $\a_5$ in the continuum branch is $\a_{5*}$ with $\a_{5,M} < \a_{5*}$, 
while now $\b_{4*}$ corresponds to the minimum value of $\b_4$ and $\b_{4,M} > \b_{4*}$ in the continuum branch.
The running of $\b_5$ is obtained by simply substituting $f/\g \to f \g$ in \eq{b4b5mu}.
Specifying to $\ve=-1$, $N_C = 2$ and $ {\cal C}_A = 2 $ we have
\be
\b_{4*} = \frac{11}{3\pi^2} \frac{f}{\g}\, ,  \hskip 1cm \b_{5*} = \frac{11}{3\pi^2} f {\g}
\label{b4*}
\ee
while the RGE of \eq{b4b5mu} gives
\bea\label{b4b5mu2}
\b_4(\m) &=& \Bigl ( - \frac{11 }{3 \pi^2} \frac{f}{\g} + \b_{4,M}  \Bigr ) \frac{M}{\m} + \frac{11}{3 \pi^2} \frac{f}{\g}\nonumber\\
\b_5(\m) &=& \Bigl ( - \frac{11 }{3 \pi^2} {f}{\g} + \b_{5,M}  \Bigr ) \frac{M}{\m} + \frac{11}{3 \pi^2} {f}{\g}\, .
\eea
As G is approached, $\m \to 0$ and $ \b_4 \to \infty $ and when $\m \to \m_* = \infty$, $\b_4 \to \b_{4*} $.
\eq{b4*} yields a phase diagram of the form on the left of \fig{b4-mu}, as anticipated.
A numerical plot of \eq{b4*} and the curve of fixed points that it generates can be seen (it is the blue curve) in Fig.  \ref{PhDi}.
The qualitative behaviour of $\b_4$ as a function of $\m$ can be seen on the right of Fig. \ref{b4-mu}. 

\begin{figure}
\begin{minipage}{.5\textwidth}
\includegraphics[width=7cm]{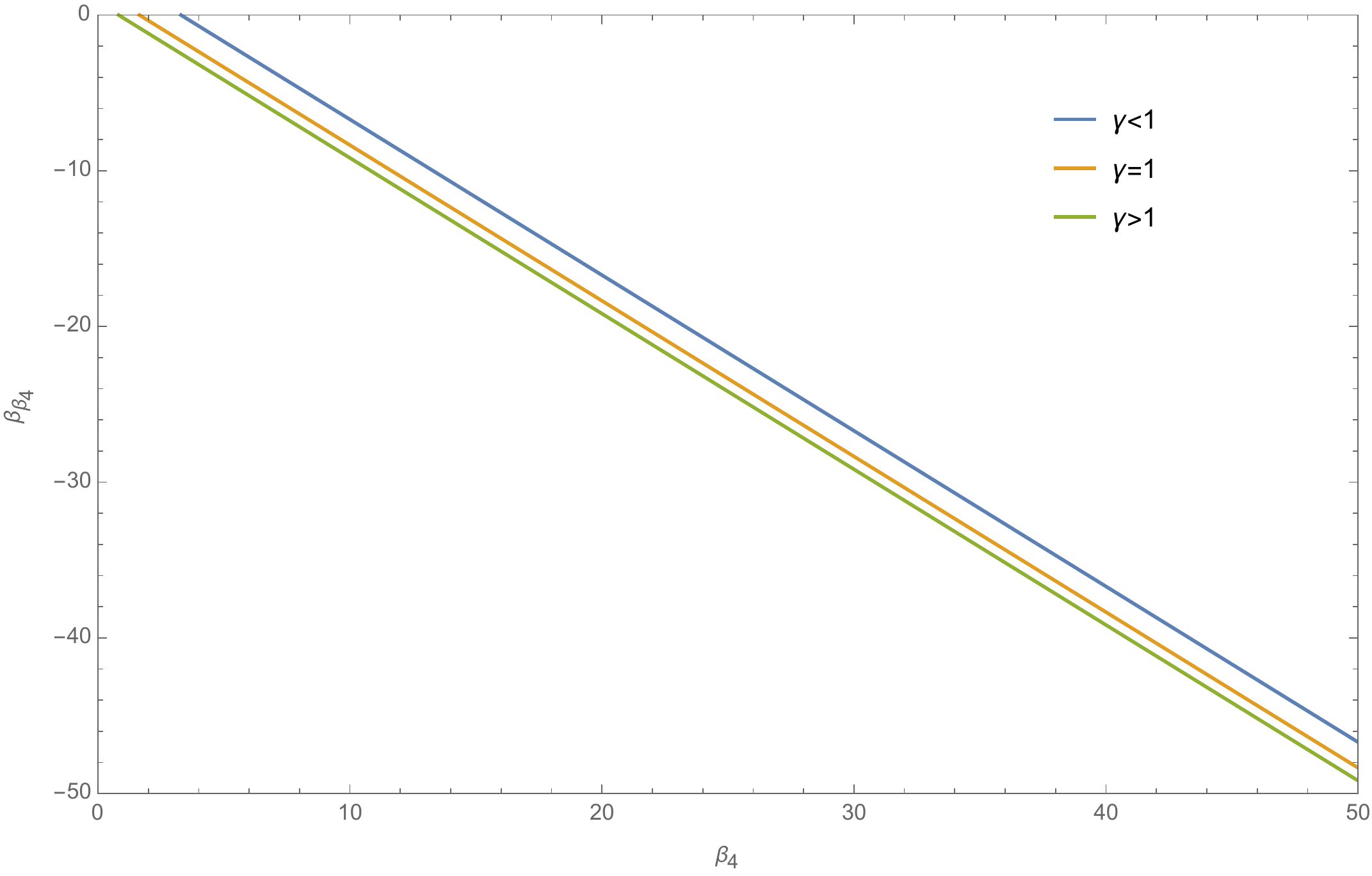}
\end{minipage}
\begin{minipage}{.5\textwidth}
\includegraphics[width=7cm]{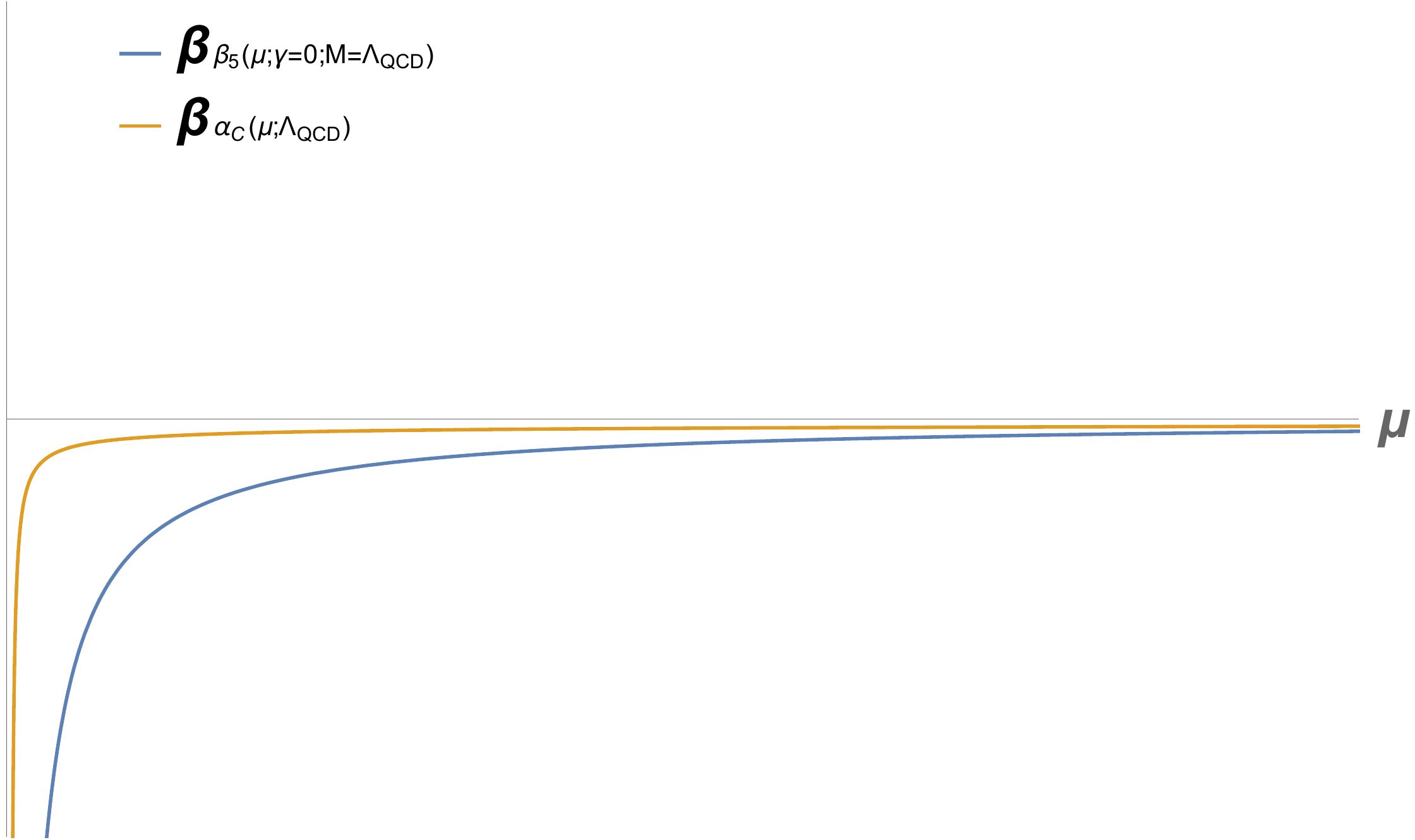}
\end{minipage}
\caption{\small Left: The RG flow of $\bb_{\b_4}$ as a function of $\b_4$ for various $\g$ in the Coulomb phase according to \eq{Bb45}. 
At G, $ \b_4(0) = \infty$ and $\bb_{\b_4} \to - \infty$.
At WF $\b_4(\m_*) = \b_{4*}$ and $\bb_{\b_4} \to 0$. As boundary conditions we have chosen
$ M = \L_{4d} \simeq 200 MeV $ and $ \b_{4,M} = 4\b_{4*}$. Right: Plot of $\bb_{\b_5}(\m;\g=0; M=\L_{4d}) = - \b_{5,\L_{4d}} \frac{\L_{4d}}{\m}$ from \eq{b4b5mu2} lower (blue) curve
and of $\bb_{\a_C}(\m; \L_{4d})$ from \eq{betagC}  upper (yellow) curve as functions of $\m$.
These illustrate that as $\m\to\infty$ both go to zero from negative values, showing the tendency of the system becoming scale invariant from both sides
along the RG flows $A$ and $B$ of Fig. \ref{PhDi}.
\label{bb45-b45}}
\end{figure}

One thing to keep in mind is that $\b_4$, thus $\b_5$ as well, decreases as the WF line is approached along constant $\g$ trajectories
and that the critical value of $\b_4$ increases as $\g$ decreases. These are all generic features of the non-perturbative phase diagram. 
%%%%%%%%%%%%%%%%%%%%%%%%%%%%%%%%%%%%%%%%%%%%
%%%%%%%%%%%%%%%%%%%%%%%%%%%%%%%%%%%%%%%%%%%%
We can perform a quantitative comparison of the phase diagrams if we fix the constant $f$. 
For this we can use the knowledge of the value $\b_{5*} = \b_{4*} = \b_* = 1.65$ for the $SU(2)$ coupling that determines the phase transition
on the isotropic lattice \cite{Creutz, KnechtliRago} and the corresponding value of the coupling on the phase transition for any $d$ \cite{Ntrekis} in the Mean-Field approximation
\be
\b_* = \frac{6.704840}{d-1}\, .
\ee
From the above and \eq{b4*} for $\g = 1$ we can fix
\be
f = \frac{d-4}{d-1} \frac{6.704840}{0.371} \simeq 4.51\, .
\ee
In our numerical analysis, we will use the value $f=4.44$ that reproduces better the Monte Carlo rather than the Mean-Field data.
In $d=5$ and for $SU(2)$ the $\b$-functions of $\b_4$ and $\b_5$ are
\bea\label{Bb45}
\bb_{\b_4} &=& - \b_4 + \frac{11}{3\pi^2} \frac{f}{\g} \nonumber\\
\bb_{\b_5} &=& - \b_5 + \frac{11}{3\pi^2} {f}{\g}\, .
\eea
The behaviour of $\bb_{\b_4}$ as a function of $\b_4$ is plotted on the left of \fig{bb45-b45} for various values of $\g$. The behaviour of $\bb_{\b_5}$ is similar.
Indeed, as the WF line is approached from the side of the Coulomb phase, the system tends towards scale invariance as the vanishing
of $\bb_{\b_4}$ (thus also of $\bb_{\b_5}$) shows.
%
%%%%%%%%%%%%%%%%%%%%%%%%%%%%%%%%%%%%%%%
%%%%%%%%%%%%%%%%%%%%%%%%%%%%%%%%%%%%%%%
%

Whether it is possible to derive the RG flows in a general 5d Confined phase using the $\ve$-expansion is not clear. In the special case where the 5d space
breaks into approximately independently fluctuating 4d planes though, called the Confined-layered phase, it is. We will assume that the layering is perfect
in which case we have in the bulk an array of non-interacting 4d $SU(2)$ gauge theories.  
This can be made exact by setting $\b_5=0$ that makes the contribution of plaquettes in the lattice having links along the fifth dimension, vanish.
Then indeed the lattice decomposes into 4d sublattices and only the coupling $\b_4$ survives. 
In this regime, with the boundary naturally decoupled from the bulk, our truncated to marginal operators classical Lagrangian is a good approximation.
The $4d$ Yang-Mills $\b$-function is known up to several loops. 
To 1-loop we can extract (and verify) the $\b$-function of the Confined phase coupling from our computation. 
Replacing $g_5 \m^{\frac{- \ve}{2}} \to g_C $ in the bulk expression \eq{bg5-ba5} gives
\be\label{betagC}
{\bb}_{g_C} = - \frac{22}{3} \frac{g_C^3}{16 \pi^2}
\ee
or equivalently ${\bb}_{\a_C} = - \frac{44}{3} \a_C^2$ 
with $g_C$ ($\a_C$) identified as the dimensionless coupling of the 4d $SU(2)$ theory on each 4d plane (we added a subscript $C$ to quantities
to remind that it is their value in the Confined-layered phase).
The only fixed point is the Gaussian fixed point $g_C=0$ ($\a_C = 0$) which in terms of the 5d couplings is at $\b_5=0$ and $\b_4=\infty$,
or at $\g=0$ and $\b_C=\infty$ (here $\b_C$ is the coupling $\b$ defined in \eq{b4b5} evaluated in the Confined phase). It is easy to see that $\b_C=2N_C/g_C^2$. 
At this point of the 5d phase diagram the WF fixed point from the point of view of the Coulomb phase
coincides with the Gaussian fixed point from the point of view of the Confined phase. It is useful to look also at $\b_C$. Its $\b$-function for $N_C=2$ is
\bea
\m \frac{d \b_C}{d \m} &=& - \frac{8}{g_C^3} \m \frac{d g_C}{d \m} \Rightarrow \nonumber\\
\bb_{\b_C} &=& \m \frac{d \b_C}{d \m} =- \frac{8}{g_C^3} \bb_{g_C} =  \frac{11}{3 \pi^2} %\frac{22}{3 \pi^2} \sqrt{\frac{1}{\b_C}}
\eea
which has the solution
\be\label{betaConfined}
\b_C(\m) =\b_{C,\L_{4d}} + \frac{11}{3 \pi^2} \ln \frac{\m}{\L_{4d}} + \cdots %\b_C^{3/2}(\m) =\b^{3/2}_{C,\L} + \frac{11}{\pi^2} \ln \frac{\m}{\L}
\ee
with $\b_{C,\L_{4d}}=\b_C(\m=\L_{4d})$ an integration constant and the dots representing higher loop corrections.
%
%%%%%%%%%%%%%%%%%%%%%%%%%%%%%%%%%%%%%%%
%%%%%%%%%%%%%%%%%%%%%%%%%%%%%%%%%%%%%%%
%
On the right of Fig. \ref{bb45-b45} we pick the RG flow line in the Coulomb phase for $\g=0$ from \eq{b4b5mu2} 
and the corresponding RG flow line in the Confined-layered phase from \eq{betaConfined}
and we plot their $\m$-dependence. We can clearly see the tendency
of the system becoming scale invariant as the fixed point where $\m_*= \infty$ is approached from either side.

A related observation is that the scale $\m$ in \eq{betaConfined} sweeps through the same values as the $\m$ that enters in the $\g=0$ Coulomb phase RGE.
This is because they originate from the fluctuations of the same lattice and in the vicinity of the UV fixed point each value of
$\m$ in one phase corresponds to a point on the other side of the phase transition with the same value of $\m$.
We can then solve \eq{betaConfined} for $\m$ 
and substitute it into the RGE of $\b_5$. Choosing $M= \L_{4d}$ we obtain
\be\label{correlatedRG}
\b_5(\m) = \b_{5, \L_{4d}} e^{-\frac{3 \pi^2}{11} \b_C(\m)} %\b_5(\m) = \b_{5, \L_{\rm QCD}} e^{-\frac{\pi^2}{11} \b^{3/2}_C(\m)}
\ee
a relation that correlates the running of the couplings on the two sides of the phase transition.
This is just one example of a generic property that the runnings of couplings on opposite sides of quantum phase transitions
seem to have \cite{IrgesCorfu16}. Notice for example that if the Landau branch were physical, the existence of a Landau pole from 
one side of the phase transition would impose a maximum cut-off on the other side. We actually expect to see a behaviour along these lines
near a first order phase transition.
On Fig. \ref{PhDi} we summarize our findings for the RG flows on the bulk phase diagram according to the $\ve$-expansion and 
in particular for the lines $A$ and $B$ under the assumption explained below \eq{mu}.

%---------------------------------------------------------------------------------------------------------------------------------------------------
\section{Scale invariance and the Stress-Energy Tensor}
%---------------------------------------------------------------------------------------------------------------------------------------------------

The easiest way to establish a connection between results from the $\ve$-expansion and scale invariance is via the Stress-Energy (S-E) tensor.
In fact, in many cases scale invariance leads to conformal invariance. 
There is a known relation between scale and conformal invariance \cite{Jackiw}, which we will use here in order to show that the fixed points that the $\ve$-expansion sees may correspond to CFTs.
Our analysis here is preliminary, as it does not take into account possible subtleties with unitarity, the gauge fixing and ghost contributions etc., but we plan to return to 
all these issues in greater detail in the future.

According to the prescription the first step is to evaluate the dilatation current by performing scale transformations of the fields. 
For a given field $\Phi_i$, the variation under dilatation transformations is
\be
\d_D \Phi_i = ( x \cdot \partial + d_{\Phi_i}  ) \Phi_i \, ,
\ee 
where the subscript $D$ stands for dilatation and $d_{\Phi_i}$ is the classical dimension of $\Phi_i$. 
Applying the above variation to a Lagrangian, ${\cal L}(\Phi_i, \partial \Phi_i)$, the dilatation current ${\cal D}^\m$ is given by 
\be\label{Dc}
{\cal D}^\m = x_\n \Theta^{\m\n}_c + \frac{\partial {\cal L}(\Phi_i, \partial \Phi_i) }{\partial (\partial^\m \Phi_i)} d_{\Phi_i}  \Phi_i \, ,
\ee
where $\Theta^{\m\n}_c$ is the canonical S-E tensor. This and all the following S-E tensors are assumed to be conserved.
In addition, we can insert the symmetric Belinfante S-E tensor, $\Theta^{\m\n}_B$, defined by\footnote{$X^{\rho\m\n}$ is antisymmetric in $\rho\m$ so that $\Theta^{\m\n}_B$ is conserved.}
\be
\Theta^{\m\n}_B = \Theta^{\m\n}_c + \partial_\r X^{\r\m\n}\nonumber
\ee
in \eq{Dc} in order to obtain the expression
\be\label{Dc2}
{\cal D}^\m = x_\n \Theta^{\m\n}_B + V^\m \, ,
\ee
where $V^\m$ is the virial field.
We now recall the two conditions which should be fulfilled by a conformally invariant theory. The first condition is that the virial field should be a total derivative 
\be
V^\m = \partial_\b \Sigma^{\m\b} \nonumber
\ee 
and the second is that the divergence of the dilatation current should vanish. 
These conditions suggest the definition of an improved S-E tensor $T^{\m\n}$ via the relation
\be\label{DC}
{\cal D}^\m = x_\n T^{\m\n} \, .
\ee
An easy way to calculate this S-E tensor is to consider a general manifold with metric 
$g_{\m\n}$ and then take the functional derivative of the action with respect to the metric evaluated in the flat limit, i.e.
\be\label{S-E}
T^{\m\n} = - \frac{2}{\sqrt{{\cal G}} }\frac{\d(\sqrt{{\cal G}} {\cal L})}{\d g_{\m\n}} \Bigl |_{g_{\m\n} \to \eta_{\m\n}}
\ee
where ${\cal G} = \det g_{\m\n}$.\\According to the above discussion if the divergence of \eq{DC}, given by 
\be\label{trT}
\partial_\m {\cal D}^\m = T^\m_\m \, ,
\ee  
vanishes then the theory is conformally invariant.
The above arguments obviously hold both at the classical and quantum levels. 
In what follows we obtain the classical and renormalized improved 
S-E tensors for the boundary and the bulk and then we evaluate their trace. 
On the fixed points we expect that the trace of the renormalized S-E tensor vanishes indicating that both theories may become CFTs at the quantum level.
The discussion has to be necessarily carried out though order by order in the $\ve$-expansion so when we speak of a CFT we really mean "conformally invariant at 1-loop in the $\ve$-expansion".
For the Gaussian fixed points this could be presumably generalized to all orders but it is not clear to us if this is the case also for fixed points of the WF type.
This is of course consistent with the spirit of Weak Asymptotic Safety as described in the Introduction.
We note that in the following evaluation of the stress-energy tensors contributions from the gauge fixing term and the Faddeev-Popov ghost fields have been neglected.

%---------------------------------------------------------------------------------------------------------------------------------------------------
\subsection{Boundary}
%---------------------------------------------------------------------------------------------------------------------------------------------------

Using \eq{a-g2} the boundary Lagrangian can be written as
\be\label{lsqed1}
{\cal L}_{\rm bound,0} =  - \frac{(4\pi)^2}{4 \a_{4,0} \g} F^{3,\m\n}_0F^3_{0,\m\n}  + |D_\m \phi_0 |^2 \, ,
\ee
The classical, improved S-E tensor then reads
\bea\label{Tsqed1}
T^{\m\n}_{\rm bound,0} &=& - \frac{2}{\sqrt{{\cal G}} }\frac{\d(\sqrt{{\cal G}} {\cal L}_{\rm bound,0})}{\d g_{\m\n}} \Bigl |_{g_{\m\n} \to \eta_{\m\n}} \nonumber\\
&=& \frac{\eta^{\m\n}(4\pi)^2}{4 \a_{4,0} \g} F^{3,\r\s}_0F^3_{0,\r\s} -  
\frac{(4\pi)^2}{\a_{4,0} \g} F^{3,\m\s}_0F^{3,\n}_{0,\s} - \eta^{\m\n} \left({D^\r \phi_0}\right)^* D_\r \phi_0 + 2 \left({D^\m \phi_0}\right)^* D^\n \phi_0 \, ,\nonumber\\
\eea
whose trace is 
\bea
\eta_{\m\n} T^{\m\n}_{\rm bound,0} &=& (4\pi)^2 \frac{4}{4 \a_{4,0} \g}  F^{3,\r\s}_0F^3_{0,\r\s} -  \frac{(4\pi)^2}{\a_{4,0} \g} F^{3,\r\s}_0F^3_{0,\r\s} 
- 4 \left({D^\r \phi_0}\right)^* D_\r \phi_0 + 2 \left({D^\r \phi_0}\right)^* D_\r \phi_0 \Rightarrow \nonumber\\
T_{b,0} &=& - 2 \left( {D^\r \phi_0}\right)^* D_\r \phi_0 = 0 \, ,
\eea 
where in the last step we used the equation of motion of the scalar field and defined $T_{b,0} \equiv \eta_{\m\n} T^{\m\n}_{\rm bound,0} $.
\eq{trT} combined with the above relation indicates that for the 4d SQED at the classical level the divergence of the dilatation current vanishes and the theory is conformally invariant. 

To extend the statement to the quantum level we must renormalize the trace of the S-E tensor. 
The trace of \eq{Tsqed1} in $d$-dimensions (using the equation of motion of $\phi$) reads
\be
T_{b,0} = \frac{d-4}{4 } \frac{(4\pi)^2}{\a_{4,0} \g \m^{\ve}} F^{3,\r\s}_0 F^3_{0,\r\s}\, .
 \ee
Using the results of \sect{ROOA} its quantum version in the $\ve$-expansion can be defined through
\bea
T_{b,0} &=& \frac{d-4}{4 } \frac{(4\pi)^2}{\a_{4} \g \m^{4-d}} Z_{\a_4}^{-1} Z_{A_\m} F^{3,\r\s} F^3_{\r\s}\,  %+ \Bigl ( -2 \d_g +  \d_{A_\m} \Bigr )\frac{d-4}{4 } \frac{(4\pi)^2}{\a_{4} \g \m^{4-d}} F^{3,\r\s} F^3_{\r\s}
\eea
by separating the finite, renormalized part $T_b$ from the rest, $\langle T_{b} \rangle$:
\bea
T_{b,0} &=& T_b + \langle T_{b} \rangle \, .
\eea
Thus, 
\bea\label{rTsqed}
T_{b} &=& \frac{d-4}{4 } \frac{(4\pi)^2}{\a_{4} \g \m^{\ve}} F^{3,\r\s} F^3_{\r\s} = \frac{-\ve}{4 } \frac{(4\pi)^2}{\a_{4} \g \m^{\ve}} F^{3,\r\s} F^3_{\r\s} \, .
\eea
Now notice that in $d$-dimensions the renormalized part of \eq{lsqed1} is
\be
{\cal L}_{\rm bound} =  - \frac{(4\pi)^2}{4\a_{4} \g \m^{\ve}}  F^{3,\m\n}F^3_{\m\n}  + |D_\m \phi |^2 \, ,\nonumber
\ee
so that \eq{rTsqed} can be rewritten in terms of ${\cal L}_{\rm bound}$ as
\bea
T_{b} &=& - \m \frac{d {\cal L}_{\rm bound}}{d\m} = - \bb_{\a_4} \frac{d {\cal L}_{\rm bound}}{d \a_4 }
\eea 
with $\bb_{\a_4}$ the $\b$-function of the boundary coupling $\a_4$. 
The above relation is known as the trace anomaly preventing the theory from being conformally 
invariant at the quantum level.
Nevertheless at the Gaussian point G where $\bb_{\a_4} = 0$, $T_{b}$ vanishes. Therefore, \eq{trT} 
implies that the divergence of the dilatation current is zero and as a consequence the boundary theory is a CFT on G.

%---------------------------------------------------------------------------------------------------------------------------------------------------
\subsection{Bulk}
%---------------------------------------------------------------------------------------------------------------------------------------------------

In the bulk, using \eq{a-g52} we start with the classical Lagrangian
\be\label{lbulk1}
{\cal L}_{\rm bulk,0} =  - \frac{1}{4 \a_{5,0} \m^{-1} \Omega} F^{A,MN}_0F^A_{0,MN} 
\ee
where $\Omega = 2N_C / (4\pi)^2$ and the S-E tensor
\bea\label{Tbulk1}
T^{MN}_{\rm bulk,0} = - \frac{2}{\sqrt{{\cal G}} }\frac{\d(\sqrt{{\cal G}} {\cal L}_{\rm bulk,0})}{\d g_{\m\n}} \Bigl |_{g_{\m\n} \to \eta_{\m\n}} =
\frac{\eta^{MN}}{4 \a_{5,0} \m^{-1} \Omega} F^{A,RS}_0F^A_{0,RS}  - \frac{1}{\a_{5,0} \m^{-1} \Omega} F^{A,MS}_0F^{A,N}_{0,S} \, ,
\eea
whose trace is
\bea
T_{B,0}\equiv \eta_{MN}T^{MN}_{\rm bulk,0} &=& \frac{5}{4 \a_{5,0} \m^{-1} \Omega} F^{A,RS}_0F^A_{0,RS}  - \frac{1}{\a_{5,0} \m^{-1} \Omega} F^{A,RS}_0F^A_{0,RS} \nonumber\\
&=& \frac{1}{4 \a_{5,0} \m^{-1} \Omega} F^{A,RS}_0F^A_{0,RS}\, ,
\eea
which is non-zero. This means that the 5d $SU(2)$ theory is neither scale nor conformally invariant at the classical level.

Now let us renormalize $T_{B,0}$. In $d$-dimensions it reads
\be
T_{B,0} = \frac{d-4}{4 \a_{5,0} \m^{4-d} \Omega} F^{A,RS}_0F^A_{0,RS} 
\ee 
and its renormalized version, according to \sect{ROOA}, can be extracted from
\bea
T_{B,0} = \frac{d-4}{4 \a_5 \m^{4-d} \Omega} Z_{\a_5}^{-1} Z_{A_M}F^{A,RS}F^A_{RS} = T_{B} + \langle T_B \rangle \, 
\eea 
where
\bea
T_B = \frac{d-4}{4 \a_5 \m^{4-d} \Omega} F^{A,RS}F^A_{RS} = \frac{-\ve}{4 \a_5 \m^{\ve} \Omega} F^{A,RS}F^A_{RS} \, .
\eea
Combining this with the renormalized part of \eq{lbulk1} 
\be
{\cal L}_{\rm bulk} =  - \frac{1}{4 \a_{5} \m^{\ve} \Omega} F^{A,MN}F^A_{MN} \, ,\nonumber
\ee
we end up with
\be
T_B = - \bb_{\a_5} \frac{d {\cal L}_{\rm bulk}}{d \a_5} \, ,
\ee
where $\bb_{\a_5}$ is the $\b$-function of the bulk coupling defined in \eq{BaG5}.
For the bulk $SU(2)$ gauge theory we found a Gaussian fixed point and a
WF fixed point, located at $\a_5 = 0$ and $ \a_{5*} = \frac{3}{11} (-\ve)$ respectively, where $\bb_{\a_5}$ vanishes. Then \eq{trT} implies
that on those points the theory becomes conformally invariant. This is an interesting result since, even though classically the bulk theory is generically not a CFT,
it flows to a CFT at the fixed points.

%---------------------------------------------------------------------------------------------------------------------------------------------------
\section{Conclusions and outlook}\label{Discussion}
%---------------------------------------------------------------------------------------------------------------------------------------------------

We have computed the RG flows in a theory that is defined by the quantization of the truncated in the lattice spacings expansion of 
a five-dimensional $SU(2)$ orbifold lattice model. Here in part I of the work we have truncated the expansion to the leading non-trivial order
where the boundary (a massless, free 4d SQED model) decouples from the bulk (a 5d $SU(2)$ gauge theory) and the two can be studied 
separately. The computational tool we used was the $\ve$-expansion. On the boundary there is a Gaussian fixed point where the theory becomes a CFT.
In the bulk where in infinite volume the phase diagram is two-dimensional there is a curve of WF fixed points. 
The resulting phase diagram agrees qualitatively with the non-perturbative phase diagram,
if we identify the WF curve with the curve of quantum phase transitions extracted from Monte Carlo simulations.
RG flow lines starting from the Gaussian fixed point
can end on the WF curve where classically marginal operators become relevant. There is a special point on the phase diagram, the one at 
$(\b_4, \b_5) = (\infty, 0)$, which is seen from the Coulomb phase as a WF fixed point and from the Confined phase as a Gaussian fixed point.
The RG flow lines (lines labelled by $A$ and $B$ on Fig. \ref{PhDi}) that end on that point can be correlated. This is possible because in this regime the Confined phase
becomes 4d-layered along the fifth dimension.

Throughout our discussion we have tried to interpret the results from the $\ve$-expansion as physical.
This has led us to a notion of Weak Asymptotic Safety, as the WF fixed points by construction are associated with a 
vanishing $\b$-function which, if generalized to all orders, implies a continuum limit that we know not to exist in this model by Monte Carlo simulations.
Instead, the Weak Asymptotic Safety point of view is that the fixed loop-order vanishing of the $\b$-function is not a signal of exact scale invariance
but an indication of the tendency of the system to become approximately scale invariant as a first order quantum phase transition is approached, 
from either side. 
Under such a perspective the WF curve of the $\ve$-expansion can be perhaps related to the first order quantum phase transition seen by the lattice.
We have given several arguments to support this picture. 
A consequence of these arguments is that near the phase transition an alternative way to describe the quantum theory is possibly via a radiatively broken CFT.

A shortcoming in our analysis is the complete decoupling of the boundary from the bulk. 
A result of the decoupling is the absence of a Higgs phase that we know however to exist non-perturbatively in the model
and it is the reason why we call the model one of Non-Perturbative Gauge-Higgs Unification.
This absence is due to the leading order truncation of the lattice spacing expansion that generates our classical action.
Truncating the lattice spacing expansion at higher order amounts to allowing classically irrelevant operators in the action.
Quantizing this action will keep the boundary coupled to the bulk and may generate large enough anomalous dimensions
to change some of the irrelevant operators to relevant ones. Then if at least one of these operators generates a mass term for the boundary scalar
and another generates a mass term for the gauge field, we will have a purely bosonic and quantum version of the Higgs mechanism.
This next step justifies the detailed exposition of the classical lattice spacing expansions, Feynman rules, diagram computations, renormalization etc. presented here, 
as in the next stage no already existing results (for $\b$-functions and 
anomalous dimensions) can be used without explicit new calculations. 
The construction of the RG flow lines itself and a matching of the kind we performed for the lines $A$ and $B$ in Fig. \ref{PhDi} 
is expected to reveal non-trivial information about the 
Higgs phase and we believe that it may offer an alternative resolution to the Higgs hierarchy problem.
This will be the topic of part II of this work.

%{\bf Acknowledgements}

%We would like to thank ..........

%\pagebreak
%---------------------------------------------------------------------------------------------------------------------------------------------------
 
\begin{appendices}

%---------------------------------------------------------------------------------------------------------------------------------------------------
\section{Couplings}\label{dim.analy.}\label{AppCoupl}
%---------------------------------------------------------------------------------------------------------------------------------------------------

There are several dimensionless couplings that can form from the dimensionful parameters of the classical 5d $SU(N_C)$ theory
$a_4$, $a_5$ and $g_5$ and the extra parameter $\m$ that appears in the quantum theory.
They are not all independent of course but one may be more convenient to use than another in specific cases.
Below we summarize the various couplings in the order they appear in the text with a pointer to their definition.

%%%%%%%%%%%%   TABLE   %%%%%%%%%%
\hskip 2cm
\begin{center}
\begin{tabular}{|c|c|c|}
\hline 
${\rm Coupling}$ & ${\rm Value}$ & ${\rm Equation}$ \\
\hline \hline   
%%%%
$\b_4$  & $\frac{2N_C a_5}{g_5^2}$ & $\eq{b4b5}$ \\ \hline
$\b_5$  & $\frac{2N_C a_4^2}{a_5 g_5^2}$ & $\eq{b4b5}$ \\ \hline
$\g$  & $\sqrt{\b_5/\b_4}=\frac{a_4}{a_5}$ & ${\rm below} \,\, \eq{b4b5}$ \\ \hline
$\b$  & $\sqrt{\b_5 \b_4}=\frac{2N_C a_4}{g_5^2}$ & ${\rm below} \,\, \eq{b4b5}$ \\ \hline
$g$  & $\frac{g_5}{\sqrt{a_4}}$ & $\eq{g-g5}$ \\ \hline
%$\L_5$  & $\frac{1}{{a_5}}$ & ${\rm below} \,\, \eq{rescaled_bulk}$ \\ \hline
%$G_4$  & $ \m^{\frac{d-4}{2}} g$ & $\eq{G40}$ \\ \hline
%$G_5$  & $ \m^{\frac{d-4}{2}} g_5$ & $\eq{G50}$ \\ \hline
$\a_4$  & $ \frac{1}{(4 \pi)^2} \m^{d-4} g^2 $ & $\eq{a-g2}$ \\ \hline
$\a_5$  & $ \frac{2 N_C}{(4 \pi)^2} \m^{d-4} g_5^2 $ & $\eq{a-g52}$ \\ \hline
\end{tabular}
%\center{TABLE 1. }
\end{center}
%%%%%%%%%%%%%%%%%%%%%%%%%%%%

%---------------------------------------------------------------------------------------------------------------------------------------------------
\section{Dimensional analysis}\label{dim.analy.}
%---------------------------------------------------------------------------------------------------------------------------------------------------

In this Appendix we review some well known facts about the dimensional analysis involved in RG flows in QFT's.
The notation $[\Phi] = d_{\Phi}$ is used for the mass dimension of the field $\Phi$. The gauge link
\bea
U_M = e^{i a g_5 \bold{A}_M}
\eea
is dimensionless which implies that
\bea
[a] + [g_5] + [\bold{A}_M] = 0\, .
\eea
Using that $[a] = [a_4] = [a_5] = -1$ and $ [g_5] = -\frac{1}{2} $ we have $ [ \bold{A}_M] = [ \bold{A}_\m] = [ \bold{A}_5] = \frac{3}{2}$. 
Moreover, since $ [\hat \Delta_M] = 1 $ it is easy to see that $[\bold{F}_{MN}] = [\bold{F}_{\m\n}] = [\bold{F}_{\m5}] = \frac{5}{2}$.

Dimensional analysis is relevant for the quantum behaviour of operators. 
We first determine the classical dimensions that are needed in \sect{Boundary Case} and \sect{Bulk Case} to be compared with 
the corresponding anomalous dimensions. Starting with the boundary part of \eq{g.f.orb.ac.} we have that in $d$-dimensions 
\bea
{\cal S}^{\rm bound.} \propto \int d^d x \Bigl [ (\partial_\m A_\n)^2 + (\partial_\m \phi)^2 + g\sqrt{\g} A_\m \phi \partial_\m \bar\phi + g^2 \g (A_\m)^2 \phi \bar\phi \Bigr ]\, ,
\eea   
from which we infer that the operator of interest is $g \sqrt{\g} {\cal O}_{A\phi \bar \phi}$ where $ {\cal O}_{A\phi \bar \phi} = A_\m \phi \partial_\m \bar\phi $.
From the kinetic terms of the gauge field we see that $-d + 2 [\partial_\m] + 2 d_{A_\m} = 0$, determining
\be
d_{A_\m} = \frac{d-2}{2}
\ee
and from the kinetic term of the scalar that $-d + 2 [\partial_\m] + 2 d_{\phi} = 0$, determining
\be
 d_{\phi} = \frac{d-2}{2} \, .
\ee
From the interaction term $g\sqrt{\g} A_\m \phi \partial_\m \bar\phi$ we get that $d_g + d_{A_\m} +1 + 2 d_{\phi} = d $ thus
\be
d_g = \frac{4-d}{2} \, .
\ee
For $d=4$ we have that $d_{A_\m} = 1$, $d_{\phi} = 1$ while the boundary coupling is dimensionless.
For the bulk part of \eq{g.f.orb.ac.} recall that there is no need to perform any rescaling and therefore, the bulk action in d-dimensions reads%we have two possibilities. The first corresponds to the bulk action before the rescaling of the gauge fields which in d-dimensions reads
\bea
{\cal S}^{\rm bulk} \propto \int d^d x  \Bigl [ (\partial_M A_N)^2 + g_5 (\partial_M A_N) A_M A_N + g_5^2 A_M A_N A_M A_N \Bigr ] \, .
\eea   
From the kinetic term we obtain again that 
\bea\label{dim.AM}
d_{A_M} = \frac{d-2}{2} \, .
\eea
Here, the operator of interest is $g_5 {\cal O}_{AAA}$ with $ {\cal O}_{AAA}  = (\partial_M A_N) A_M A_N $ that determines
\bea\label{dim.g5}
d_{g_5} = \frac{4-d}{2} \, .
\eea
In $d=5$, $ d_{A_M} =\frac{3}{2} $ and $ d_{g_5} = -\frac{1}{2} $.
%The second case corresponds to the bulk action after the rescaling $A_M \to \sqrt{\L_5} A_M$.
%Now $ d_{A_M} = 1$ and $d_{g\sqrt{\g}} = ({5-d})/{2}$. 

For a dimensionless in $d$-dimensions coupling $G_{i,0} = \m^{ d_{g_i} } g_{i,0} =  \m^{ d_{g_i} } ( g_i + \d g_i ) $ holds the RG equation
\bea\label{gbf0}
\m \frac{d G_{i,0} }{d \m} &=& \m \frac{d [ \m^{ d_{g_i} } ( g_i + \d g_i ) ] }{d \m} = 0 \Rightarrow \nonumber\\
\bb_{g_i}(g_i) &\equiv& \m \frac{d g_i}{d\m} = - d_{g_i} ( g_i + \d g_i ) +  d_{g_i} g_i \partial_{g_i} \d g_i \, ,
\eea  
with $d_{g_i}$ the mass dimension of $g_i$. Equivalently, the above relation can be modified to
\bea\label{gbf1}
\bb_{g_i}(g_i) &=& ( d_{{\cal O}_i} - d ) g_i + \bb^1_{g_i}(g_i)\, ,
\eea  
where $ d_{{\cal O}_i} + d_{g_i} = d $. $d_{{\cal O}_i}$ is the classical mass dimension of the operator associated with $g_i$, while
\bea
\bb^1_{g_i} = - d_{g_i} \d g_i + d_{g_i} g_i \partial_{g_i} \d g_i
\eea
is the one-loop part of the corresponding $\b$-function.
The above relations indicate the existence of a set of couplings, $g_{i*}$ for which $\bb_{g_i}$ is zero. The points on a phase diagram where this happens may
indicate fixed points.
An important implication of the fixed points is that they can show us if the couplings $g_i$ under discussion and their 
associated operators, are relevant, marginal or irrelevant. 
To see this choose a fixed point $g_i = g_{i*} $ and define a small area around it, $\d \sigma_i$ so that $ g= g_{i*} + \d \sigma_i $. 
The $\b$-function then deforms as $\bb_{g_i}(g_{i*} + \d \sigma_i)$.
Performing an expansion around $g_{i*}$ to linear order in $\d \sigma_i$ we have that
\bea\label{dsigma1}
\m \frac{d (g_{i*} + \d \sigma_i)}{d\m} &=& ( d_{{\cal O}_i} - d ) (g_{i*} + \d \s )  + \bb^1_{g_i}(g_{i*} + \d \s) \Rightarrow \nonumber\\
\m \frac{d \d \sigma_i }{d\m} &=& \Bigl [ ( d_{{\cal O}_i} - d )g_{i*} + \bb^1_{g_i}(g_{i*}) \Bigr] + \Bigl [ ( d_{{\cal O}_i} - d )  + \partial_{g_i} \bb^1_{g_i}(g_{i*})  \Bigr] \d \sigma_i + {\cal O}(\d \sigma_i^2) \Rightarrow \nonumber\\
\m \frac{d \d \sigma_i }{d\m} &=& ( \Delta_{{\cal O}_i} - d ) \d \sigma_i + {\cal O}(\d \sigma_i^2) \, ,
\eea
where we used that by definition $( d_{{\cal O}_i} - d )g_{i*} + \bb^1_{g_i}(g_{i*}) = \bb_{g_i}(g_{i*}) = 0 $
and 
\bea\label{gOidOi}
\Delta_{{\cal O}_i}  = d_{{\cal O}_i} + \gg_{{\cal O}_i} \, ,
\eea
is the quantum scaling dimension of the operator ${\cal O}_i $ associated with $g_i$ and the anomalous dimension of the operator ${\cal O}_i$ is defined as
\bea\label{gOi}
\gg_{{\cal O}_i} =  \partial_{g_i} \bb^1_{g_i}(g_{i*}) \, .
\eea
At a Gaussian fixed point $G$ where $g_i=0$ the anomalous dimensions vanish and $\Delta_{{\cal O}_i} = d_{{\cal O}_i}$. 

Solving \eq{dsigma1} we obtain
\bea\label{dsigma2}
\d \sigma_i (\m) = \Bigr (\frac{\m}{M} \Bigl )^{\Delta_{{\cal O}_i} - d} \d \sigma_i (M)\, ,
\eea
with $M$ an arbitrary mass scale, which shows that for $\Delta_{{\cal O}_i} < d $, $\d \sigma_i (\m) >> (<<)\, 1$ as $\m \to 0\, (\infty)$, 
for $ \Delta_{{\cal O}_i} > d $, $\d \sigma_i (\m) << (>>)\, 1$ as $\m \to 0 \, (\infty)$ while for $ \Delta_{{\cal O}_i} = d $, $\d \sigma_i (\m) = \d \sigma_i (M)$. 
The above cases correspond to an IR relevant (UV irrelevant), an IR irrelevant (UV relevant) and a marginal coupling $g_i$, respectively.  
The same terminology is used for the operator $g_i {\cal O}_i$.

Finally, the anomalous dimension of the field $\Phi$ is defined
%as \bea
%\gg_\Phi = - \frac{d {\cal M}_\Phi}{d p^2}
%\eea
%with ${\cal M}_\Phi$ the field's 2-point function. Equivalently $\gg_\Phi$ is given
by the Callan-Symanzik equation of a renormalized n-point Green function of a field $\Phi$ and a coupling $g$, reading
\be
\Bigl ( \m \frac{\partial}{\partial \m} + \b(g) \frac{\partial}{\partial g} + \frac{n}{2} \gg_\Phi  \Bigr) G^{(n)}(p_1, \cdot \cdot \cdot , p_n  ) = 0 \, , 
\ee
where
\be
\gg_\Phi = \frac{\m}{Z_{\Phi}} \frac{d Z_{\Phi}}{d \m}
\ee
The sum of the anomalous dimensions of the fields contained in an operator is not equal in general to the anomalous dimension of the operator.

%---------------------------------------------------------------------------------------------------------------------------------------------------
\section{Global symmetries}\label{G.T.C.O.}
%---------------------------------------------------------------------------------------------------------------------------------------------------

In this Appendix we discuss the global symmetries involved in the construction.
At the level of the lattice formulation the global symmetries were identified in \cite{Symmetries}.
Here we extend the discussion by considering the action of the global symmetries Parity (P), charge conjugation (C) and Stick symmetry ($\cal S$) on the continuum fields.

%---------------------------------------------------------------------------------------------------------------------------------------------------
\subsection{Parity P}\label{Parity}
%---------------------------------------------------------------------------------------------------------------------------------------------------

The Parity transformation acts on the lattice coordinates and the links as \cite{Symmetries}
\bea
P n_M = P (n_0, \vec n , n_5  ) = (n_0, - \vec n , n_5  )  \equiv \bar n_M \nonumber
\eea 
and
\bea
P U(n_M,i) = U^\dag(\bar n_M - \hat i,i), \hskip .1cm P U(n_M,0) = U(\bar n_M,0), \hskip .1cm P U(n_M,5) = U(\bar n_M,5) \nonumber
\eea
respectively.
In the lattice space expansion this gives
\bea\label{PUi}
P U(n_M,i) &=& P \Bigl [ 1 + i a g_5 \bold{A}_i(n_M) + O(a^2) + \cdot \cdot \cdot \Bigr ] \nonumber\\
P U(n_M,\{0,5\}) &=& P \Bigl [ 1 + i a g_5 \bold{A}_{\{0,5\}}(n_M) + O(a^2) + \cdot \cdot \cdot \Bigr ]
\eea
where $\bold{A}_M \equiv A^A_M T^A$.
This implies an action on the continuum fields
\bea\label{PAi}
P : {A}^A_i(x) &\to& - {A}^A_i(\bar x) \nonumber\\
P : {A}^A_{\{0, 5\}}(x) &\to& {A}^A_{\{0, 5\}}(\bar x)\, .
\eea 
Scalars and ghosts that do not carry a space-time index are invariant under $P$.
The boundary and bulk actions are both invariant under $P$.

%---------------------------------------------------------------------------------------------------------------------------------------------------
\subsection{Charge conjugation C}\label{char.conj.}
%---------------------------------------------------------------------------------------------------------------------------------------------------

The action of charge conjugation on the gauge links is given by 
\bea\label{CU_M}
C U(n_M,N) = U^*(n_M,N)\, .
\eea
In the lattice spacing expansion we have
\bea
C U(n_M,N) &=& C \Bigl [ 1 + i a g_5 \bold{A}_N(n_M) + O(a^2) + \cdot \cdot \cdot \Bigr ] \nonumber
\eea
from which we extract that 
\bea\label{CA_M}
C \bold{A}_M (x) = -(\bold{A}_M (x))^*\, ,
\eea
which implies the action $C T^A = - (T^A)^*$ at the level of the Lie algebra. For $SU(2)$ and transferring the action on the fields, we have
\bea\label{CA123}
C: {A}^1_M &\to& - {A}^1_M \nonumber\\
C: {A}^2_M &\to&  {A}^2_M \nonumber\\
C: {A}^3_M &\to& - {A}^3_M
\eea 
Regarding the boundary complex scalar $\phi=1/\sqrt{2}(A_5^1 + i A_5^2)$ the above determines the transformation
\be
C: \phi \to - {\phi^*}
\ee
Ghosts transform as in \eq{CA123} according to their gauge index.
The boundary and bulk actions are both invariant under $C$.

%---------------------------------------------------------------------------------------------------------------------------------------------------
\subsection{Stick symmetry $\cal S$}\label{stick}
%---------------------------------------------------------------------------------------------------------------------------------------------------

The stick symmetry $\cal S$ is a global symmetry particular to the lattice orbifold \cite{Murata}.
In \cite{Symmetries} it was identified as the symmetry governing the Higgs mechanism, therefore it is of special importance.
The stick transformation acts only on the boundary and hybrid links. In particular, restricting to the left boundary of the orbifold lattice, its action is
\bea\label{U5Um}
U((n_\m,0),5) \to g_{\cal S}^{-1} U_5((n_\m,0),5), \hskip .3cm U((n_\m,0),\m) \to g_{\cal S}^{-1} U((n_\m,0),\m) g_{\cal S}\, ,
\eea  
where the above relation shows clearly that the hybrid links transform as matter fields under stick transformations.
Without loss of generality we can take $g_{\cal S} = - i \s^2$.
For the boundary links we therefore have
\bea\label{SUA}
{\cal S}\, U((n_\m,0),\m) &=& g^{-1}_{\cal S}  \Bigl [ 1 + i a_4 g_5 {A}^3_\m(n_\m,0)) T^3 + O(a_4^2) + \cdot \cdot \cdot \Bigr ] g_{\cal S}
\eea
while for hybrid links
\bea\label{SU5}
{\cal S}\, U((n_\m,0),5) &=& g^{-1}_{\cal S}  \Bigl [ 1 + i a_5 g_5 \sum_{\hat \a = 1,2} {A}^{\hat \a}_5(n_\m,0)) T^{\hat \a} + O(a_5^2) + \cdot \cdot \cdot \Bigr ]\, .
\eea
This implies that
\bea\label{S.A3}
{\cal S} : {A}^3_\m &\to& - {A}^3_\m \, .
\eea 
The action on $\phi$ is a non-trivial rotation 
\be
{\cal S} :
\begin{pmatrix} 
0 & -\phi^* \\
\phi & 0 
\end{pmatrix}
\longrightarrow 
\begin{pmatrix} 
\phi & 0 \\
0 & -\phi^* 
\end{pmatrix}
\ee
such that the kinetic term of $\phi$ is invariant.

%---------------------------------------------------------------------------------------------------------------------------------------------------
\section{Feynman rules}\label{F.rules}
%---------------------------------------------------------------------------------------------------------------------------------------------------

Here we follow the usual procedure in order to evaluate the Feynman rules of the continuum orbifold action \eq{g.f.orb.ac.}. 
We split the action into bulk and boundary parts as follows:
\bea
S_{S^1/\mathbb{Z}_2} = S^1 +S^2 \nonumber
\eea
with
\bea\label{s_1}
S^1 &=& \int d^5 x P (x_5) \Biggl [ -\frac{1}{4} F^A_{MN}F^A_{MN} - \frac{1}{2\xi} (\partial_M A_M^A)^2 + \partial_M \bar c^C D^{CB}_M c^B \Biggr ]
\eea
and
\bea\label{s_2}
S^2 = \int d^5 x  \d(x_5)   \Biggl \{ {\cal L}_{\rm bound}  - \frac{1}{2\xi} (\partial_\m A^3_\m)^2 + \partial_\m \bar c^3 \partial_\m c^3   \Biggr \}\, .
\eea
Starting with the bulk action \eq{s_1} and expanding the field strength and the covariant derivative we have 
\bea
S^1 &=& \int d^5 x P (x_5) \Biggl [ -\frac{1}{4}\Bigl \{ \partial_M A_N^A - \partial_N A_M^A - g_5  f^{ABC} A^B_M A^C_N \Bigr \}^2  
- \frac{1}{2\xi} (\partial_M A_M^A)^2 +  \partial_M \bar c^B \partial_M c^B \nonumber\\
&+&  g_5 f^{CBA} \partial_M \bar c^C  c^B A_M^A\Biggr ] \Leftrightarrow \nonumber\\
&=& \int d^5 x P (x_5) \Biggl [ -\frac{1}{4}\Bigl ( \partial_M A_N^A - \partial_N A_M^A \Bigr )^2 - \frac{1}{2\xi} (\partial_M A_M^A)^2  
+ \frac{g_5}{2} \Bigl ( \partial_M A_N^A - \partial_N A_M^A \Bigr )  f^{ABC} A^B_M A^C_N  \nonumber\\
&-& \frac{g_5^2}{4}  (f^{ABC} A^B_M A^C_N)(f^{ADE} A^D_M A^E_N)  + \partial_M \bar c^B \partial_M c^B + g_5 f^{CBA} \partial_M \bar c^C  c^B A_M^A \Biggr ]\, .
\eea     
Massaging the kinetic part gives
\bea
S^1_{kin} &=& \int d^5 x d^5 y \delta(x-y) \delta_{AB} \Biggl [ \frac{1}{2} P(x_5)  A_M^A(x) \Bigl \{ g^{MN} \Box_y 
+ \Bigl ( \frac{1}{\xi} - 1 \Bigr ) \partial_M \partial_{N,y}   \Bigr \}  A_N^B(y) \nonumber\\
&-&  \bar c^A \Box c^B    \Biggr ]  \nonumber\\
&=& \int d^5 x d^5 y \delta(x-y) \delta_{AB} \Biggl [ \frac{1}{2} P(x_5)  A_M^A(x) \Bigl \{ g^{MN} \Box_y + \Bigl ( \frac{1}{\xi} - 1 \Bigr ) 
\partial_M \partial_{N,y}   \Bigr \} P^{-1}(y_5) P(y_5)  A_N^B(y)  \nonumber\\
&-&  \bar c^A \Box  c^B    \Biggr ] \Leftrightarrow \nonumber\\
S^1_{kin} &=& \int d^5 x d^5 y \d_{xy} \delta_{AB} \Biggl [ \frac{1}{2} A_M^{'A}(x) \Bigl \{g^{MN} \Box_y + 
\Bigl ( \frac{1}{\xi} - 1 \Bigr ) \partial_M \partial_{N,y}   \Bigr \} P^{-1}(y_5) A_N^{'B}(y)  -  \bar c^A \Box  c^B    \Biggr ] \nonumber
\eea
where $\d(x-y)=\d_{xy}$ and we have defined the projected gauge field $ A_M^{'A}(x) = P(x_5)  A_M^A(x) $.
We Fourier transform to momentum space using
\bea
A_M^{'A} (x) = \int \frac{d^5p}{(2\pi)^5} A_M^{'A}(p) e^{-i p \cdot x} \nonumber
\eea
and then $S^1_{kin}$ becomes 
\bea
S^1_{kin} &=& \int d^5 x d^5 y \frac{d^5p}{(2\pi)^5} \frac{d^5q}{(2\pi)^5}\frac{\delta_{xy}\delta_{AB}}{2}  A_M^{'A}(p) e^{-i p \cdot x} 
\Bigl \{ g^{MN} \Box_y + \Bigl ( \frac{1}{\xi} - 1 \Bigr ) \partial_M \partial_{N,y}   \Bigr \} e^{-i q \cdot y} P^{-1}(q_5) A_N^{'B}(q)   \nonumber\\
&+& ghosts \nonumber\\
&=& \int \frac{d^5p}{(2\pi)^5} \frac{d^5q}{(2\pi)^5} (2\pi)^5 \delta(p+q) \frac{\delta_{AB}}{2} A_M^{'A}(p)  \Bigl \{ - g^{MN} q^2 
- \Bigl ( \frac{1}{\xi} - 1 \Bigr ) q_M q_N   \Bigr \} P^{-1}(q_5) A_N^{'B}(q) + ghosts \nonumber
\eea
where we have used that $ \int d^5 x d^5 y   \delta_{xy} e^{-i p \cdot x} e^{-i  q \cdot y} =  \int d^5 x   e^{-i (p + q) \cdot x}  
= (2\pi)^5 \delta^{(5)}(p+q)  $.
Using the $\delta$-function to eliminate the $q$-integral and setting $q =-p$ we get
\bea
S^1_{kin} = \int \frac{d^5p}{(2\pi)^5} \frac{1}{2} A_M^{'A}(p) M_{AB}^{MN} A_N^{'B}(-p) + ghosts
\eea
with $ M_{AB}^{MN} =\delta_{AB} \left(- g^{MN} p^2 + \Bigl (  1 - \frac{1}{\xi} \Bigr ) p^M p^N \right) P^{-1}(-p_5)  $.  
Inverting the matrix we obtain the gauge boson propagator that respects $M_{AB}^{MN} \Pi^{BC}_{N R} = \delta^{AC} \delta_{M R} $:
\bea\label{g.b.prop.}
\Pi^{AB}_{M N} &=& P(p_5) \frac{\delta^{AB}}{p^2} \Biggl (- g_{MN}  +  (1- \xi)\frac{p_M p_N}{p^2} \Biggr ).
\eea
Following similar arguments for the ghost part, we obtain the ghost propagator
\bea
G^{AB} = P(p_5) \frac{\delta^{AB}}{p^2}\, .
\eea 
From the rest of \eq{s_1} we read off the interaction vertices which contain two self-interacting gauge boson 
vertices and one ghost-gauge boson vertex. These interactions corresponds to the terms 
 $  \partial_M A_N^{'A}  A^{'B}_M A^{'C}_N $, $ A^{'B}_M A^{'C}_N A^{'D}_M A^{'E}_N$ and $ \partial_M \bar c^C  c^B A_M^{'A}$.%, with $ A_M = \sqrt{\L_5} A_M$, $ c' = \sqrt{\L_5} c$. %or equivalently to $ \frac{1}{\sqrt{\L_5}} \partial_M \sqrt{\L_5} A_N^A  \sqrt{\L_5} A^B_M \sqrt{\L_5} A^C_N $, $ \frac{1}{\sqrt{\L_5}} \sqrt{\L_5} A^B_M \sqrt{\L_5} A^C_N \sqrt{\L_5} A^D_M \sqrt{\L_5} A^E_N$ and $ \frac{1}{\sqrt{\L_5}} \partial_M \sqrt{\L_5} \bar c^C \sqrt{\L_5}  c^B A_M^A$.

The procedure for the boundary action \eq{s_2} is analogous.  
We have
\bea\label{S2}
S^2 &=& \int d^5 x  \d(x_5)   \Biggl \{ -\frac{1}{4}F^3_{\m\n} F^3_{\m\n} + \partial_\m \bar \phi \partial_\m  \phi   - \frac{1}{2\xi} (\partial_\m A^3_\m)^2 + \partial_\m \bar c^3 \partial_\m c^3  \nonumber\\
&+& i g \sqrt{\g} A_\m^3  \Bigl (  \phi \partial_\m \bar \phi  -   \bar \phi  \partial_\m \phi  \Bigr) + g^2 \g  (A_\m^3)^2 \bar \phi \phi  \Biggr \}
\eea
so that the propagator of $A_\m^3$ is
\bea
\Pi^3_{\m\n} =\frac{\delta_{p_5,0}}{p^2} \Biggl (- g_{\m\n}  +  (1- \xi)\frac{p_\m p_\m}{p^2} \Biggr )\, ,
\eea
the propagator of the scalar field, $\phi$ is
\bea
\Pi^{\phi} = \frac{\delta_{p_5,0} }{ p^2}\, ,
\eea
and the propagator of the ghost, $c^3$ is
\bea
\Pi^{\phi} = \frac{\delta_{p_5,0} }{ p^2}
\eea
The boundary vertices are determined by the interaction terms $A_\m^3  \Bigl (  \partial_\m \bar \phi  \phi -   \bar \phi  \partial_\m \phi  \Bigr)$ and $(A_\m^3)^2 \bar \phi \phi$.

We summarize the Feynman rules below:
\begin{itemize}
\item Gauge boson Propagator
%-------------------------------------
\begin{center}
\begin{tikzpicture}[scale=0.8]
\draw[photon] (0,-0.19)--(2.5,-0.19) ;
\node at (0,0.2) {$M,A$};
\node at (2.5,0.2) {$N,B$};
\node at (3.3,-0.2) {$=$};
\node at (10,-0.1) {$i \Pi^{AB}_{MN}(p_M,q_\m) =\displaystyle
i P(p_5) \frac{ \delta^{AB}}{p^2} \Biggl (- g_{MN}  +  (1- \xi)\frac{p_M p_N}{p^2} \Biggr )$};
\node at (10.5,-2) {$\displaystyle
+ i\frac{\delta_{p_5,0}\delta_{M,\m}\delta_{N,\n}}{q^2} \Biggl (- g_{\m\n}  +  (1- \xi)\frac{q_\m q_\n}{q^2} \Biggr )
+ i\frac{\delta_{p_5,0} \delta_{M,5}\delta_{N,5}}{ q^2} $};
\end{tikzpicture}
\end{center}
%------------------------------------- 
\item Ghost Propagator
%-------------------------------------
\begin{center}
\begin{tikzpicture}[scale=0.8]
\draw[] (0,-0.19)--(2.5,-0.19) ;
\node at (0,0.1) {$A$};
\node at (2.5,0.1) {$B$};
\draw[->] (0,-0.2)--(1.25,-0.2);
\node at (7,-0.1) {$=\displaystyle
i P(p_5) \frac{\delta^{AB}}{p^2} + i \frac{\d^{A3}\d^{B3}\delta_{p_5,0} }{ q^2}$};
\end{tikzpicture}
\end{center}
%------------------------------------- 
\item three-point self interaction vertex
%-------------------------------------
\begin{center}
\begin{tikzpicture}[scale=0.7]
\draw [photon] (-2.5,1.5)--(-1,0);
\draw [photon] (-2.5,-1.5)--(-1,0);
\draw [<-] (-1.5,0.8)--(-2.2,1.5);
\node at (-2.6,1.1) {$p_3$};
\node at (-2.7,1.8) {$M,A$};
\draw [<-] (-1.5,-0.8)--(-2.2,-1.5);
\node at (-2.6,-1) {$p_2$};
\node at (-2.8,-1.8) {$R,C$};
\draw [<-] (-0.5,0.2)--(0.3,0.2);
\node at (0,-0.5) {$p_1$};
\node at (1,0.4) {$N,B$};
\draw[photon] (-1,0)--(1,0) ;
\node at (7,0) {$=  \displaystyle P(p_5) g \sqrt{\g} f^{ABC} G_{NRM} $};
\node at (7,-1.5) {$ \displaystyle + i \delta_{p_5,0} g\sqrt{\g} Q_{NRM}  $};
\end{tikzpicture}
\end{center}
 
\item four-point self interaction vertex
%-------------------------------------
\begin{center}
\begin{tikzpicture}[scale=0.7]
\draw [photon] (0,0)--(1.5,1.4);
\draw [photon] (0,0)--(1.5,-1.4);
\draw [photon] (-1.5,1.4)--(0,0);
\draw [photon] (-1.5,-1.4)--(0,0);
\draw [<-]  (0.7,0.3)--(1.3,0.9);
\node at (0.8,1.4) {$p_2$};
\node at (2.1,1.7) {$N,C$};
\draw [<-]  (0.9,-0.4)--(1.5,-1);
\node at (1.2,-1.7) {$p_3$};
\node at (2.3,-1.7) {$R,D$};
\draw [<-] (-0.9,0.3)--(-1.6,1);
\node at (-1,1.4) {$p_1$};
\node at (-2.2,1.7) {$M,B$};
\draw [<-] (-0.8,-0.4)--(-1.5,-1);
\node at (-0.8,-1.4) {$p_4$};
\node at (-2.1,-1.7) {$S,E$};
\node at (7,0) {$=\displaystyle - i P(p_5) g^2 \g  K^{BCDE}_{MNRS}  $};
\node at (8,-1.5) {$\displaystyle + 2i  \delta_{p_5,0} g^2 \g g_{\m\n} \Delta_{MNRS}  $};
\end{tikzpicture}
\end{center}
%------------------------------------- 
\item Ghost-gauge boson vertex
%-------------------------------------
\begin{center}
\begin{tikzpicture}[scale=0.7]
\draw [] (-2.5,1.5)--(-1,0);
\draw [] (-2.5,-1.5)--(-1,0);
\draw [<-] (-1.5,0.8)--(-2.2,1.5);
\node at (-2.7,1.1) {$q$};
\node at (-2.7,1.8) {$B$};
\node at (-2.7,-1.8) {$C$};
\draw [<-] (-1.5,-0.8)--(-2.2,-1.5);
\draw [->, very thick] (-1.9,-0.9)--(-1.7,-0.7);
\draw [<-, very thick] (-1.9,0.9)--(-1.7,0.7);
\node at (-2.5,-1) {$q'$};
\draw [<-] (-0.5,0.3)--(0.3,0.3);
\node at (0,-0.5) {$p$};
\node at (1.8,0) {$M,A$};
\draw[photon] (-1,0)--(1,0) ;
\node at (6,0) {$=  \displaystyle  - P(p_5) g \sqrt{\g} f^{ABC} q_M$\, ,};
\end{tikzpicture}
\end{center}
%------------------------------------- 

\end{itemize}
where we have defined that 
\bea
G_{NRM} &\equiv& G_{NRM}(p_{12},p_{23},p_{31}  ) = g_{NR}( p_1 - p_2  )_M + g_{RM}( p_2 - p_3  )_N \nonumber\\
&+& g_{MN}( p_3 - p_1  )_R \label{GMNR} \\
L^{ABC}_{NRM} &=& f^{ABC} G_{NRM} \label{LMNR} \\
Q_{MNR} &=& ( q_2 - q_1 )_\m( \d_{M,5}\delta_{N,5}\delta_{R,\rho} + \d_{N,5}\delta_{R,5}\delta_{M,\m} + \d_{R,5}\delta_{M,5}\delta_{N,\n} ) \label{QMNR} \\
K^{BCDE}_{MNRS} &=& f^{ABC}f^{AED} ( g_{MS}g_{NR} - g_{MR}g_{NS}   ) + f^{ABE}f^{ACD} ( g_{MN}g_{SR} - g_{MR}g_{NS}   ) \nonumber\\
&+& f^{ABD}f^{ACE} ( g_{MN}g_{SR} - g_{MS}g_{NR}   ) \label{KBCDE} \\
\Delta_{MNRS} &=& (\delta_{M,\m}\delta_{N,\n}\delta_{S,5}\delta_{R,5} + \delta_{M,5}\delta_{N,5}\delta_{S,\sigma}\delta_{R,\rho}) \label{DMNRS}
\eea
with $ p_{ij} =p_i - p_j $.

%---------------------------------------------------------------------------------------------------------------------------------------------------
\section{Scaleless Integrals}\label{R.tools}
%---------------------------------------------------------------------------------------------------------------------------------------------------

Here we present a discussion which affects both the boundary and the bulk renormalization procedure. 
In particular, recall that in both cases the fields are massless and as a consequence the one-loop diagrams, 
involve massless integrals. On-shell, both the external momenta and the masses go to zero. These integrals are called scaleless and their general form is
\bea\label{k2a}
{\cal FD} = \int \frac{d^4 k}{(2 \pi)^4 } \frac{1}{k^{2a}} \, ,
\eea
with $a$ an integer.
In $d$-dimensions the above integral becomes
\bea
{\cal FD} = \int \frac{d^d k}{(2 \pi)^d } \frac{1}{k^{2a}} \equiv \int^\infty_{-\infty} d k k^{d -(1+2a)}
\eea
and in DR it is zero.
The simplest example of this type \cite{Schwartz} is $B_0(q,q)$, which is of the form
\bea
B_0(q,q) = \int \frac{d^4 q}{(2 \pi)^4 } \frac{1}{q^4} \nonumber
\eea
and in DR it does not converge in any dimension, in fact it has both IR and UV divergences. The way to regularize it is by separating the IR from the UV divergent part. 
$B_0(q,q)$ in Euclidean space becomes
\bea
\int \frac{d^4 q}{(2 \pi)^4 } \frac{1}{q^4} &\equiv& \int d^d q_E \frac{1}{q_E^4} \Rightarrow \nonumber\\
\int \frac{d^4 q}{(2 \pi)^4 } \frac{1}{q^4} &\equiv& \Omega_d \int^M_0 d q_E q_E^{d -5} + \Omega_d \int^\infty_M d q_E q_E^{d -5}\Rightarrow \nonumber\\
\int \frac{d^4 q}{(2 \pi)^4 } \frac{1}{q^4} &\equiv& \Omega_d \Bigl ( \ln M - \frac{1}{\ve_{IR}} \Bigr) + \Omega_d \Bigl( - \ln M + \frac{1}{\ve_{UV}} \Bigr)
\eea
which shows that only if we consider $\ve_{IR} = \ve_{UV} $, $B_0(q,q)$ really goes to zero. Here $M$ is an arbitrary scale. 
When we are interested only in the UV case we can extract from the above scaleless integral the $\ve_{UV}$ part obtaining that
\bea\label{scaleless.int.}
B_0(q,q) = \Bigl [ \int \frac{d^d q}{(2 \pi)^d } \frac{1}{q^4} \Bigr]_{UV} = \frac{1}{16 \pi^2} \frac{2}{\ve_{UV}} \, .
\eea 
The same arguments hold also for the scaleless tensor $B$-integrals and for the massless $B_0(q,q + l)$ integrals.

Now, let us deal with two more examples of scaleless integrals, corresponding to the massless limit of scalar integrals $C_0$ and $D_0$. The general form of these massless integrals is 
\bea
C_0(q,q+L_1,q+L_2) &=& \int \frac{d^4 q}{(2 \pi)^4 } \frac{1}{q^2(q+L_1)^2(q+L_2)^2} \nonumber
\eea
and
\bea
D_0(q,q+L_a,q+L_b,q+L_c) &=& \int \frac{d^4 q}{(2 \pi)^4 } \frac{1}{q^2(q+L_a)^2(q+L_b)^2(q+L_c)^2}\nonumber
\eea  
for a triangle and a square loop respectively.
In order to evaluate them it is useful to separate them into two regions, regarding the loop and the external momentum, obtaining the following cases ($i = 1,2$ and $j = a,b,c$):
\begin{itemize}
\item Case $q \textgreater \textgreater L_i, L_j$. 
In this limit $C_0$ becomes 
\bea\label{cql}
C_0(q,q+L_1,q+L_2) \equiv \int \frac{d^4 q}{(2 \pi)^4 } \frac{1}{q^6} \nonumber
\eea
while $D_0$ becomes
\bea\label{dql}
D_0(q,q+L_a,q+L_b,q+L_c) \equiv \int \frac{d^4 q}{(2 \pi)^4 } \frac{1}{q^8} \, .\nonumber
\eea  
The above integrals correspond to \eq{k2a}, for $a = 3,4$ respectively, thus they should vanish in DR.
\item Case $L_i, L_j \textgreater \textgreater q $.
In this limit $C_0$ reads
\bea\label{clq}
C_0(q,q+L_1,q+L_2) \equiv \frac{1}{L_1^2 L_2^2} \int \frac{d^4 q}{(2 \pi)^4 } \frac{1}{q^2} \nonumber
\eea
and $D_0$
\bea\label{dlq}
D_0(q,q+L_a,q+L_b,q+L_c) \equiv \frac{1}{L_a^2 L_b^2L_c^2} \int \frac{d^4 q}{(2 \pi)^4 } \frac{1}{q^2}\, . \nonumber
\eea
Therefore, according again to \eq{k2a} for $a=1$ these integrals superficially vanish in DR.
\end{itemize} 
A comment regarding the second case is that in the on-shell scheme, which coincides with the zero-momentum scheme in a massless theory, 
then $l_i^2, l_j^2 \to 0$ and $ \frac{1}{L_1^2 L_2^2}, \hskip .1cm \frac{1}{L_a^2 L_b^2L_c^2} \to \frac{1}{0} $.
As a consequence, both $C_0$ and $D_0$ have a $ \frac{0}{0}$ form so they can be rendered to be equal to
\bea
C_0(q,q+L_1,q+L_2) \equiv C_{s}\, ,
\eea
and
\bea
D_0(q,q+L_a,q+L_b,q+L_c) \equiv D_{s}
\eea
with $C_{s}, D_{s}$ undetermined constants.
The arguments presented above hold also for the massless tensor $C$- and $D$-integrals.

%---------------------------------------------------------------------------------------------------------------------------------------------------
\section{Box Diagrams on the Boundary }\label{BDS}
%---------------------------------------------------------------------------------------------------------------------------------------------------

The boundary theory contains at one-loop level
also contributions to the $4A_\m$ and $4\phi$ vertices. These diagrams are divergent and here we present their calculation.
Let us first make a qualitative study of the possible one-loop contributions to the four point function, usually called Boxes. 
Box diagrams, denoted collectively as ${\cal B}^i_j$, are separated into reducible and irreducible Boxes and there are three possible categories, of the form
\bea
%-------------------------------------
\begin{tikzpicture}[scale=0.7]
\draw [] (1,0)--(3,-1.5);
\draw [] (1,0)--(3,1.5);
\draw [] (0,0) circle [radius=1];
\draw [] (-3,-1.5)--(-1,0);
\draw [] (-3,1.5)--(-1,0);
\end{tikzpicture}
%-------------------------------------
\, , \hskip .3 cm
%-------------------------------------
\begin{tikzpicture}[scale=0.7]
\draw [] (0.9,-0.5)--(3,-1.5);
\draw [] (0.9,0.5)--(3,1.5);
\draw [] (0,0) circle [radius=1];
\draw [] (-3,-1.5)--(-1,0);
\draw [] (-3,1.5)--(-1,0);
\end{tikzpicture}
%-------------------------------------
\, , \hskip .3 cm
%-------------------------------------
\begin{tikzpicture}[scale=0.7]
\draw [] (0.9,-0.5)--(3,-1.5);
\draw [] (0.9,0.5)--(3,1.5);
\draw [] (-0.9,0.5)--(-3,1.5);
\draw [] (-0.9,-0.5)--(-3,-1.5);
\draw [] (0,0) circle [radius=1];
\end{tikzpicture}
%-------------------------------------
\nonumber
\eea
corresponding to \textit{C-Boxes} (or \textit{Candies}), \textit{T-Boxes} and \textit{S-Boxes} respectively \cite{IrgesFotis}.
Since the boundary theory at this order is just a massless and free SQED (no non-zero 4-vertex at the classical level) 
the above diagrams will contribute as quantum corrections to three processes. 
The first set of one-loop four-point functions corrects the $A_\m$-$A_\m$-$\phi$-$\phi$ vertex and is given by the following diagrams:
\bea
%-------------------------------------
\begin{tikzpicture}[scale=0.7]
\draw [dashed] (1,0)--(2.3,1);
\draw [photon] (1,0)--(2.3,-1);
\draw [dashed] (-1,0)--(1,0);
%%%
\draw [dashed] (-1,0)--(-2.3,1);
\draw [photon] (-1,0)--(-2.3,-1);
%%%
\draw [<-, very thick ] (-1.65,0.45)--(-1.85,0.65);
\draw [->, very thick ] (0,0) -- (0.2,0);
\draw [->, very thick ] (1.65,0.45)--(1.85,0.65);
\draw  [photon] (-1,0) .. controls (-1,0.555) and (-0.555,1) .. (0,1)
.. controls (0.555,1) and (1,0.555) .. (1,0);
\end{tikzpicture}
\, , \hskip .3 cm
%-------------------------------------
\begin{tikzpicture}[scale=0.7]
\draw [dashed] (-1,0) -- (1,1);
\draw [dashed] (-1,0) -- (1,-1);
\draw [photon] (1,1) -- (1,-1);
%%%
\draw [dashed] (1,1) -- (2.5,1);%p3
\draw [dashed] (1,-1) -- (2.5,-1);%p4
%%%
\draw [photon] (-3,-1.5)--(-1,0);
\draw [photon] (-3,1.5)--(-1,0);
%%%
\draw [->, very thick] (0.2,0.6) -- (0,0.5);
\draw [<-, very thick] (0.2,-0.6) -- (0,-0.5);
\draw [->, very thick] (1.7,1) -- (1.5,1);
\draw [->, very thick] (1.5,-1) -- (1.7,-1);
\end{tikzpicture}
%-------------------------------------
\, , \hskip .3 cm
%-------------------------------------
\begin{tikzpicture}[scale=0.7]
\draw [dashed] (-1,1) -- (1,1);
\draw [photon] (1,1) -- (1,-1);
\draw [dashed] (1,-1) -- (-1,-1);
\draw [dashed] (-1,-1) -- (-1,1);
%%%
\draw [photon] (-2,2) -- (-1,1);%p1
\draw [photon] (-2,-2) -- (-1,-1);%p2
\draw [dashed] (2,2) -- (1,1);%p3
\draw [dashed] (2,-2) -- (1,-1);%p4
%%%
\draw [<-, very thick] (-0.2,1) -- (0,1);
\draw [->, very thick] (-0.2,-1) -- (0,-1);
\draw [->, very thick] (-1,0.2) -- (-1,0);
\draw [->, very thick] (1.6,1.6) -- (1.4,1.4);
\draw [<-, very thick] (1.6,-1.6) -- (1.4,-1.4);\nonumber
%
%%%
%%%
\end{tikzpicture}
%-------------------------------------
\nonumber
\eea
Note that for $ {\cal B}^{C}_{A\phi A\phi} $ and $ {\cal B}^{T}_{A\phi A\phi} $ there are two possible channels, 
while for $ {\cal B}^{S}_{A\phi A\phi} $ there is only one. Apart from that, all of them are divergent and thus they have a non-trivial 
contribution to the renormalization procedure. Nevertheless, there is no need to calculate them explicitly here. The reason is that the 
renormalization of the theory allows us to fix the counter-term of the four-vertex in terms of counter-terms of lower dimensional vertices
(by Ward identities). 

It is legitimate to ask whether there are Box diagrams that contribute to the four-photon 
scattering similarly to the spinor QED. For the boundary theory the Feynman rules show that indeed such diagrams exist and they are of the form
\bea\label{g.s.a.}
%-------------------------------------
\begin{tikzpicture}[scale=0.7]
\draw [photon] (1,0)--(3,-1.5);
\draw [photon] (1,0)--(3,1.5);
\draw [dashed] (0,0) circle [radius=1];
\node at (0,1) {$ > $};
\node at (0,-1) {$ < $};
\draw [photon] (-3,-1.5)--(-1,0);
\draw [photon] (-3,1.5)--(-1,0);
\end{tikzpicture}
\, , \hskip .3 cm
%-------------------------------------
\begin{tikzpicture}[scale=0.7]
\draw [dashed] (-1,0) -- (1,1);
\draw [dashed] (-1,0) -- (1,-1);
\draw [dashed] (1,1) -- (1,-1);
%%%
\draw [photon] (1,1) -- (2.5,1);%p3
\draw [photon] (1,-1) -- (2.5,-1);%p4
%%%
\draw [photon] (-3,-1.5)--(-1,0);
\draw [photon] (-3,1.5)--(-1,0);
%%%
\draw [->, very thick] (0.2,0.6) -- (0,0.5);
\draw [<-, very thick] (0.2,-0.6) -- (0,-0.5);
\draw [->, very thick] (1,-0.1) -- (1,0.1);
\end{tikzpicture}
%-------------------------------------
\, , \hskip .3 cm
%-------------------------------------
\begin{tikzpicture}[scale=0.7]
\draw [dashed] (-1,1) -- (1,1);
\draw [dashed] (1,1) -- (1,-1);
\draw [dashed] (1,-1) -- (-1,-1);
\draw [dashed] (-1,-1) -- (-1,1);
%%%
\draw [photon] (-2,2) -- (-1,1);%p1
\draw [photon] (-2,-2) -- (-1,-1);%p2
\draw [photon] (2,2) -- (1,1);%p3
\draw [photon] (2,-2) -- (1,-1);%p4
%%%
\draw [<-, very thick] (-0.2,1) -- (0,1);
\draw [->, very thick] (-0.2,-1) -- (0,-1);
\draw [->, very thick] (-1,0.2) -- (-1,0);
\draw [->, very thick] (1,-0.1) -- (1,0.1);
%
%%%
%%%
\end{tikzpicture}
%-------------------------------------
\eea
The explicit form of the four-photon scattering amplitudes is given below.
There are two main differences between this case and the spinor QED. The first one is that here there are three classes 
of four-photon diagrams, while in spinor QED only one, the ${\cal B}^{S}_{4ph.,\m\n\a\b}$. The second is that ${\cal B}^{S}_{4ph.,\m\n\a\b}$ 
is finite in spinor QED, while in SQED all of the above diagrams are divergent. Actually, this is a severe problem since SQED is an Abelian 
gauge theory and there is no $4-A_\m$ vertex to be renormalized in order to absorb the corresponding divergences.

An analogous question concerns the boundary 1-loop diagrams that could contribute a $\phi^4$ term to the scalar potential
(we remind that terms proportional to $\phi$ or $\phi^3$ are prohibited for various reasons).
Such diagrams indeed exist (they do not break any symmetry) and are of the form
\bea\label{phi.s.a.}
%-------------------------------------
\begin{tikzpicture}[scale=0.7]
\draw [dashed] (1,0)--(3,-1.5);
\draw [dashed] (1,0)--(3,1.5);
\draw [photon] (0,0) circle [radius=1];
\draw [dashed] (-3,-1.5)--(-1,0);
\draw [dashed] (-3,1.5)--(-1,0);
\draw [<-, very thick ] (-1.65,0.45)--(-1.85,0.65);
\draw [->, very thick ] (-1.65,-0.45)--(-1.85,-0.65);
\draw [->, very thick ] (1.65,0.45)--(1.85,0.65);
\draw [<-, very thick ] (1.65,-0.45)--(1.85,-0.65);
\end{tikzpicture}
\, , \hskip .3 cm
%-------------------------------------
\begin{tikzpicture}[scale=0.7]
\draw [photon] (-1,0) -- (1,1);
\draw [photon] (-1,0) -- (1,-1);
\draw [dashed] (1,1) -- (1,-1);
\draw [<-, very thick ] (-1.65,0.45)--(-1.85,0.65);
\draw [->, very thick ] (-1.65,-0.45)--(-1.85,-0.65);
%%%
\draw [dashed] (1,1) -- (2.5,1);%p3
\draw [dashed] (1,-1) -- (2.5,-1);%p4
%%%
\draw [dashed] (-3,-1.5)--(-1,0);
\draw [dashed] (-3,1.5)--(-1,0);
%%%
\draw [->, very thick] (1,-0.1) -- (1,0.1);
\draw [<-, very thick] (1.7,1) -- (1.5,1);
\draw [<-, very thick] (1.5,-1) -- (1.7,-1);
\end{tikzpicture}
%-------------------------------------
\, , \hskip .3 cm
%-------------------------------------
\begin{tikzpicture}[scale=0.7]
\draw [dashed] (-1,1) -- (1,1);
\draw [photon] (1,1) -- (1,-1);
\draw [dashed] (1,-1) -- (-1,-1);
\draw [photon] (-1,-1) -- (-1,1);
%%%
\draw [dashed] (-2,2) -- (-1,1);%p1
\draw [dashed] (-2,-2) -- (-1,-1);%p2
\draw [dashed] (2,2) -- (1,1);%p3
\draw [dashed] (2,-2) -- (1,-1);%p4
%%%
\draw [->, very thick] (-0.2,1) -- (0,1);
\draw [<-, very thick] (-0.2,-1) -- (0,-1);
%
%%%
\draw [<-, very thick] (-1.5,1.5) -- (-1.7,1.6);
\draw [->, very thick] (1.5,1.5) -- (1.7,1.6);
\draw [->, very thick] (-1.5,-1.5) -- (-1.7,-1.6);
\draw [<-, very thick] (1.5,-1.5) -- (1.7,-1.6);
%%%
\end{tikzpicture}
%-------------------------------------
\eea
Their explicit form is given in Sect. \ref{4phi}.
For the $ {\cal B}^{C}_{4\phi} $ and $ {\cal B}^{T}_{4\phi} $ diagrams there are two possible channels, 
while for $ {\cal B}^{S}_{4\phi} $ there is only one. Now notice that even though the boundary theory does not contain a scalar potential 
at the classical level, this seems to imply that at the 1-loop level a $\phi^4$ term appears. One could therefore argue that this radiative scalar 
potential is similar to the Coleman-Weinberg potential and a Higgs mechanism could be triggered, perhaps at two loops. As we show in Sect. \ref{4phi} 
all of the above $4\phi$ diagrams are divergent and we hit on the same issue that came up in the $4A_\m$ case:
there is no tree level $\phi^4$ term to be renormalized and a corresponding counter-term to absorb the 
divergences, leaving only a finite term which could then play the role of a Higgs potential.
So, both the $4\g$ and the $4\phi$ diagrams seem to break the renormalizability of the massive, free SQED. As far as we know this was first noticed by Salam in \cite{salam}.
In our case where we consider the massless limit we are able to deal with this problem and to keep the theory finite, as we show in the Renormalization section.

%---------------------------------------------------------------------------------------------------------------------------------------------------
\subsection{Four-photon Diagrams }\label{4g}
%---------------------------------------------------------------------------------------------------------------------------------------------------

Let us start with the one-loop diagrams contributing to the $\g \g \to \g \g$ scattering amplitude. This set contains the following diagrams 
\bea
%-------------------------------------
\begin{tikzpicture}[scale=0.7]
\draw [photon] (1,0)--(3,-1.5);
\draw [photon] (1,0)--(3,1.5);
\draw [dashed] (0,0) circle [radius=1];
\node at (0,1) {$ > $};
\node at (0,-1) {$ < $};
\draw [photon] (-3,-1.5)--(-1,0);
\draw [photon] (-3,1.5)--(-1,0);
\end{tikzpicture}
\, , \hskip .3 cm
%-------------------------------------
\begin{tikzpicture}[scale=0.7]
\draw [dashed] (-1,0) -- (1,1);
\draw [dashed] (-1,0) -- (1,-1);
\draw [dashed] (1,1) -- (1,-1);
%%%
\draw [photon] (1,1) -- (2.5,1);%p3
\draw [photon] (1,-1) -- (2.5,-1);%p4
%%%
\draw [photon] (-3,-1.5)--(-1,0);
\draw [photon] (-3,1.5)--(-1,0);
%%%
\draw [<-, very thick] (0.2,0.6) -- (0,0.5);
\draw [->, very thick] (0.2,-0.6) -- (0,-0.5);
\draw [<-, very thick] (1,-0.1) -- (1,0.1);
\end{tikzpicture}
%-------------------------------------
\, , \hskip .3 cm
%-------------------------------------
\begin{tikzpicture}[scale=0.7]
\draw [dashed] (-1,1) -- (1,1);
\draw [dashed] (1,1) -- (1,-1);
\draw [dashed] (1,-1) -- (-1,-1);
\draw [dashed] (-1,-1) -- (-1,1);
%%%
\draw [photon] (-2,2) -- (-1,1);%p1
\draw [photon] (-2,-2) -- (-1,-1);%p2
\draw [photon] (2,2) -- (1,1);%p3
\draw [photon] (2,-2) -- (1,-1);%p4
%%%
\draw [->, very thick] (-0.2,1) -- (0,1);
\draw [>-, very thick] (-0.2,-1) -- (0,-1);
\draw [<-, very thick] (-1,0.2) -- (-1,0);
\draw [<-, very thick] (1,-0.1) -- (1,0.1);\nonumber
%
%%%
%%%
\end{tikzpicture}
%-------------------------------------
\eea
where $C$- and $S$-Boxes come in three versions, corresponding to the usual $s$, $t$ and $u$ channels, where
\bea
s=(l_1+l_2)^2 \hskip 1cm t=(l_1+l_3)^2 \hskip 1cm u=(l_1+l_4)^2\, ,
\eea
while, the $T$-Boxes come in six versions since they are not invariant under a reflection with respect to the axis passing through the centre 
of the loop in the diagram. In particular, there are two independent topologies and each one comes with the three known channels.
Here we follow the procedure and the notation of \cite{IrgesFotis}. So, starting with the Candy topology notice that its generic 
momentum dependence is ${\cal B}^C(L_1)$ with $L_1 = \sqrt{s}, \sqrt{t}, \sqrt{u}$ thus for the $s$-channel we get that 
\vskip .5cm
\begin{center}
\begin{tikzpicture}[scale=0.7]
\draw [photon] (0.9,0)--(2.5,1.5);
\draw [photon] (0.9,0)--(2.5,-1.5);
\draw [photon] (-2.5,1.5)--(-0.9,0);
\draw [photon] (-2.5,-1.5)--(-0.9,0);
\node at (2.7,1.1) {$l_3$};
\node at (2.7,-1.1) {$l_4$};
\node at (-2.5,1.1) {$l_1$};
\node at (-2.5,-0.9) {$l_2$};
\node at (0,0.9) {$ > $};
\node at (0,-0.9) {$ < $};
\draw [dashed,thick] (0,0) circle [radius=0.9];
\node at (0,1.4) {$q+L_1$};
\node at (0,-1.4) {$q$};
\node at (-1,0.6) {$\m$};
\node at (-1,-0.6) {$\n$};
\node at (1,0.6) {$\a$};
\node at (1,-0.6) {$\b$};
\node at (5,0) {$=\, \,i {\cal B}^{\phi \bar \phi,s}_{4 ph.,\m\n\a\b}\, , \hskip 0.5 cm S^A_{{\cal B}^C}= 1 $  \, .};
\end{tikzpicture}
\end{center}
%-------------------------------------
Using the Feynman rules from Appendix \ref{F.rules}, its explicit form reads
\bea
i {\cal B}^{\phi \bar \phi,s}_{4 ph.,\m\n\a\b} &=& (2i g^2\g g_{\m\n}) (2i g^2\g g_{\a\b}) \int \frac{d^4 q}{(2\pi)^4} \frac{i}{q^2}\frac{i}{(q+L_1)^2} \Rightarrow \nonumber\\
{\cal B}^{\phi \bar \phi,s}_{4 ph.,\m\n\a\b} &=& 4 g^4\g^2 g_{\m\n} g_{\a\b} \int \frac{d^4 q}{(2\pi)^4 i} \frac{1}{q^2(q+L_1)^2} \, ,
\eea
where $L_1 = l_1 + l_2$.
Using PV reduction formulae the above integral becomes
\bea\label{s.candy}
{\cal B}^{\phi \bar \phi,s}_{4 ph.,\m\n\a\b} = 4 g^4\g^2 g_{\m\n} g_{\a\b} B_0(q, q+L_1)\, .
\eea
Adding up the three channels we get the complete $C$-Box contribution given by
\bea\label{f.candies}
{\cal B}^{C}_{4 ph.,\m\n\a\b} = 4 g^4\g^2 g_{\m\n} g_{\a\b} \Bigl [ B_0(q, q+\sqrt{s}) + B_0(q, q+\sqrt{t}) + B_0(q, q+\sqrt{u})  \Bigr ]\, .
\eea
Next, we consider the $T$-Boxes which are determined by two linear combination of the external momenta $L_1$ and $L_2$. 
A consistent choice for $(L_1,L_2)$ for the channels $T_{1,\cdot \cdot \cdot,6}$ is
$T_1: ( l_1+l_2, l_1+l_2+l_3 )$, $T_2: ( l_1+l_3, l_1+l_3+l_4 )$, $T_3: ( l_1+l_4, l_1+l_3+l_4 )$, $T_4: (l_2, l_1+l_2 )$, $T_5: ( l_1,l_1 +l_3 )$ and $T_6: ( l_1,l_1 +l_4 )$.
Since the generic $T$-Box is of the form ${\cal B}^T(L_1,L_2)$ it is enough to calculate only one diagram and then take the sum 
over the six different pairs $(L_1,L_2)$. Here we choose to evaluate the following $s$-channel diagram:
%-------------------------------------
\vskip .5cm
\begin{center}
\begin{tikzpicture}[scale=0.7]
\draw [dashed,thick] (-1,0) -- (1,1);
\draw [dashed,thick] (-1,0) -- (1,-1);
\draw [dashed,thick] (1,1) -- (1,-1);
%%%
\draw [photon] (-2,1) -- (-1,0);%p1
\draw [photon] (-2,-1) -- (-1,0);%p2
\node at (-1,0.5) {$\m$};
\node at (-1,-0.5) {$\n$};
\node at (1,1.5) {$\a$};
\node at (1,-1.5) {$\b$};
\draw [photon] (1,1) -- (2.5,1);%p3
\draw [photon] (1,-1) -- (2.5,-1);%p4
%%%
\node at (1.7,1.5) {$l_3$};
\node at (1.7,-1.5) {$l_4$};
\node at (-1.9,1.5) {$l_1$};
\node at (-1.9,-1.50) {$l_2$};
%%%
\node at (-0.2,1.3) {$q+L_1$ };
\node at (0,-0.9) {$q$};
\node at (2.0,0) {$q+L_2$};
%%%
%%%
\draw [<-, very thick] (0.2,0.6) -- (0,0.5);
\draw [->, very thick] (0.2,-0.6) -- (0,-0.5);
\draw [<-, very thick] (1,-0.1) -- (1,0.1);
\node at (6.5,0) {$=\,\, i{\cal B}^{\phi \phi \phi}_{4ph.,\m\n\a\b}\, , \hskip 0.5cm  S^A_{{\cal B}^T} = 1 $ \, .};
\end{tikzpicture}
\end{center}
%-------------------------------------
It is equal to
\bea
i{\cal B}^{\phi \phi \phi,s}_{4ph.,\m\n\a\b} &=& (2i g^2\g g_{\m\n}) (i g \sqrt{\g} ) (i g \sqrt{\g} ) \int \frac{d^4 q}{(2\pi)^4} \frac{i(2q + L_1 +L_2)_\a}{q^2}\frac{i(2q +L_2)_\b}{(q+L_1)^2}\frac{i}{(q+L_2)^2} \Rightarrow \nonumber\\
{\cal B}^{\phi \phi \phi,s}_{4 ph.,\m\n\a\b} &=& -2 g^4\g^2 g_{\m\n} \int \frac{d^4 q}{(2\pi)^4 i} \frac{(2q + L_1 +L_2)_\a (2q +L_2)_\b}{q^2(q+L_1)^2(q+L_2)^2} \, ,
\eea
and using PV reduction formulae it becomes
\bea\label{s.triangle}
{\cal B}^{\phi \phi \phi,s}_{4ph.,\m\n\a\b} &=& -2 g^4\g^2 g_{\m\n} \Bigl [ 4 C_{\a\b}(q,q+L_1,q+L_2) + (2 L_1 + 4 L_2 )_\a C_{\b}(q,q+L_1,q+L_2)  \nonumber\\
&+& L_{2,\a} ( L_1 + L_2  )_\b C_0(q,q+L_1,q+L_2)    \Bigr ]\, .
\eea  
Considering the sum over the six different topologies, the complete $T$-Box contribution is given by
\bea\label{f.Tboxes}
{\cal B}^{T}_{4ph.,\m\n\a\b} = \sum_{(L_1,L_2)} {\cal B}^{\phi \phi \phi,s}_{4ph.,\m\n\a\b}(L_1,L_2)\, .
\eea
Regarding the four-photon scattering amplitude there is one more set of one-loop diagrams, the $S$-Boxes. 
These are determined by three linear combinations of the external momenta, $(L_1,L_2,L_3)$ and in this case there are 
only three topologies corresponding to the usual $s$, $t$ and $u$-channels. So, a proper choice for the $(L_1,L_2,L_3)$ 
for the three channels $S_{1,2,3}$ is $ S_1 = (l_1,l_1+l_3,l_1+l_3+l_4) $, $ S_2 = (l_3,l_1+l_3,l_1+l_3+l_4) $ and 
$ S_3 = (l_1,l_1+l_4,l_1+l_3+l_4) $.
In order to take into account all the channels it is enough to calculate one of 
them and then take the sum over the three momenta, $(L_1,L_2,L_3)$. Let us start by evaluating the $s$-channel $S$-Box which is given by
%-------------------------------------
\vskip .5cm
\begin{center}
\begin{tikzpicture}[scale=0.7]
\draw [dashed] (-1,1) -- (1,1);
\draw [dashed] (1,1) -- (1,-1);
\draw [dashed] (1,-1) -- (-1,-1);
\draw [dashed] (-1,-1) -- (-1,1);
%%%
\draw [photon] (-2,2) -- (-1,1);%p1
\draw [photon] (-2,-2) -- (-1,-1);%p2
\draw [photon] (2,2) -- (1,1);%p3
\draw [photon] (2,-2) -- (1,-1);%p4
%%%
%%%
\draw [->, very thick] (-0.2,1) -- (0,1);
\draw [>-, very thick] (-0.2,-1) -- (0,-1);
\draw [<-, very thick] (-1,0.2) -- (-1,0);
\draw [<-, very thick] (1,-0.1) -- (1,0.1);\nonumber
%
%%%
\node at (-1.4,0.9) {$\m$};
%
%%%
\node at (-1.4,-0.9) {$\n$};
%
%%%
\node at (1.4,0.9) {$\a$};
%
%%%
\node at (1.4,-0.9) {$\b$};
%
%%%
\node at (-2.3,2.2) {$l_1$};
\node at (-2.3,-2.2) {$l_2$};
\node at (2.3,2.2) {$l_3$};
\node at (2.3,-2.2) {$l_4$};
%
%%%
\node at (0,1.5) {$q+L_1$ };
\node at (-1.3,0) {$q $};
\node at (0,-1.4) {$q +L_3$};
\node at (2,0) {$q+ L_2$};
%%%
\node at (7,0) {$=\,\, i{\cal B}^{\phi \phi \phi \phi}_{4ph.,\m\n\a\b}\, , \hskip 0.5cm  S^A_{{\cal B}^S} = 1 \, ,$};
\end{tikzpicture}
\end{center}
%-------------------------------------
and its explicit form reads
\bea
i{\cal B}^{\phi \phi \phi \phi,s}_{4ph.,\m\n\a\b} &=& (i g \sqrt{\g} ) (i g \sqrt{\g} ) (i g \sqrt{\g} ) (i g \sqrt{\g} ) \int \frac{d^4 q}{(2\pi)^4} \frac{i(2q + L_1)_\m}{q^2}\frac{i(2q +L_1 + L_2)_\a}{(q+L_1)^2} \nonumber\\
&\times& \frac{i(2q + L_2 +L_3)_\b}{(q+L_2)^2}\frac{i(2q +L_2)_\n}{(q+L_3)^2} \Rightarrow \nonumber\\
{\cal B}^{\phi \phi \phi \phi,s}_{4ph.,\m\n\a\b} &=& g^4\g^2 \int \frac{d^4 q}{(2\pi)^4 i} \frac{(2q + L_1)_\m(2q + L_1 +L_2)_\a(2q + L_2 +L_3)_\b (2q +L_3)_\n}{q^2(q+L_1)^2(q+L_2)^2(q+L_3)^2} \, .\nonumber\\
\eea
Using the PV reduction, the above integral becomes 
\bea\label{s.Rec.}
{\cal B}^{\phi \phi \phi \phi,s}_{4ph.,\m\n\a\b} &=& g^4\g^2 \Bigl [ 16 D_{\m\n\a\b}(q,q+L_1,q+L_2,q+L_3) +8  L_{3,\a} D_{\b\m\n}(q,q+L_1,q+L_2,q+L_3) \nonumber\\
&+&  8 ( L_2 + L_3)_\b D_{\m\n\a}(q,q+L_1,q+L_2,q+L_3) + 8 L_{1,\m} D_{\b\n\a}(q,q+L_1,q+L_2,q+L_3) \nonumber\\
&+&  8( L_1 + L_2)_\n D_{\b\m\a}(q,q+L_1,q+L_2,q+L_3) \nonumber\\
&+& 4 ( L_{3,\a}L_{3,\b} + L_{2,\a}L_{3,\b}  ) D_{\m\n}(q,q+L_1,q+L_2,q+L_3)  \nonumber\\
&+& 4 ( L_{1,\m}L_{3,\a} + L_{1,\m}L_{2,\a}  ) D_{\n\b}(q,q+L_1,q+L_2,q+L_3) \nonumber\\
&+& 4 L_{1,\m}L_{3,\b} D_{\a\n}(q,q+L_1,q+L_2,q+L_3) \nonumber\\
&+& 4 ( L_{1,\a}L_{3,\n} + L_{2,\a}L_{3,\n} + L_{1,\a}L_{2,\n} + L_{2,\a}L_{2,\n}  ) D_{\m\b}(q,q+L_1,q+L_2,q+L_3) \nonumber\\
&+& 4 ( L_{1,\n}L_{3,\b} + L_{2,\n}L_{3,\b}  ) D_{\m\a}(q,q+L_1,q+L_2,q+L_3) \nonumber\\
&+& 4 ( L_{1,\m}L_{1,\n} + L_{1,\m}L_{2,\n}  ) D_{\a\b}(q,q+L_1,q+L_2,q+L_3) \nonumber\\
&+& 2 ( L_{1,\m}L_{2,\a}L_{3,\b} + L_{1,\m}L_{3,\b}L_{3,\a}  ) D_{\n}(q,q+L_1,q+L_2,q+L_3) \nonumber\\
&+& 2 ( L_{1,\n}L_{3,\b}L_{3,\a} + L_{2,\n}L_{3,\b}L_{3,\a} + L_{1,\n}L_{2,\a}L_{3,\b} + L_{2,\a}L_{2,\n}L_{3,\b}  ) D_{\m}(q,q+L_1,q+L_2,q+L_3) \nonumber\\
&+& 2 ( L_{1,\m}L_{1,\n}L_{3,\a} + L_{1,\m}L_{2,\n}L_{3,\a} + L_{1,\m}L_{1,\n}L_{2,\a} + L_{1,\m}L_{2,\n}L_{2,\a}  ) D_{\b}(q,q+L_1,q+L_2,q+L_3) \nonumber\\
&+& 2 ( L_{1,\m}L_{1,\n}L_{3,\b} + L_{1,\m}L_{2,\n}L_{3,\b} ) D_{\a}(q,q+L_1,q+L_2,q+L_3) \nonumber\\
&+& L_{1,\m}L_{3,\b} ( L_{1,\n}L_{2,\a} + L_{2,\n}L_{2,\a} + L_{1,\n}L_{3,\a} + L_{2,\n}L_{3,\a}  )D_0(q,q+L_1,q+L_2,q+L_3)  \Bigr] \, .
\eea
The complete $S$-Box contribution comes by summing the three different channels and is given by
\bea\label{f.Rboxes}
{\cal B}^{S}_{4ph.,\m\n\a\b} = \sum_{(L_1,L_2,L_3)} {\cal B}^{\phi \phi \phi \phi,s}_{4ph.,\m\n\a\b}(L_1,L_2,L_3) \, ,
\eea
while the full contribution to the $\g \g \to \g \g$ scattering amplitude is given by adding the $C$-, $T$- and $S$-Boxes and reads
\bea\label{ctr.box.g}
{\cal B}_{4ph.,\m\n\a\b} = {\cal B}^{C}_{4ph.,\m\n\a\b} + {\cal B}^{T}_{4ph.,\m\n\a\b} + {\cal B}^{S}_{4ph.,\m\n\a\b}\, .
\eea

%---------------------------------------------------------------------------------------------------------------------------------------------------
\subsection{Four-scalar Diagrams }\label{4phi}
%---------------------------------------------------------------------------------------------------------------------------------------------------

Apart from the four-photon scattering amplitudes, the boundary action contains also a set of one-loop diagrams
contributing to the $\phi \bar \phi \to \phi \bar \phi $ process. In particular, this set contains the following diagrams
\bea
%-------------------------------------
\begin{tikzpicture}[scale=0.7]
\draw [dashed] (1,0)--(3,-1.5);
\draw [dashed] (1,0)--(3,1.5);
\draw [photon] (0,0) circle [radius=1];
\draw [dashed] (-3,-1.5)--(-1,0);
\draw [dashed] (-3,1.5)--(-1,0);
\draw [<-, very thick ] (-1.65,0.45)--(-1.85,0.65);
\draw [->, very thick ] (-1.65,-0.45)--(-1.85,-0.65);
\draw [->, very thick ] (1.65,0.45)--(1.85,0.65);
\draw [<-, very thick ] (1.65,-0.45)--(1.85,-0.65);
\end{tikzpicture}
\, , \hskip .3 cm
%-------------------------------------
\begin{tikzpicture}[scale=0.7]
\draw [photon] (-1,0) -- (1,1);
\draw [photon] (-1,0) -- (1,-1);
\draw [dashed] (1,1) -- (1,-1);
\draw [<-, very thick ] (-1.65,0.45)--(-1.85,0.65);
\draw [->, very thick ] (-1.65,-0.45)--(-1.85,-0.65);
%%%
\draw [dashed] (1,1) -- (2.5,1);%p3
\draw [dashed] (1,-1) -- (2.5,-1);%p4
%%%
\draw [dashed] (-3,-1.5)--(-1,0);
\draw [dashed] (-3,1.5)--(-1,0);
%%%
\draw [->, very thick] (1,-0.1) -- (1,0.1);
\draw [<-, very thick] (1.7,1) -- (1.5,1);
\draw [<-, very thick] (1.5,-1) -- (1.7,-1);
\end{tikzpicture}
%-------------------------------------
\, , \hskip .3 cm
%-------------------------------------
\begin{tikzpicture}[scale=0.7]
\draw [dashed] (-1,1) -- (1,1);
\draw [photon] (1,1) -- (1,-1);
\draw [dashed] (1,-1) -- (-1,-1);
\draw [photon] (-1,-1) -- (-1,1);
%%%
\draw [dashed] (-2,2) -- (-1,1);%p1
\draw [dashed] (-2,-2) -- (-1,-1);%p2
\draw [dashed] (2,2) -- (1,1);%p3
\draw [dashed] (2,-2) -- (1,-1);%p4
%%%
\draw [->, very thick] (-0.2,1) -- (0,1);
\draw [<-, very thick] (-0.2,-1) -- (0,-1);
%
%%%
\draw [<-, very thick] (-1.5,1.5) -- (-1.7,1.6);
\draw [->, very thick] (1.5,1.5) -- (1.7,1.6);
\draw [->, very thick] (-1.5,-1.5) -- (-1.7,-1.6);
\draw [<-, very thick] (1.5,-1.5) -- (1.7,-1.6);\nonumber
%%%
\end{tikzpicture}
%-------------------------------------
\eea
where in this case, $C$- and $T$-Boxes come in two possible channels, $s$ and $t$, while the $S$-Boxes come in only one channel, $s$. 
The reason for this is that the external legs are particle-antiparticle pairs so they cannot be interchanged.
Now, notice here that there are four $T$-Boxes, since they have two independent topologies and each
comes with the two channels mentioned previously. Moreover, notice that also for the $S$-Boxes there are two independent topologies, 
since there are two possible ways to arrange the propagators inside the loop, while each of them comes in one channel. 
So, with these in mind let us start with the $s$-channel of the $C$-Box, which diagrammatically reads 
\vskip .5cm
\begin{center}
\begin{tikzpicture}[scale=0.7]
\draw [dashed] (0.9,0)--(2.5,1.5);
\draw [dashed] (0.9,0)--(2.5,-1.5);
\draw [dashed] (-2.5,1.5)--(-0.9,0);
\draw [dashed] (-2.5,-1.5)--(-0.9,0);
\node at (2.7,1.1) {$l_3$};
\node at (2.7,-1.1) {$l_4$};
\node at (-2.5,1.1) {$l_1$};
\node at (-2.5,-0.9) {$l_2$};
\draw [photon] (0,0) circle [radius=0.9];
\node at (0,1.4) {$q+L_1$};
\node at (0,-1.4) {$q$};
\node at (-1,0.6) {$\m$};
\node at (-1,-0.6) {$\n$};
\node at (1,0.6) {$\a$};
\node at (1,-0.6) {$\b$};
\draw [<-, very thick ] (-1.65,0.65)--(-1.85,0.85);
\draw [->, very thick ] (-1.65,-0.65)--(-1.85,-0.85);
\draw [->, very thick ] (1.65,0.65)--(1.85,0.85);
\draw [<-, very thick ] (1.65,-0.65)--(1.85,-0.85);
\node at (5,0) {$=\, \,i {\cal B}^{A A,s}_{4\phi}\, , \hskip 0.5 cm S^\phi_{{\cal B}^C}= 1 $  \, .};
\end{tikzpicture}
\end{center}
%-------------------------------------
Its explicit form is given by
\bea
i {\cal B}^{A A,s}_{4\phi} &=&  (2i g^2\g g_{\m\n}) (2i g^2\g g_{\a\b}) \int \frac{d^4 q}{(2\pi)^4} \frac{-i g^{\m\a}}{q^2}\frac{-ig^{\n\b}}{(q+L_1)^2} \Rightarrow \nonumber\\
{\cal B}^{A A,s}_{4\phi} &=& 4 d g^4\g^2 \int \frac{d^4 q}{(2\pi)^4 i} \frac{1}{q^2(q+L_1)^2} \, ,
\eea
while in DR the above integral reads
\bea
{\cal B}^{A A,s}_{4\phi} &=& 4 d g^4\g^2 B_0(q,q+L_1)\, .
\eea
The complete $C$-Box contribution comes by summing over the two channels giving
\bea\label{c.box.phi}
{\cal B}^{C}_{4\phi} &=& 4 d g^4\g^2 \Bigl [  B_0(q,q+\sqrt{s}) + B_0(q,q+\sqrt{t})  \Bigr ]\, .
\eea
Next, we move on to the $T$-Boxes which, in this case, are determined by two pairs of two linear combinations of the external momenta, 
$(L_1,L_2)$ and $(L_A,L_B)$. A good choice for these pairs for the channels $T_{1,\cdots,4}$ is the following
\bea\label{t1234}
T_1 &:& (l_1+l_2, l_1 +l_2 + l_3)\, , \hskip .5 cm (l_1 + l_2 +2 l_3,l_1 +l_2 +l_3 -l_4) \nonumber\\
T_2 &:& (l_1+l_3, l_1 +l_3 + l_4)\, , \hskip .5 cm (l_1 + l_3 +2 l_4,0) \nonumber\\
T_3 &:& (l_2, l_1 +l_2)\, , \hskip .5 cm (2 l_2 ,l_2 -l_1) \nonumber\\
T_4 &:& (l_1, l_1 + l_3)\, , \hskip .5 cm (2l_1,l_1 +l_3 )\, .
\eea
The first $s$-channel here is given by
%-------------------------------------
\vskip .5cm
\begin{center}
\begin{tikzpicture}[scale=0.7]
\draw [photon] (-1,0) -- (1,1);
\draw [photon] (-1,0) -- (1,-1);
\draw [dashed] (1,1) -- (1,-1);
%%%
\draw [dashed] (-2,1) -- (-1,0);%p1
\draw [dashed] (-2,-1) -- (-1,0);%p2
\node at (-1,0.5) {$\m$};
\node at (-1,-0.5) {$\n$};
\node at (1,1.5) {$\a$};
\node at (1,-1.5) {$\b$};
\draw [dashed] (1,1) -- (2.5,1);%p3
\draw [dashed] (1,-1) -- (2.5,-1);%p4
%%%
\node at (1.7,1.5) {$l_3$};
\node at (1.7,-1.5) {$l_4$};
\node at (-1.9,1.5) {$l_1$};
\node at (-1.9,-1.50) {$l_2$};
%%%
\node at (-0.2,1.3) {$q+L_1$ };
\node at (0,-0.9) {$q$};
\node at (2.0,0) {$q+L_2$};
%%%
\draw [<-, very thick ] (-1.65,0.65)--(-1.85,0.85);
\draw [->, very thick ] (-1.65,-0.65)--(-1.85,-0.85);
\draw [<-, very thick] (1.7,1) -- (1.5,1);
\draw [<-, very thick] (1.5,-1) -- (1.7,-1);
%%%
\draw [->, very thick] (1,-0.1) -- (1,0.1);
\node at (6.5,0) {$=\,\, i{\cal B}^{A A \phi,s}_{4\phi}\, , \hskip 0.5cm  S^\phi_{{\cal B}^T} = 1 $ \, ,};
\end{tikzpicture}
\end{center}
%-------------------------------------
and it is equal to
\bea
i{\cal B}^{A A \phi,s}_{4\phi} &=& (2i g^2\g g_{\m\n}) (i g \sqrt{\g} ) (i g \sqrt{\g} ) \int \frac{d^4 q}{(2\pi)^4} \frac{-ig^{\m\a}(q + L_A)_\a}{q^2}\frac{-ig^{\n\b}(q +L_B)_\b}{(q+L_1)^2}\frac{i}{(q+L_2)^2} \Rightarrow \nonumber\\
{\cal B}^{A A \phi,s}_{4\phi} &=& -2 g^4\g^2 \int \frac{d^4 q}{(2\pi)^4 i} \frac{(q + L_A) \cdot (q +L_B)}{q^2(q+L_1)^2(q+L_2)^2} \, .
\eea
In DR the above integral becomes
\bea
{\cal B}^{A A \phi,s}_{4\phi} &=& -2 g^4\g^2  \Bigl [ B_0(q,q+L_2-L_1) + ( L_A + l_B )_\m C^\m(q,q+L_1,q+L_2) \nonumber\\
&+& ( L_A \cdot L_B )C_0(q,q+L_1,q+L_2)      \Bigr ]\, .
\eea
The complete $T$-Box contribution is given by
\bea\label{t.box.phi}
{\cal B}^{T}_{4\phi} = \sum_{(L_A,L_B)} \sum_{(L_1,L_2)} {\cal B}^{A A \phi,s}_{4\phi} ( L_1,L_2,L_A,L_B ) \, . 
\eea
The last set of diagrams contains the $S$-Boxes. On the one hand there 
is only one channel and as a consequence we have a unique choice for $(L_1,L_2,L_3)$, say $( l_1,l_1+l_3, l_1 +l_3 + l_4 )$. On the other hand 
these $S$-Boxes are determined by four linear combinations of the external momenta, $L_A$, $L_B$, $L_C$ and $L_D$. 
Thus, for the two possible topologies that we have here a proper choice for $L_{A,\cdot \cdot \cdot D}$ is the following
\bea
L_A &=& 2 l_1  \, ,  \hskip .5cm  -l_1 \nonumber\\ 
L_B &=& l_1 -l_3  \, , \hskip .5cm  l_1  \nonumber\\ 
L_C &=& l_1 +l_3 + 2 l_4  \, , \hskip .5cm l_1 + l_3 + l_4 \nonumber\\ 
L_D &=&  l_1 + l_3 + l_4 - l_2 \, , \hskip .5cm  l_2 \, . 
\eea  
So, the first diagram is
%-------------------------------------
\vskip .5cm
\begin{center}
\begin{tikzpicture}[scale=0.7]
\draw [dashed] (-1,1) -- (1,1);
\draw [photon] (1,1) -- (1,-1);
\draw [dashed] (1,-1) -- (-1,-1);
\draw [photon] (-1,-1) -- (-1,1);
%%%
\draw [dashed] (-2,2) -- (-1,1);%p1
\draw [dashed] (-2,-2) -- (-1,-1);%p2
\draw [dashed] (2,2) -- (1,1);%p3
\draw [dashed] (2,-2) -- (1,-1);%p4
%%%
%%%
\draw [->, very thick] (-0.2,1) -- (0,1);
\draw [<-, very thick] (-0.2,-1) -- (0,-1);
%
%%%
\node at (-1.4,0.9) {$\m$};
%
%%%
\node at (-1.4,-0.9) {$\n$};
%
%%%
\node at (1.4,0.9) {$\a$};
%
%%%
\node at (1.4,-0.9) {$\b$};
%
%%%
\node at (-2.3,2.2) {$l_1$};
\node at (-2.3,-2.2) {$l_2$};
\node at (2.3,2.2) {$l_3$};
\node at (2.3,-2.2) {$l_4$};
%
%%%
\node at (0,1.5) {$q+L_1$ };
\node at (-1.3,0) {$q $};
\node at (0,-1.4) {$q +L_3$};
\node at (2,0) {$q+ L_2$};
%%%
%%%
\draw [<-, very thick] (-1.5,1.5) -- (-1.7,1.65);
\draw [->, very thick] (1.5,1.5) -- (1.7,1.65);
\draw [->, very thick] (-1.5,-1.5) -- (-1.7,-1.65);
\draw [<-, very thick] (1.5,-1.5) -- (1.7,-1.65);\nonumber
%%%
\node at (7,0) {$=\,\, i{\cal B}^{A \phi A \phi,s}_{4\phi}\, , \hskip 0.5cm  S^\phi_{{\cal B}^S} = 1 \, ,$};
\end{tikzpicture}
\end{center}
%-------------------------------------
and it is equal to
\bea
i{\cal B}^{A \phi A \phi,s}_{4\phi} &=& (i g \sqrt{\g} ) (i g \sqrt{\g} ) (i g \sqrt{\g} ) (i g \sqrt{\g} ) \int \frac{d^4 q}{(2\pi)^4} \frac{-i g^{\a\b} (q + L_A)_\m}{q^2}\frac{i(-q - L_B)_\a}{(q+L_1)^2} \nonumber\\
&\times& \frac{-ig^{\m\n}(q + L_C)_\b}{(q+L_2)^2}\frac{i(-q - L_D)_\n}{(q+L_3)^2} \Rightarrow \nonumber\\
{\cal B}^{A \phi A \phi,s}_{4\phi} &=& g^4\g^2 \int \frac{d^4 q}{(2\pi)^4 i} \frac{(q + L_A)\cdot(q +L_D) (q + L_B)\cdot(q + L_C) }{q^2(q+L_1)^2(q+L_2)^2(q+L_3)^2} \, .
\eea
The above integral in DR becomes
\bea
{\cal B}^{A \phi A \phi,s}_{4\phi} &=& g^4\g^2 \Bigl [ g_{\m\n}g_{\a\b} D^{\m\n\a\b}(q,q+L_1,q+L_2,q+L_3) \nonumber\\
&+& g_{\m\n}( L_A +L_B +L_C +L_D )_\a D^{\m\n\a}(q,q+L_1,q+L_2,q+L_3) \nonumber\\ 
&+&g_{\m\n}( L_B\cdot L_C + L_A\cdot L_D ) D^{\m\n}(q,q+L_1,q+L_2,q+L_3) \nonumber\\ 
&+& ( L_{C,\a}L_{D,\b} + L_{B,\a}L_{D,\b} + L_{A,\a}L_{C,\b} + L_{A,\a}L_{B,\b}) D^{\a\b}(q,q+L_1,q+L_2,q+L_3) \nonumber\\
&+&(L_B\cdot L_C L_{D,\m} + L_A\cdot L_D L_{C,\m} + L_A\cdot L_D L_{B,\m} + L_B\cdot L_C L_{A,\m}  )D^{\m}(q,q+L_1,q+L_2,q+L_3) \nonumber\\
&+& L_A\cdot L_D L_B\cdot L_C D_0(q,q+L_1,q+L_2,q+L_3)      \Bigr ]\, ,
\eea 
while the complete $S$-Box contribution is taken by considering the sum over the two possible topologies:
\bea\label{r.box.phi}
{\cal B}^{S}_{4\phi} = \sum_{(L_A,L_B,L_C,L_D)} {\cal B}^{A \phi A \phi,s}_{4\phi}(L_A,L_B,L_C,L_D)\, .
\eea
The full one-loop contribution to the scalar potential is given by adding the $C$-, $T$- and $S$-Boxes:
\bea\label{ctr.box.phi}
{\cal B}_{4\phi} = {\cal B}^{C}_{4\phi} + {\cal B}^{T}_{4\phi} + {\cal B}^{S}_{4\phi}\, .
\eea

\end{appendices} 
%---------------------------------------------------------------------------------------------------------------------------------------------------

\end{document}